%% file: nn_to_bind_or_not.tex
\documentclass[aps,prd,twocolumn,tightenlines,preprintnumbers,showpacs,superscriptaddress,notitlepage,nofootinbib,eqsecnum,floatfix,longbibliography,10pt]{revtex4-1}

\input{preamble}

\input{def}

\input{affiliations}

\newcount\hour \newcount\hourminute \newcount\minute
\hour=\time \divide \hour by 60
\hourminute=\hour \multiply \hourminute by 60
\minute=\time \advance \minute by -\hourminute

\begin{document}

\title{Di-nucleons do not form bound states at heavy pion mass}
\author{John Bulava}
\affiliation{\bochum}

\author{M.A. Clark}
\affiliation{\nvidia}

\author{Arjun S. Gambhir}
\affiliation{\llnldesign}

\author{Andrew~D.~Hanlon}
\affiliation{\kentState}
\affiliation{\cmu}

\author{Ben H\"{o}rz}
\affiliation{\intelDEUTCH}

\author{B\'alint~Jo\'o}
\affiliation{\nvidia}

\author{Christopher~K\"orber}
\affiliation{\bochum}

\author{Ken McElvain}
\affiliation{\ucb}

\author{Aaron S. Meyer}
\affiliation{\llnl}

\author{Henry Monge-Camacho}
\affiliation{\ornl}

\author{Colin~Morningstar}
\affiliation{\cmu}

\author{Joseph Moscoso}
\affiliation{\unc}
\affiliation{\lblnsd}

\author{Amy~Nicholson}
\affiliation{\unc}

\author{Fernando Romero-L\'opez}
\affiliation{\bern}

\author{Ermal~Rrapaj}
\affiliation{\lblnersc}

\author{Andrea Shindler}
\affiliation{\aachen}
\affiliation{\ucb}
\affiliation{\lblnsd}

\author{Sarah Skinner}
\affiliation{\cmu}

\author{Pavlos M. Vranas}
\affiliation{\llnl}
\affiliation{\lblnsd}

\author{Andr\'{e} Walker-Loud}
\affiliation{\lblnsd}
\affiliation{\ucb}

\collaboration{for the Baryon Scattering (BaSc) Collaboration}

\date{\today}

%\author{some authors}
%\affiliation{\ucb}
%\affiliation{\unc}

\begin{abstract}
We perform a high-statistics lattice QCD calculation of the low-energy two-nucleon scattering amplitudes.  In order to address discrepancies in the literature, the calculation is performed at a heavy pion mass in the limit that the light quark masses are equal to the physical strange quark mass, $m_\pi = m_K \simeq 714 $ MeV.  Using a state-of-the-art momentum space method, we rule out the presence of a bound di-nucleon in both the isospin 0 (deuteron) and 1 (di-neutron) channels, in contrast with many previous results that made use of compact hexaquark creation operators.  In order to diagnose the discrepancy, we add such hexaquark interpolating operators to our basis and find that they do not affect the determination of the two-nucleon finite volume spectrum, and thus they do not couple to deeply bound di-nucleons that are missed by the momentum-space operators.  Further, we perform a high-statistics calculation of the HAL QCD potential on the same gauge ensembles and find qualitative agreement with our main results.  We conclude that di-nucleons do not form bound states at heavy pion masses and that previous identification of deeply bound di-nucleons must have arisen from a misidentification of the spectrum from off-diagonal elements of a correlation function.
\end{abstract}

% make title
\maketitle
\preprint{LLNL-JRNL-2011688}

%\tableofcontents

%-------------------------------------------------------------------------------
%  Introduction
\section{Introduction\label{sec:intro}}
Our ability to predict the low-energy, two-nucleon (NN) forces directly from the underlying theory of strong interactions, Quantum Chromodynamics (QCD), remains an outstanding theoretical challenge.  Achieving such quantitative predictions from QCD will refine our description of matter in the universe, enabling us to improve our microscopic description of the reactions that power the stars and disentangle effects seen in laboratory detectors when rare processes occur.  

In order to carry out such predictions, lattice QCD (LQCD) can be used to make systematically improvable calculations of the properties and interactions of nucleons directly from the quark and gluon degrees of freedom of QCD.  Such calculations are carried out in Euclidean spacetime, where Monte Carlo methods can be used to stochastically evaluate QCD correlation functions as ratios of path integrals.
For single hadrons, the spectrum is easily determined from the long-time behavior of the correlation function.  However, for two hadrons, their scattering amplitudes are not directly accessible due to both the finite volume and the Euclidean metric. 

To access the two-particle scattering amplitudes, the finite-volume formalism in the form of the two-body Quantization Condition (QC2)~\cite{Luscher:1986pf,Luscher:1990ux} is used to relate the two-particle spectrum in finite volume to the infinite volume scattering amplitudes.  Specifically, what is required is a precise and accurate determination of the interacting energy of the system.  A significant challenge for two-nucleon (NN) calculations is that this interaction energy is $\mathrm{O}(0.1\%)$ of the total energy of the system.  Because LQCD is a relativistic quantum field theory (QFT), and calculations are carried out with quark and gluon degrees of freedom, we must compute the total energy of the system rather than having direct access to these small interacting energies.
This is not a problem in principle, but LQCD calculations of temporal correlation functions involving nucleons also suffer from an exponentially bad signal-to-noise (S/N) problem~\cite{Parisi:1983ae,Lepage:1989hd}: in the large time limit, the S/N degrades roughly as ${\rm S/N}\sim\sqrt{N_{\rm sample}}\,{\rm exp}(-A(m_N -\frac{3}{2}m_\pi)t)$ where $A$ is the number of nucleons.

For small Euclidean times, correlation functions receive significant contributions from excited states.  For single hadrons, the excited state gap is roughly set by $2m_\pi$. Thus, the ground state is resolvable over a Euclidean time of roughly $1-2$~fm.  In contrast, for NN calculations, the excited state gap is roughly set by the quantized momentum modes of the nucleon.  With periodic spatial boundary conditions, these are $\D E_n=2\sqrt{m_N^2+p_n^2}-2m_N\approx p_n^2/m_N$ with $p_n=\sqrt{n}(2\pi/L)$ for integer $n$.  For typical sizes of the lattice $L$, these energy gaps are $\mathrm{O}(20-50)$~MeV and thus, without a suppression of the excited states, the ground state is not resolvable until $4-10$~fm in time while the S/N for NN calculations has degraded around $t\sim2$~fm.
Thus, for correlation functions constructed from a single creation/annihilation operator, there is no region in time where the ground state dominates the signal and the S/N is sufficiently large that the ground state interaction energy can be precisely or accurately determined.

This S/N problem has contributed to discrepancies in the literature, even at heavy pion masses of ${m_\pi \approx 800}$~MeV, where the S/N challenge is significantly less severe. Notably, different groups report qualitatively conflicting results on whether di-nucleons form bound states, for example, see the 2012 and 2013 reviews from the annual lattice conference~\cite{Doi:2012ab,Walker-Loud:2014iea}.
In particular, after the first dynamical LQCD calculation of the NN system was performed~\cite{Beane:2006mx}, an alternative method was proposed to determine the NN interactions~\cite{Ishii:2006ec} which is known as the HAL QCD Potential~\cite{Ishii:2012ssm,Aoki:2020bew}.  Earlier calculations of NN systems indicated deeply bound di-nucleons~\cite{NPLQCD:2011naw,NPLQCD:2012mex,Yamazaki:2012hi,Yamazaki:2015asa,Berkowitz:2015eaa,Beane:2017edf,Wagman:2017tmp} while results with the HAL QCD potential found no bound states~\cite{Iritani:2016jie,Iritani:2017rlk,Iritani:2018vfn}.

This discrepancy was believed to be due to uncontrolled systematic uncertainties with the HAL QCD Potential method~\cite{Detmold:2007wk,Beane:2008ia,Beane:2008dv,Beane:2010em,Walker-Loud:2014iea,Savage:2016egr}.
At the same time, subsequent work by HAL QCD raised serious concerns about the calculations that observed bound states~\cite{Iritani:2016jie,Iritani:2017rlk,Aoki:2017byw,Iritani:2018vfn}.  This was followed up by two-baryon calculations that utilized a set of operators that enabled the use of momentum space creation operators, in contrast to the local hexa-quark (HX) creation operators used in Ref.~\cite{NPLQCD:2011naw,NPLQCD:2012mex,Yamazaki:2012hi,Yamazaki:2015asa,Berkowitz:2015eaa,Beane:2017edf,Wagman:2017tmp} (these operators will be defined subsequently).
These new methods found significantly reduced binding energies in the case of the $H$ dibaryon~\cite{Francis:2018qch,Green:2021qol}, and a lack of bound states in the case of di-nucleons~\cite{Horz:2020zvv,Amarasinghe:2021lqa}.
This discrepancy has long plagued progress toward the physical point and led to a lack of confidence in results reported for further two- and more-nucleon scattering amplitudes~\cite{Drischler:2019xuo,Tews:2022yfb}.

In this work, we address the systematic uncertainty arising from the methods used to determine the spectrum and the subsequent amplitudes by utilizing all classes of methods in the literature on the same set of LQCD gauge configurations, isolating this source of uncertainty.  
The calculation is carried out at a heavy pion mass in the $SU(3)$ flavor limit with $m_\pi=m_K\simeq714$~MeV, minimizing the S/N challenge as well as mimicking the pion mass point of previous calculations where discrepancies occur.
We find that di-nucleons do not form bound states on this ensemble with $\gtrsim5\s$ confidence for the deuteron and $\gtrsim2.5\s$ for the di-neutron.

%-------------------------------------------------------------------------------

%-------------------------------------------------------------------------------
%  NN with sLapH
\section{NN spectrum and amplitudes\label{sec:slaphnn}}

Scattering amplitudes are generally not directly accessible in LQCD calculations (though see Refs.~\cite{Bulava:2019kbi,Patella:2024cto} for a proposal to do so with smeared spectral functions).  Instead, the two-particle finite volume quantization condition~\cite{Luscher:1986pf,Luscher:1990ux} (QC2), often referred to as ``the L\"uscher method'', provides a mapping between the two-particle spectrum and the infinite volume scattering amplitudes: in essence, the two-particle interactions cause the energies to be shifted, in a finite-volume, from their corresponding non-interacting values providing a relationship between the scattering phase shifts and geometric functions that depend upon the spatial geometry and boundary conditions. 
Therefore, the essential ingredient in determining the scattering amplitudes with LQCD is the spectrum.  In particular, the small shift of the two-particle energy from the non-interacting levels must be determined.  For NN systems, $\D E_{NN} = E_{NN} - 2m_N$ is $\sim\mathrm{O}(0.1\%)$ of the total energy, presenting significant challenges for an accurate and precise determination given the known excited state contamination and S/N degradation challenges in LQCD calculations.

An emerging conclusion from recent literature~\cite{Francis:2018qch,Horz:2020zvv,Amarasinghe:2021lqa,Nicholson:2021zwi,Green:2022rjj,Tews:2022yfb,Green:2025rel} is that, to accurately identify the spectrum, it is necessary to use a basis of interpolating fields that includes momentum-space creation operators as opposed to the method of only using a spatially local HX like creation operator that was common initially.  All results in the literature have made use of momentum space annihilation operators.  Therefore, using the HX creation operators leads to off-diagonal correlators in the space of a creation/annihilation operator basis.  The complication with such correlation functions is that the overlap factors are not guaranteed to be positive.  This can lead to subtle cancellations between excited states and the ground state contributions to the correlation function such that a ``false plateau'' in the corresponding effective energy is identified as the lowest-lying state~\cite{Iritani:2016jie}.

There are several methods of constructing momentum-space creation operators which all require quark propagators from all spatial sites on the source timeslice to all spatial sites on the sink timeslice, so called all-to-all propagators.
The most common method in the literature is known as the distillation framework~\cite{HadronSpectrum:2009krc}.
A variant of this method known as the stochastic Laplacian Heaviside (sLapH) method~\cite{Morningstar:2011ka} is what we use in this work.
A different momentum sparsening method~\cite{Detmold:2019fbk} is utilized by NPLQCD~\cite{Amarasinghe:2021lqa,Detmold:2024iwz}.

We construct a set of NN operators as described in Ref.~\cite{Morningstar:2013bda}.
We use multi-hadron operators which are linear combinations of products of the individually momentum-projected constituent hadron operators.
Each multi-hadron operator has a given total momentum, and therefore the constituent hadrons in each term must have momentum that sums to this total.
The particular linear combination is determined by constructing an overall operator that transforms in a given irreducible representation (irrep) $\Lambda$ of the little group of the total momentum.
This allows a hadron in one term to carry a different momentum than in another term, so long as they are all related by rotations within the little group and sum to the same total momentum.  In \tabref{tab:nn_irreps}, we list the different irreps used for the isoscalar, spin triplet (deuteron) channel and the isovector, spin-singlet (dineutron) channel as well as the number of independent operators used for each irrep.

%------------------------------------------------------------------------------
% TABLE irreps
%------------------------------------------------------------------------------
\begin{table}[t]
\caption{\label{tab:nn_irreps}For a given total momentum $\mathbf{P}_{\rm tot} = \mathbf{d}_{\rm tot}\frac{2\pi}{L}$, where $\mathbf{d}_{\rm tot}$ is a vector of integers, we list the cubic irreps that use in this work that are expected to have a large overlap with S-wave scattering in the deuteron channel (left) and the dineutron channel (right).  The $g$ label denotes irreps of definite positive parity while the $u$ label denotes irreps of definite negative parity.  We also list the number of operators used for that irrep in order to build the correlator matrices and determine the spectrum.}
\begin{ruledtabular}
\begin{tabular}{ccccc}
&\multicolumn{2}{c}{deuteron}& \multicolumn{2}{c}{dineutron}\\
\cline{2-3}\cline{4-5}
$\mathbf{d}_{\rm tot}^2$& irrep& ${\rm N}_{\rm op}$& irrep& ${\rm N}_{\rm op}$\\
\hline
0& $T_{1g}$& 15& $A_{1g}$& \phantom{1}6 \\
1& $A_2$   & 10& $A_1$   & 10 \\
1& $E$     & 18\\
2& $A_2$   & 15& $A_1$   & 21 \\
2& $B_1$   & 19\\
2& $B_2$   & 21\\
3& $A_2$   & \phantom{1}9& $A_1$& \phantom{1}9 \\
3& $E$     & 17\\
4& $A_2$   & \phantom{1}7& $A_1$& 10\\
4& $E$     & 15
\end{tabular}
\end{ruledtabular}
\end{table}
%------------------------------------------------------------------------------

The energies of the stationary states in finite-volume are discrete and are determined using the
standard Hermitian correlator matrix 
method~\cite{Fox:1981xz,Ishikawa:1982tb,Michael:1982gb,Michael:1985ne,Luscher:1990ck}
in lattice QCD.  Given a set of $N_{\rm op}$ operators, an $N_{\rm op}\times N_{\rm op}$ matrix of two-point 
correlation functions is evaluated for a range of temporal separations $t$ using a Markov-chain Monte Carlo method.
The elements of the Hermitian correlator matrix have spectral representations given by
\begin{align}
\label{eq:Cij}
C_{ij}(t) &= \langle \Omega| O_i(t) O^\dagger_j(0) |\Omega \rangle,
\nonumber\\&= \sum_n Z_i^{(n)} Z_j^{\dagger (n)}e^{-E_n t} ,
\end{align}
where $|\O\rangle$ is the vacuum, the time-independent overlap factors are $Z_i^{(n)} = \langle \O| O_i|n\rangle$, and the sum runs over eigenstates of the Hamiltonian that have nonzero overlap with the states created by the operator basis.  Note that negligible temporal wrap-around contributions are omitted in Eq.~(\ref{eq:Cij}). It is not
practical to fit our results directly using Eq.~(\ref{eq:Cij}).  Instead, the correlator matrix is rotated to be as 
diagonal as possible.
The simplest way to do this, which we call the single pivot method, is to solve the generalized
eigenvalue problem (GEVP) using the correlator matrix at two different times
\begin{align}
C(t_d)\ V  = C(t_0)\ V\ \Lambda_d,
\label{eq:gevp1}
\end{align}
where the columns of $V$ contain the eigenvectors and $\Lambda_d$ is the diagonal matrix containing the corresponding eigenvalues.
The matrix $V$ should be normalized such that $V^\dagger C(t_0) V = I$, where $I$ is the identity matrix.  The matrix $V$ is determined only once using the mean values of the matrix elements of $C(t_d)$ and $C(t_0)$.
The following single-pivoted matrix $D(t)$ is then formed using
\begin{align}
 D(t) = V^\dagger C(t)  V\, .
 \label{eq:spivC}
\end{align}
The choice of the two times $t_d$ and $t_0$ must be made judiciously to ensure that $D(t)$ is diagonal
within statistical errors for a large range in $t$ well beyond $t_d$.  If this can be achieved, then
the single energy $E_l$ of the $l$-th eigenstate ($l=0,1,2,\dots$) of those which have overlap with the states created by the operator basis
can be obtained by a simple fit to the single diagonal element $D_{ll}(t)$, which takes the 
form
\begin{equation}
\label{eq:Dcorr}
D_{ll}(t) = |\tilde{Z}_{l}^{(l)}|^2 e^{-E_l t}\, +
    \sum_{n=N_{\rm op}}^\infty |\tilde{Z}_{l}^{(n)}|^2 e^{-E_n t}\, ,
\end{equation}
noting that the leading excited-state contamination comes from the $N_{\rm op}+1$ eigenstate that overlaps the states created by the operator basis~\cite{Blossier:2009kd}.  The overlap factors for the original operators are then given by
\begin{equation}
\label{eq:Z_jn}
Z_i^{(l)} = C_{ik}(t_0)\ V_{kl}\ \tilde{Z}_{l}^{(l)}\, .
\end{equation}
These overlap factors provide helpful information for identifying the eigenstates.

As we will discuss in more detail in \secref{sec:conspiracy}, the diagonal correlators in \eqnref{eq:Dcorr} still have non-trivial excited state contamination.  These effects must be included in the correlator analysis to obtain a robust extraction of the spectrum that incorporates systematic uncertainty associated with the excited state contamination.  Our analysis suggests they can be accurately modeled as inelastic excitations of the single nucleons used to construct the NN correlators.

Two other methods besides the single-pivot method are in common use.  One, known here as the rolling-pivot method, applies the GEVP of Eq.~(\ref{eq:gevp1}), for every time slice $t$, fixing only $t_0$, 
\begin{align}
C(t)\ V_R(t)  = C(t_0)\ V_R(t)\ \Lambda_{R}(t),
\label{eq:gevp2}
\end{align}
yielding a rolling-pivoted correlator matrix
\begin{equation}\label{eq:rolling_pivot_DR}
  D_R(t) = V_R^\dagger(t)\ C(t)\ V_R(t).
\end{equation}
The matrix $V_R(t)$ is determined only once for each $t$ using the mean values of the matrix elements of $C(t)$ and $C(t_0)$.  A similar but slightly different method, here known as the principal correlator method, uses the eigenvalues in $\Lambda_R(t)$ as the diagonal correlators to fit.  The GEVP must be solved for each bootstrap resampling to obtain error estimates of the eigenvalues.  Both the rolling-pivot and principal correlator methods suffer from the possible need to carry out eigenvector pinning to ensure a consistent identification of the eigenstates as both time and bootstrap resamplings change. Although we use the much simpler single-pivot method for our final results, we do also employ the rolling pivot to check for agreement of the results obtained with the single-pivot method.

The Hermitian correlator matrix method described above is sometimes referred to as a ``variational'' method in the literature. However, we emphasize that this should not be confused with the variational method in quantum mechanics, which is an approximation scheme used to obtain an energy bound by minimizing the expectation value of the Hamiltonian in a parametrized wave function. The variational method in quantum mechanics does not guarantee convergence to the lowest-lying energies of the system when using a fixed number of wave-function ansätze. By contrast, the pivoted matrix introduced above can be interpreted as a change of operator basis to one that is, in a sense, optimal from a variational perspective. In this case, however, Euclidean time projection ensures that, as long as the overlap of the operator basis with the lowest-lying state is non-zero, that state will be isolated at sufficiently large Euclidean time separations. Moreover, the functional form of the excited-state contamination is now explicitly determined by the Euclidean time dependence, and it is exponential with a large energy gap that depends on the size of the operator basis. Thus, the systematics of the two methods are quite different: the Hermitian correlator matrix method does not yield upper bounds, but rather estimates whose excited-state contamination can be systematically understood and suppressed via Euclidean time evolution. Furthermore, increasing the size of the operator basis improves control over this contamination by enlarging the energy gap that governs its suppression.

%------------------------------------------------------------------------------
\subsection{Details of the lattice calculation\label{sec:lqcd_details}}
%------------------------------------------------------------------------------
%------------------------------------------------------------------------------
\subsubsection{Lattice Action}
%------------------------------------------------------------------------------

This calculation uses the CLS lattice action~\cite{Bruno:2014jqa}, which is an $N_f=2+1$ isotropic clover-Wilson action that is non-perturbatively $\mathrm{O}(a)$ improved, and the L\"uscher-Weisz improved gluon action with tree-level coefficients was used
with periodic boundary conditions in all directions.  The lattice spacing is $a\approx0.086$~fm with a lattice volume of $48^3\times96$.  The three quark masses are all approximately equal to the physical strange quark mass, i.e. $m_u=m_d=m_s\approx m_s^{\rm phys}$, which leads to $m_\pi L\approx15$.
This ensemble was generated in Ref.~\cite{Horz:2020zvv} and named C103.%
%FOOTNOTE
\footnote{In the CLS naming scheme settled after they were generated, this ensemble would be named C150.} 
%-------------------------
A total of 1490 configurations are used from four replica.
The choice of quark masses was made to roughly match the SU(3) symmetric calculation of Ref.~\cite{NPLQCD:2012mex} where it was claimed that deeply bound di-nucleons existed~\cite{NPLQCD:2012mex,Berkowitz:2015eaa,Wagman:2017tmp}.

%------------------------------------------------------------------------------
\subsubsection{Interpolating operator construction}
%------------------------------------------------------------------------------

%----------------------------------------------------------------------------------
\begin{figure*}
\includegraphics[width=0.32\textwidth]{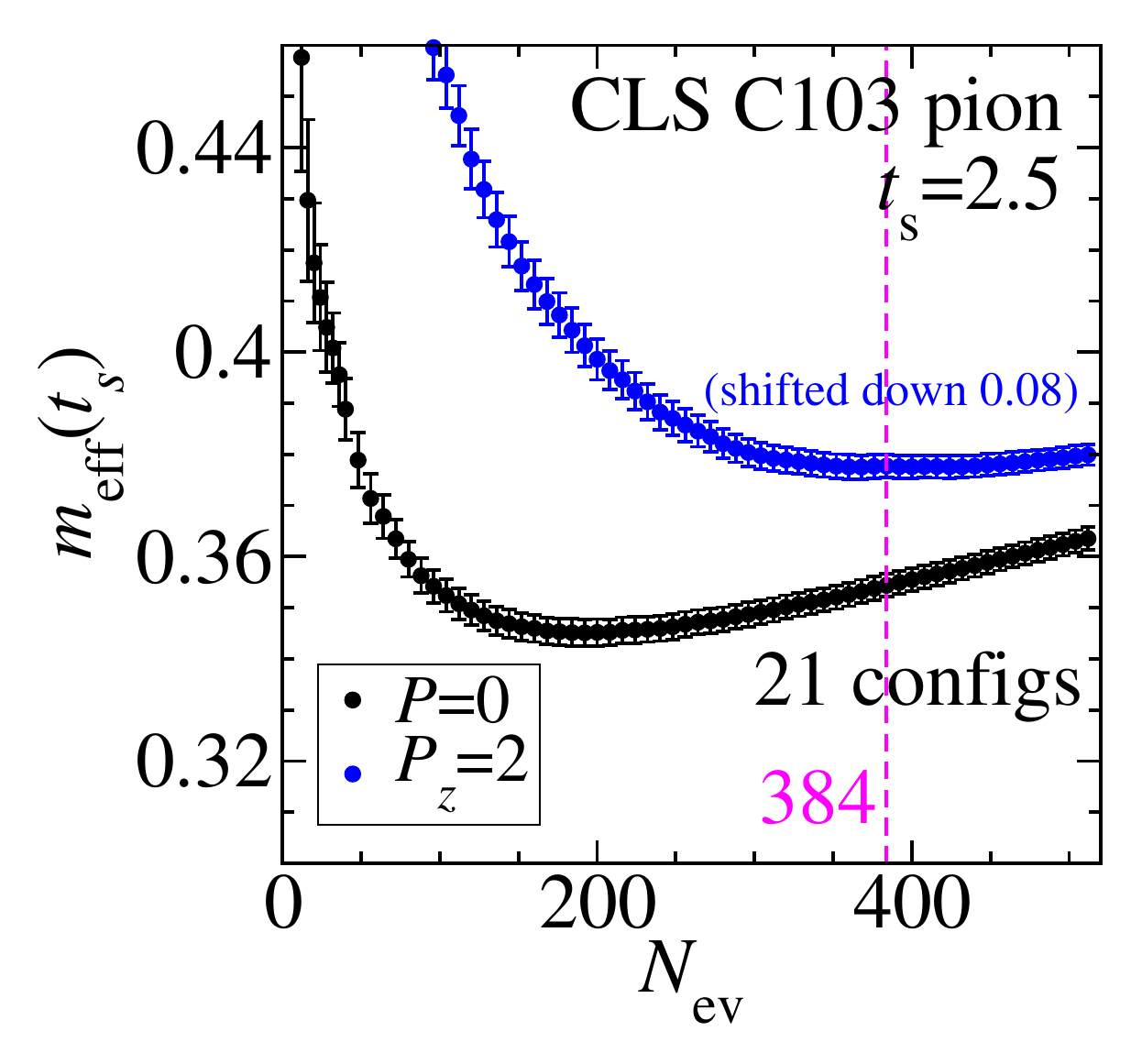}
\includegraphics[width=0.32\textwidth]{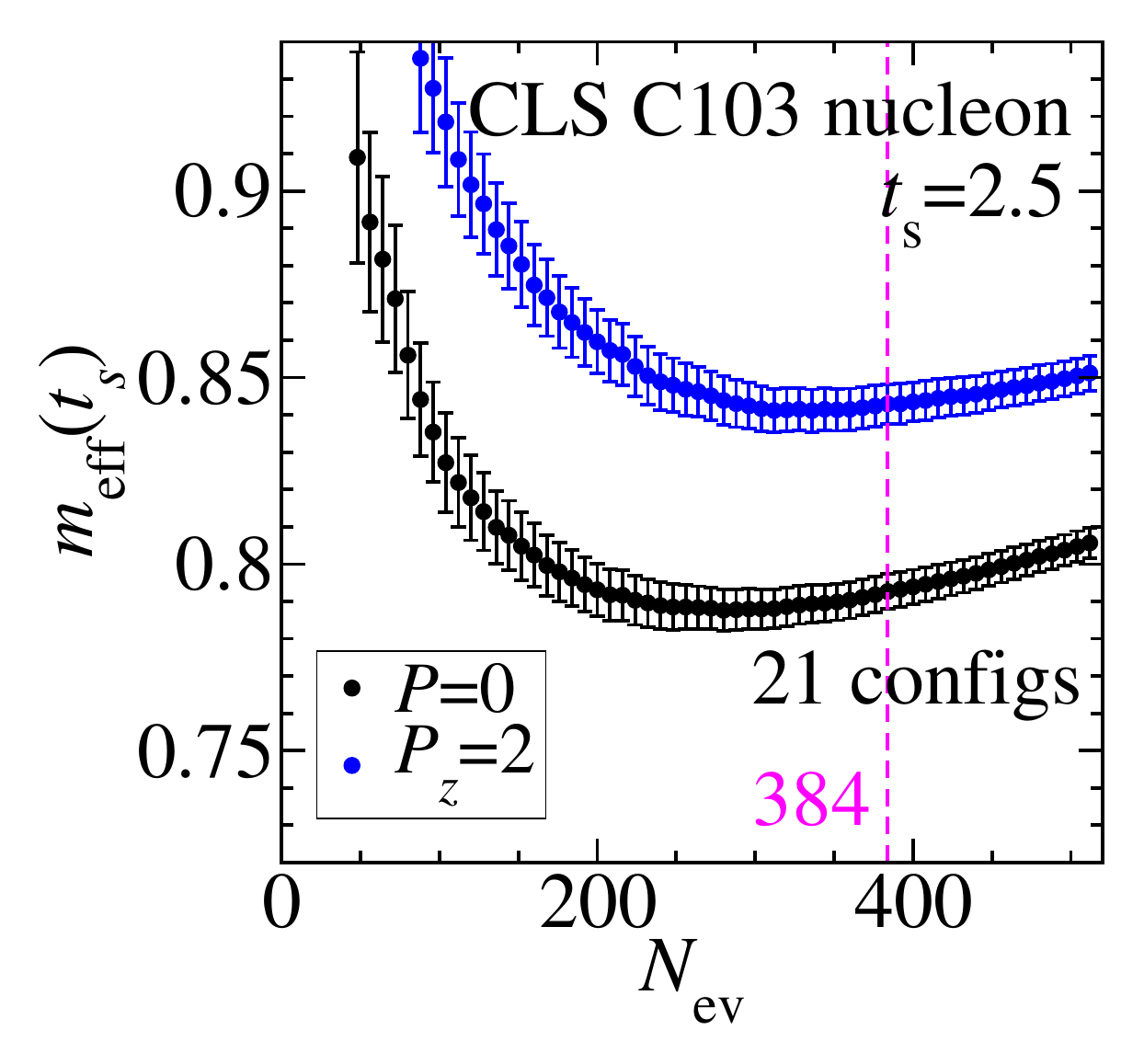}
\includegraphics[width=0.32\textwidth]{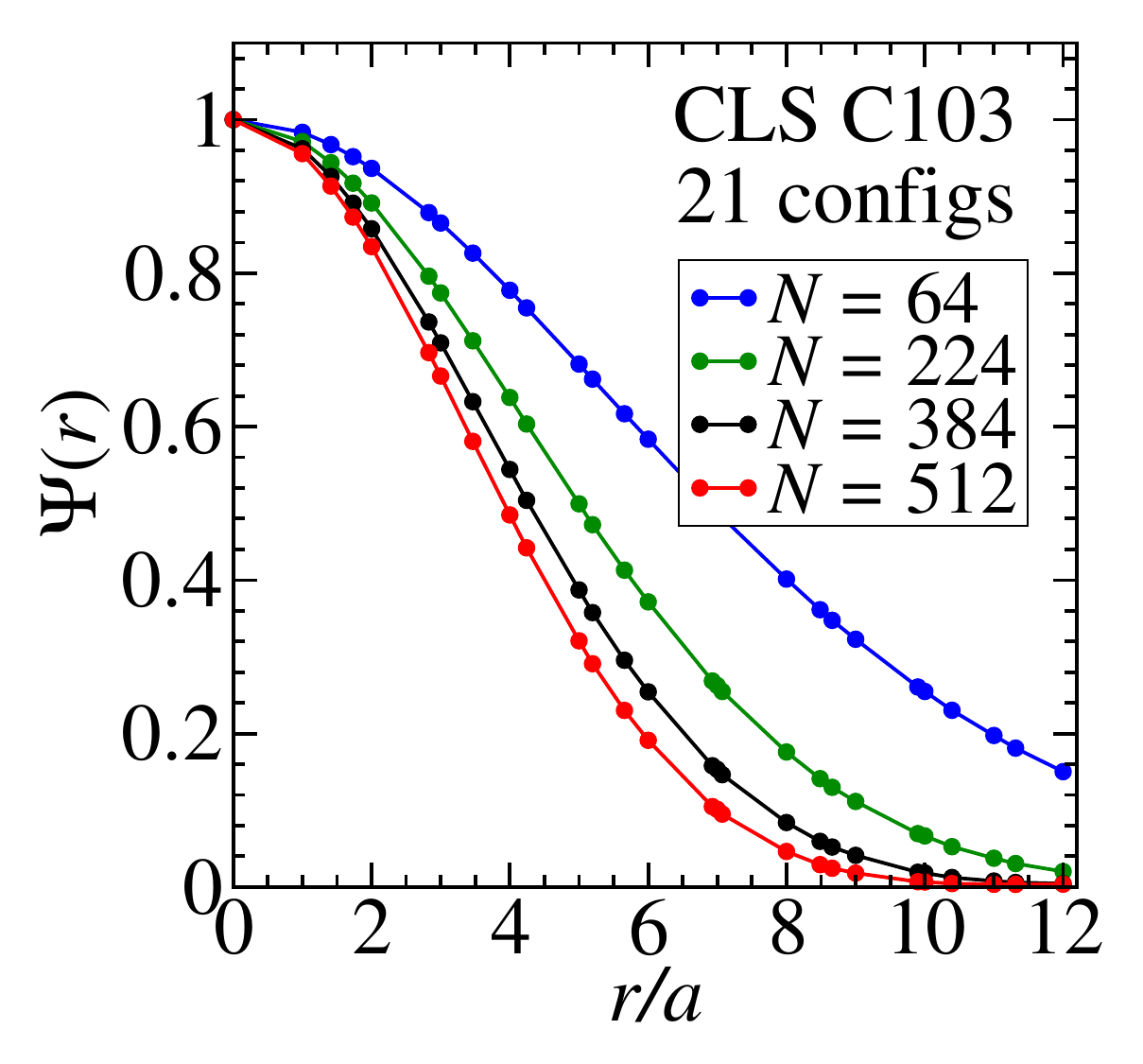}
\caption{\label{fig:smearing}
Effective mass of the pion (left) and nucleon (middle) as a function of $N_{\rm ev}$ at two different choices of total momentum of the correlators.  The effective mass is constructed as $m_{\rm eff}(t_s) = \ln(C(t_s-0.5) / C(t_s+0.5))$ in both cases for $t_s=2.5$.  The smearing profile of the quark source is plotted in the right panel for different numbers of $N_{\rm ev}$.
}
\end{figure*}
%----------------------------------------------------------------------------------

This study uses the stochastic Laplacian Heaviside (sLapH) method~\cite{Morningstar:2011ka} to construct
single- and two-nucleon operators such that the individual nucleons are projected to a definite momentum in both the source and sink operators.
A first step in utilizing the distillation or sLapH method is to choose how many eigenvectors ($N_{\rm ev}$) of the stout-smeared, three-dimensional, gauge-covariant Laplacian to use from which to build the sources (and sinks), for which larger $N_{\rm ev}$ amounts to less smearing of the quark sources/sinks.  It is well known that more smearing reduces contributions from excited states but also increases the stochastic noise of correlation functions.
In \figref{fig:smearing}, we show the effective mass of the pion and nucleon at two different values of total momentum as a function of $N_{\rm ev}$.  We also plot the smearing profile for various choices of $N_{\rm ev}$.  Balancing the reduction of excited state contamination without increasing the stochastic noise leads us to choose $N_{\rm ev}=384$ in this work.  
Stout-link smearing~\cite{Morningstar:2003gk} with staple weight $0.10$ with $20$ iterations is used for the gauge-field when constructing these sources.
Variance reduction is achieved by dilution of the noise with full spin dilution and interlace-$6$ dilution in the LapH eigenvector indices.  Further details of our implementation of the sLapH method are presented in Ref.~\cite{Horz:2020zvv}.

When $m_u=m_d=m_s$ and QED interactions are ignored, QCD has an $SU(3)$ flavor symmetry.  $SU(3)$ flavor basis states $\vert \Lambda, I,I_3,Y\rangle$ can be labeled by irrep $\Lambda$ (the quadratic and cubic Casimirs), isospin $I$, isospin projection $I_3$, and hypercharge $Y$.  Note that $Y=S+B$, where $S$ is strangeness and $B$ is baryon number. The baryons $N^0$, $N^+$, $\Lambda^0$, $\Sigma^-$, $\Sigma^0$, $\Sigma^+$, $\Xi^-$, $\Xi^0$ then all have the same mass since they all reside in a flavor $\bm{8}$ irrep.  For $NN$ scattering at the $SU(3)$ symmetric point, we must consider the $\bm{8}\times\bm{8}$ flavor irrep, whose decomposition into irreps is
\begin{equation}
\bm{8}\otimes \bm{8}=\bm{1}\oplus \bm{8} \oplus \bm{8}\oplus \bm{10} \oplus\overline{\bm{10}}\oplus \bm{27}.
\end{equation}
The deuteron is a neutron-proton pair with $I=0$ and $Y=2$ and it occurs in the $\overline{\bm{10}}$ flavor irrep.  The dineutron is a neutron-neutron pair with $I=1$ and $Y=2$ and it occurs in the $\bm{27}$ irrep.
    
The majority of our calculations in our other studies are carried out using $SU(2)$ flavor symmetry since we usually employ the more realistic approximation $m_u=m_d$ with $m_s$ being much heavier.  A  significant amount of work has gone into designing the $SU(2)$ operators and into the software evaluating the correlations of such operators.  Hence, we utilize our $SU(2)$ operators even for this calculation at the $SU(3)$ symmetric point.  The states produced by our $SU(2)$ nucleon operators should have substantial overlap with those produced by $SU(3)$ nucleon operators and should lead to the same lowest-lying energy in each isospin channel.  Another way to view our use of $SU(2)$ nucleon operators is to note that we have not initially block-diagonalized the spectrum as much as could be done with $SU(3)$ operators, but the same spectrum should be obtained in the space spanned by the states created by the operators.

%------------------------------------------------------------------------------
\subsection{Correlator Analysis: Conspiracy Model\label{sec:conspiracy}}
%------------------------------------------------------------------------------
In all previous LQCD calculations of two-baryons, one consistent observation is that the excited state contamination of the two-baryon correlation function is nearly the same as that of the single-baryons.
For example, the effective mass of the ratio correlation function exhibits mild excited state contamination and often plateaus earlier in time than either the two or single baryon correlators.  
This has led to a common strategy of fitting the ratio correlation functions to extract the interaction energies, $\Delta E = E_{N_1N_2} - E_{N_1} - E_{N_2}$~\cite{Aoki:2009ji,Berkowitz:2015eaa,Iritani:2018vfn,Green:2021qol,Horz:2020zvv,Green:2022rjj}.  Such a strategy is criticized, for example in Refs.~\cite{Wagman:2017tmp,Amarasinghe:2021lqa}, because even if the two-nucleon correlator is positive-definite, and thus monotonically approaches the large time limit, the ratio correlation function is not guaranteed to do so.

We addressed this issue in Ref.~\cite{Horz:2020zvv} by constructing a fit function that was also a ratio
\begin{equation}\label{eq:ratio_NN}
R(t) = \frac{C_{N_1N_2}(t)}{C_{N_1}(t)C_{N_2}(t)}\, ,
\end{equation}
where both the numerator and denominator were described with a sum of exponentials with positive overlap factors, $r_n = z_n / z_0$ with $z_n = \langle n|O|\O\rangle$, $A_n = |z_n|^2$,
\begin{equation}\label{eq:correlator_model}
C(t) = A_0 e^{-E_0 t}\bigg[ 1 + \sum_{n\geq 1}^{N_{max}} |r_n|^2 e^{-\Delta_{n0} t} \bigg]\, ,
\end{equation}
and $\Delta_{n0} = E_n - E_0$.
A global fit was performed to $R(t)$ and the individual nucleons $N_i(t)$.
Stability of the ground state energies was studied by varying the time-range used in the analysis as well as the number of excited states used in the numerator and denominator functions.

We improve upon this strategy in two ways:
\begin{enumerate}
\item We directly fit the two-nucleon correlation functions, as well as the single-nucleon ones, rather than fitting the ratio correlation function, for which Eq.~(\ref{eq:correlator_model}) is a robust model for all numerical results.

\item \textit{Conspiracy model:} Motivated by the observed cancellation of excited states in $R(t)$, we build a model of the NN correlator that reflects this observation, deriving the number of states used to describe the NN correlator from the number used to describe the single-nucleon correlators, so that the NN excited states have the freedom to {\it conspire} to cancel with those in the individual nucleons.  We further elaborate on this model in the following.
\end{enumerate}
If the two-nucleons did not interact with each other, then the NN correlation function could be described as the product of the two single nucleon correlators:
\begin{align}
\label{eq:NN_nonInteract}
C_{NN}(t) = \left[ \sum_{n_1=0}^{N_1-1} A_n^{(1)}e^{-E_n^{(1)}t}\right]
\left[ \sum_{n_2=0}^{N_2-1} A_n^{(2)}e^{-E_n^{(2)}t}\right]\, .
\end{align}
In the limit that the two nucleons are the same, such as for the case when the total momentum of the NN correlator is zero, the NN correlator would then be described by the square of the single-nucleon correlator.

For two nucleons, which are strongly interacting, such a model may not provide a good estimate for the overlap factors, but it is still a good starting point for the NN energies since the interacting energy is of $\mathrm{O}(0.1\%)$ of the total energy of the system.
Therefore, the conspiracy model is constructed as follows:
\begin{enumerate}
\item If $N$ states are used for the single nucleons, then $N^2$ states are used to describe the NN correlator in the general case, and $N\frac{N+1}{2}$ in the case where the two nucleons have the same magnitude of momentum;

\item The overlap factors of the NN correlator are parameterized independently from the single nucleon correlators;

\item The ground and excited state energies of the NN correlator are parameterized as
\begin{equation}\label{eq:conspiracy_energies}
E^{N^n_1 N^m_2} = E_n^{N_1} + E_m^{N_2} + \delta E_{nm}\, ,
\end{equation}
where $E_n^{N_i} = E_0^{N_i} +\Delta_{n0}^{N_i}$ is the $n^{th}$ energy of nucleon $i$ and $\delta E_{nm}$ is an interaction energy between the $n^{th}$ state of nucleon-1 ($N_1$) and the $m^{th}$ state of nucleon-2 ($N_2$).

\item A global analysis is performed to simultaneously constrain the single nucleon ground and excited state energies, $E_n^{N_i}$, along with the two-nucleon interacting energies, $\delta E_{nm}$ as well as the overlap factors for all correlators in a given irrep.
\end{enumerate}
There is freedom to pick which pair of single nucleons to use as the reference state when building the conspiracy model, provided the correct center-of-mass energies are constructed for the QC2 analysis of the amplitude.  In order to automate the choice of $N_1$ and $N_2$ for the analysis of a given principle correlator in an irrep, we first examine the relative overlaps of a given NN operator with the various principle correlators, Eq.~(\ref{eq:Z_jn}).  Often, a given operator predominantly couples to a single level, leading to a one-to-one mapping between the operators and the levels.
When this is not the case, we use the effective energies of the single nucleons and the corresponding NN correlator to pick a reference set of single nucleons by minimizing ${\rm abs}(E_{NN} - E_{N_1} - E_{N_2})$ across the set of $N_1$ and $N_2$ states that can be used to make an operator in the given NN irrep.  

We contrast this conspiracy model with an {\it agnostic model} in which the number of excited states in the NN correlator are not fixed by the choice of the single nucleon correlators, nor are the energies coupled as in \eqnref{eq:conspiracy_energies}.  Our previous work implemented such an agnostic model (and also fit the ratio correlation functions).  In this work, we show that the conspiracy models are favored over the agnostic models as measured by the Bayes Factor in a Bayesian analysis.

%------------------------------------------------------------------------------
\subsubsection{Estimation of priors for the conspiracy model}
%------------------------------------------------------------------------------
The fit priors must be carefully chosen to avoid biasing the posterior extraction of model parameters while maintaining the stability of the multi-exponential fit.   The priors, fit windows, and number of excited states are derived from pre-set criteria applied uniformly rather than hand-selecting these choices for each correlator.

The first step we take is to use \eqnref{eq:correlator_model} to describe each of the two-point functions.  For the single nucleons, the effective mass and effective overlap factors
\begin{align*}
m_{\rm eff}(t) &= \ln \left( \frac{C(t)}{C(t+1)} \right)\, , \\
A_{\rm eff}(t) &= e^{+m_{\rm eff}(t)\, t}\ C(t)\, ,
\end{align*}
can be used to estimate the ground state priors for a given correlator.
In this work, we use a reference time to estimate the Gaussian priors as
\begin{align}
\tilde{E}_0 &= m_{\rm eff}(t_{\rm ref})\times \mathcal{N}(1, 0.1)\, ,
\nonumber\\
\tilde{A}_0 &= A_{eff}(t_{\rm ref})\times \mathcal{N}(1, 1)\, ,
\end{align}
where $\mathcal{N}(\mu, \sigma)$ is a Gaussian distribution of mean $\mu$ and width $\sigma$.  Examining the effective mass of the nucleon, we choose a reference time of $t_{\rm ref}=10$ as it is sufficiently close to the region of ground state saturation to provide a good estimate of these parameters.

For the excited state parameters of the single nucleon, we use a Lognormal distribution to prior each mass gap, such that if $X\sim{\rm Lognormal}(\mu,\sigma)$ then $\ln(X)\sim\mathcal{N}(\mu,\sigma)$.
Each mass gap is estimated to be $2m_\pi$ and the ratio overlap factors are priored with the square of a normal distribution,
\begin{align}
\tilde{\Delta}_{n,n-1} &= \mathrm{Lognormal}(2m_\pi, m_\pi)\, ,
\nonumber\\
\tilde{r}_n &= \mathcal{N}(1, 0.25)\, ,
\end{align}
with $m_\pi$ determined in Ref~\cite{Horz:2020zvv} and our updated $m_N$
\begin{align}\label{eq:mpi}
&am_\pi = 0.310810(95)\, ,&
&am_N = 0.70267(40)& .
\end{align}
This ensures the ordering of excited state energies according to their $n$ label and the resulting ratio overlap factors are positive definite and of $\mathrm{O}(1\pm0.5)$.

For the parameters of the NN correlation functions, the priors of the overlap factors are estimated in the same way as described above.  Since we model the NN energies as in \eqnref{eq:conspiracy_energies}, we need estimates for the $\delta E_{nm}$ parameters.  For the ground state interaction energy, we use the effective mass of the ratio correlation function at $t_{\rm ref}$ to estimate the prior mean and we choose to give it a 100\% width
\begin{equation}
\tilde{\delta E}_{00} = m_{\rm eff}^{R}(t_{\rm ref}) \times \mathcal{N}(1,1)\, ,
\end{equation}
where the superscript $R$ indicates \eqnref{eq:ratio_NN} is used to construct the effective mass.  For the excited state interaction energies, we use this effective mass to estimate a width but set the mean to zero
\begin{equation}
\tilde{\delta E}_{mn} = \tilde{\delta E}_{nm} = \mathcal{N}(0,m_{\rm eff}^{R}(t_{\rm ref}))\, .
\end{equation}

This construction of priors, with generously wide widths, ensures that the numerical data, when it is informative, is the only significant ingredient informing the posterior ground state parameters of interest.

%------------------------------------------------------------------------------
\subsubsection{Comparison with other analysis strategies\label{sec:conspiracy_compare}}
%------------------------------------------------------------------------------

A comparison with other analysis strategies in the literature provides some of the reasons for our use of this conspiracy model. The analysis procedures used in Refs.~\cite{Fleming:2004hs,Fleming:2009wb,Beane:2009kya,Aubin:2010jc,Aubin:2011zz,Cushman:2019tcv,Romiti:2019qim,Cushman:2019hfh,Fischer:2020bgv,Fleming:2023zml} are based on the Prony Method~\cite{prony}, the Generalized Pencil of Function (GPoF) method~\cite{gpof,gpof2}, or some combination of these two methods.
A recent interest has arisen in using the Lanczos algorithm with such methods~\cite{Wagman:2024rid,Hackett:2024nbe} which have been shown to be equivalent to the Prony-like methods~\cite{Ostmeyer:2024qgu,Abbott:2025yhm}.  There has also been a suggestion of using Laplace filters~\cite{Portelli:2025lop}.
Common to all of these strategies is an attempt to isolate individual eigenvalues of the correlation functions and model their contributions by single-exponential forms.  The large number of eigenstates which contribute to the spectral representation of any correlation function usually makes such isolation difficult and prone to unknown systematic errors.
As with multi-exponential fits to any correlation functions, especially if the single nucleons are constructed with only simple 3-quark interpolating operators, the higher-lying exponential decay rates extracted should not be interpreted as stationary-state eigenvalues of QCD as the operator basis is not sufficient to reliably isolate these individual eigenstates.  In all of these strategies, the higher-lying decay rates should be viewed as just nuisance parameters used to model the excited-state contamination, which is a sum of an infinite number of exponentials, by a sum of a finite (usually small) number of exponentials.

Another important point to keep in mind is the fact that the conspiracy model is applied after the use of the GEVP, which is also utilized to removed excited state contamination arising from the lower-lying NN interacting modes.  The observed remaining excited state contamination comes from higher lying modes, $n\geq N_{\rm op}$ in \eqnref{eq:Dcorr}.  These other methods will also have to contend with this excited state contamination, which as we show below, most likely arises from excited states of the single nucleons.

Rather than trying to isolate an eigenstate that can be described by a single exponential, the conspiracy model is designed to capture the apparent physics contained in the NN correlators after the GEVP is used to diagonalize the basis of operators: the diagonal NN correlators are well described by \eqnref{eq:NN_nonInteract} with small corrections induced by the interactions of these states.  The conspiracy model not only allows for a systematic way to account for this excited-state contamination and its impact on the extraction of the ground-state energies, but also takes advantage of the positive-definite nature of these diagonal correlators to utilize regions in Euclidean time where the signal is not lost to the noise.  In future work, it will be interesting to quantitatively compare the conspiracy model with the other analysis strategies mentioned above.

%------------------------------------------------------------------------------
\subsection{Deuteron channel results\label{sec:results_deuteron}}
%------------------------------------------------------------------------------

%------------------------------------------------------------------------------
\subsubsection{Correlator analysis and the NN spectrum}
%------------------------------------------------------------------------------
A QC2 determination of scattering amplitudes depends on an accurate and precise spectrum. The deuteron channel is isospin-0 and spin-1, yielding numerous irreps that overlap with the S-wave at low energy.   We list them in \tabref{tab:nn_irreps} along with the total number of NN operators we use in each irrep.  This work considers the lowest one or two levels in each irrep after diagonalizing the correlator matrices.

%------------------------------------------------------------------------------
\begin{figure}
\includegraphics[width=\columnwidth]{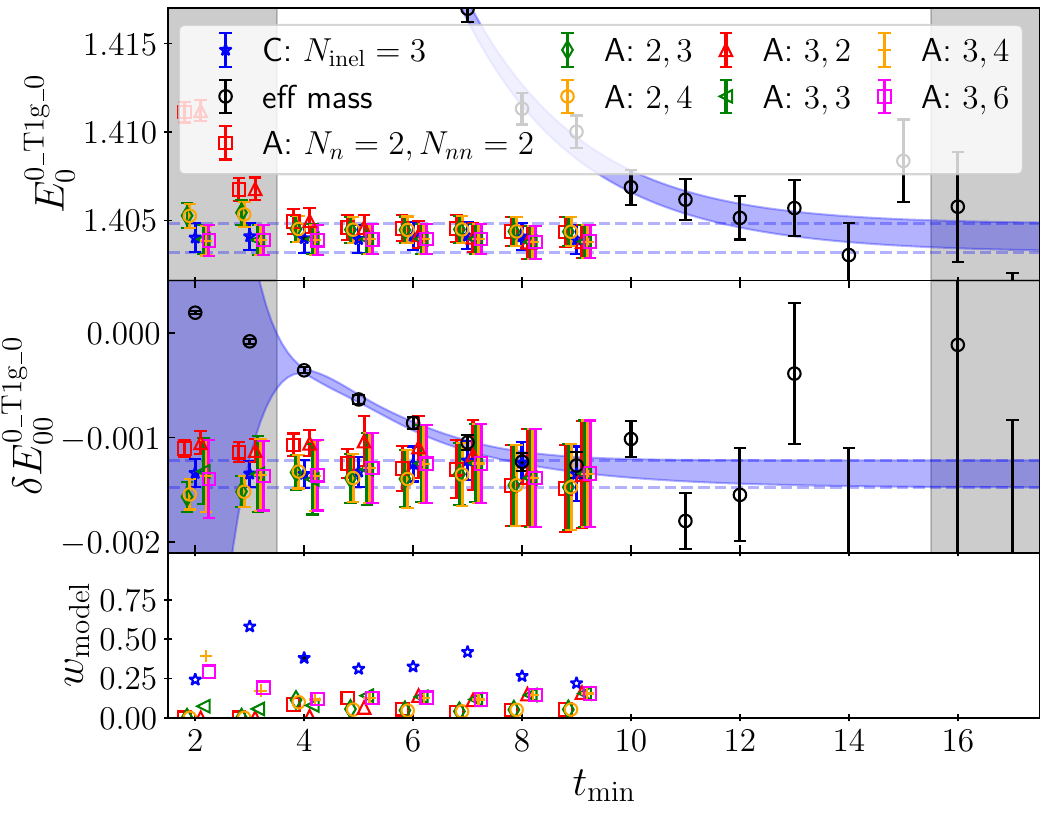}
\includegraphics[width=\columnwidth]{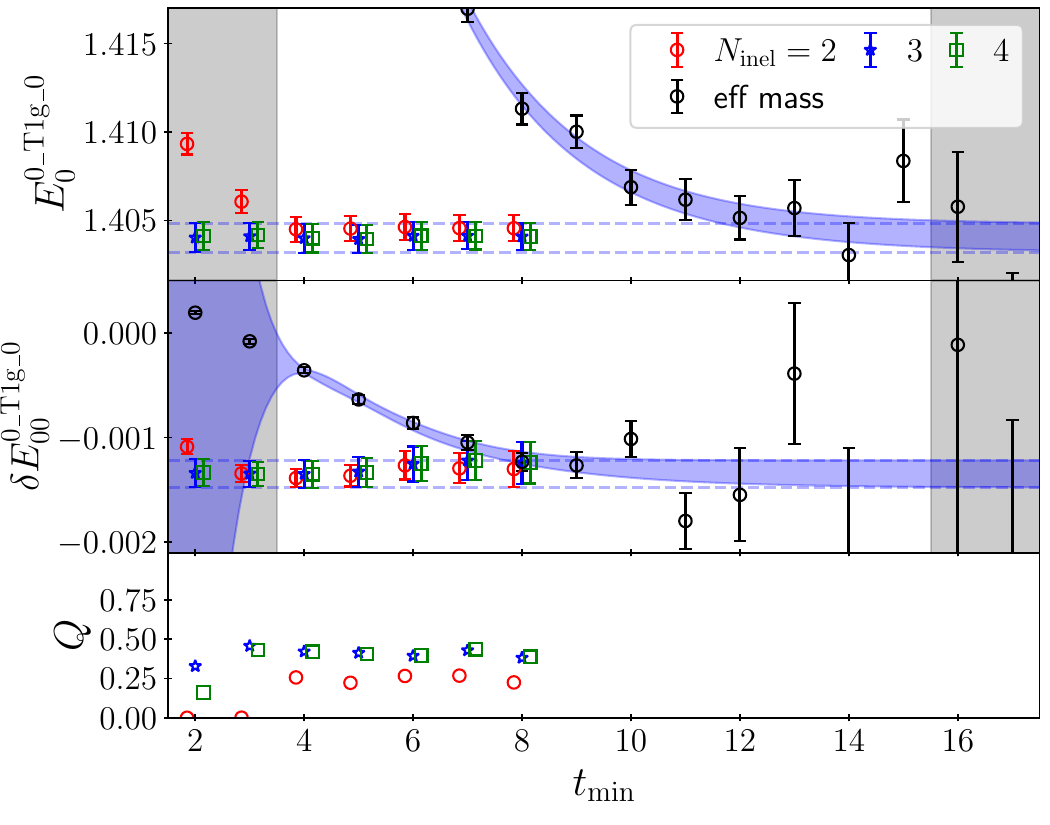}
\caption{\label{fig:conspiracy_agnostic}
Comparison of the ground state NN energy and energy shift in the $T_{1g}$ channel using the \textit{conspiracy} (C) and \textit{agnostic} (A) models (top).  For the former, we show the preferred final model while for the latter, we vary the number of exponentials used to model the single ($n$) and NN ($nn$) correlators.  The band and dashed lines are from the conspiracy model with 3-states for the single nucleon, to guide the eye.  The relative probability for a given model ($w_{\rm model})$ at each $t_{\rm min}$ is determined from the Bayes Factor.
We also show stability of the \textit{conspiracy} model with respect to the number of states used to describe the single nucleon correlators (bottom).  The $Q$-value factor is similar to the frequentist $P$-value but uses the augmented $\chi^2/{\rm dof}$~\cite{Lepage:2001ym}.
}
\end{figure}
%------------------------------------------------------------------------------

It is essential to ensure that all the systematic uncertainty associated with various choices that go into the analyzing the correlators are accounted for, including:
\begin{enumerate}[itemsep=0ex]
\item The choice of model to describe the correlation functions, such as the \textit{conspiracy} or \textit{agnostic} model;
\item The choice of GEVP times, $t_0$ and $t_d$;
\item The number of excited states used for the single and NN correlators;
\item The choice of $t_{\rm min}^{N}$ for the single nucleon analysis;
\item The choice of $t_{\rm min}^{NN}$ and $t_{\rm max}^{NN}$ for the NN analysis;
\end{enumerate}
Our final chosen fit model is a \textit{conspiracy} model with
\begin{itemize}
%\item \textit{conspiracy} model;
\item $t_0=5$ and $t_d=10$ for the GEVP times;
\item $N_b=8$ blocking;
\item 3-state model for single nucleons, $N_n = 3$;
\item $t_{\rm min}^N = t_{\rm min}^{NN} = 4$ for the single and NN correlators;
\item $t_f^N=20$ and $t_f^{NN}=15$ for the single and NN correlators respectively.
\end{itemize}
This model is applied uniformly for all the 15 irreps/levels considered in this work.

We begin by comparing \textit{conspiracy} and \textit{agnostic} models.
In this and the following, we plot effective mass bands and ground state energies from the chosen fit posterior.
We compare various agnostic fits with either 2 or 3 exponentials for the single nucleon, and 2, 3, 4 and 6 exponentials for the NN correlators.
With fixed data, we can use the Bayes Factor to determine the relative probability of a model given the data.

We show the results of this comparison for the $T_{1g}$ irrep in the top panel of \figref{fig:conspiracy_agnostic}.
To guide the eye, we plot the posterior effective mass of the NN correlator (top panel) and the ratio correlator (middle panel).  The ratio correlator's effective mass is constructed by taking the ratio of the correlated posterior of the NN and single nucleons.
The bottom panel lists the normalized weight factors, $w_{\rm model}$, normalized over the set of 8 models used in this comparison.
This analysis shows that:
\begin{enumerate}[label=\roman*)]
\item For this irrep (and the others), the conspiracy model is favored over the agnostic models by the Bayes Factor weight;

\item Fits with three or more states for the NN correlator are all consistent, while those with only 2 states are consistent at the 1-2 sigma level but systematically different for all $t_{\rm min}$;

\item For the agnostic model, we observe that increasing the number of exponentials in the NN correlator 
significantly increases the uncertainty in determining the interacting energies.
\end{enumerate}
In addition, we will show that with the conspiracy model, the resulting values and uncertainties of these $\delta E_{00}$ values are more stable with respect to variation of the number of exponentials, the value of $t_{\rm min}^{N,NN}$ and other user choices.

The stability under variation of the number of states used in the single-nucleon fits for the conspiracy model is shown in the bottom panel of \figref{fig:conspiracy_agnostic}.  The stability with respect to the choice of GEVP times, the blocking/binning size, the final time used for the NN correlator, and the minimum time used for the N correlator, can be found in Appendix~\ref{sec:correlator_analysis}.

\begin{figure}
\includegraphics[width=\columnwidth]{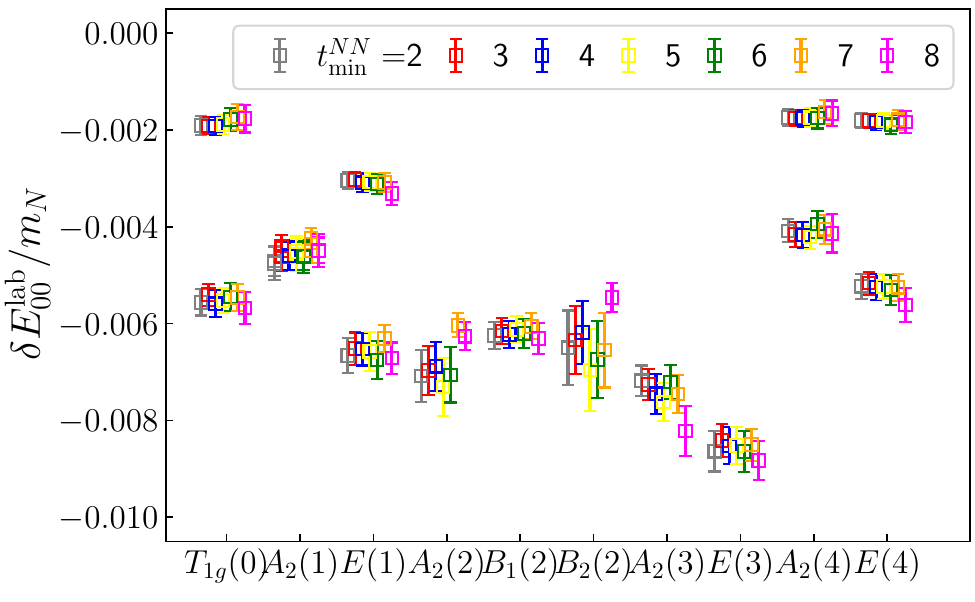}
\includegraphics[width=\columnwidth]{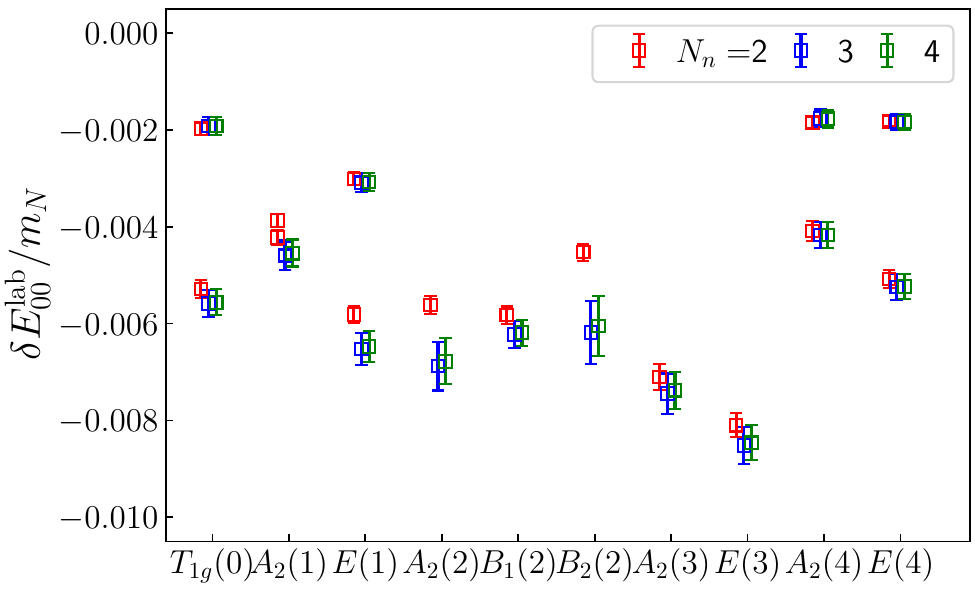}
\caption{\label{fig:deuteron_dE}
Stability of the ground state interaction energy, $\d E_{00}$, for all 15 irreps/levels used in this work for the deuteron.
Note that the two levels in $A_2(1)$ have overlapping values of $\d E_{00}$.
}
\end{figure}

Overall, for all these variations, we find that the conspiracy model provides a stable extraction of the ground-state NN energy and interaction energy.  We have shown these stability plots for the lowest level in the $T_{1g}$ irrep, and we find similar stability for all irreps considered in this work.
In \figref{fig:deuteron_dE}, we plot the stability of $\d E_{00}$ in all 15 irreps/levels considered in this work for two of these systematics, and the remainder are plotted in Appendix~\ref{sec:correlator_analysis}.

%------------------------------------------------------------------------------
%------------------------------------------------------------------------------

%------------------------------------------------------------------------------
\subsubsection{Amplitude analysis\label{sec:amplitude_deuteron}}
%------------------------------------------------------------------------------
%------------------------------------------------------------------------------
\begin{table}[ht]
\caption{\label{tab:N_disp}
We list the energy levels of the single nucleon where $n$ denotes the squared integer of momentum, $E_n^2 = M_N^2 + n (2\pi/L)^2$.  We also list what the energy would be if the continuum dispersion relation were satisfied.
}
\begin{ruledtabular}
\begin{tabular}{r|ccccc}
& $n=0$& $n=1$& $n=2$& $n=3$& $n=4$\\
\hline
%$E_n$& 0.70267(40)& 0.71423(38)& 0.72609(36)& 0.73741(37)& 0.74888(36)\\
$E_n$& 0.7027(4)& 0.7142(4)& 0.7261(4)& 0.7374(4)& 0.7489(4)\\
cont.& 0.7027(4)& 0.7148(4)& 0.7266(4)& 0.7383(4)& 0.7499(4)
\end{tabular}
\end{ruledtabular}
\end{table}
%------------------------------------------------------------------------------

Given a spectrum, we turn to applying the single-channel QC2 for which there is a one-to-one mapping between each energy level and a value of the amplitude.
As in Ref.~\cite{Horz:2020zvv}, we use the ``spectrum method'', as described in Ref.~\cite{Morningstar:2017spu}, to determine parameters of $q\cot\d$ from the resulting energy levels.  This is done by constructing the $\chi^2$ loss function in terms of $q^2_{\rm cm}$.
With our increased precision, the results have become sensitive to two sources of systematic uncertainty as compared to Ref.~\cite{Horz:2020zvv}:
\begin{enumerate}[label=\roman*)]
\item The final NN energy is sensitive to whether it is taken from the fit, or reconstructed using $\d E_{00}$ and the continuum dispersion relation of the single nucleons;

\item In some of the boosted channels, our results have become sensitive to $S-D$ mixing.
\end{enumerate}
Regarding i), our single nucleon energy levels are statistically different from what would be predicted from a continuum dispersion relation.  We list the energy levels in \tabref{tab:N_disp} along with the continuum dispersion value.  A fit to the spectrum using
\begin{equation}\label{eq:nucleon_dispersion}
E_n^2 = M_N^2 + \xi n \left(\frac{2\pi}{L}\right)^2\, ,
\end{equation}
yields the values of
\begin{align}
\label{eq:mN}
M_N &= 0.7027(4)\, ,
\\\label{eq:xi}
\xi &= 0.9790(53)\, .
\end{align}

It is expected that the leading discretization effects to the spectrum cancel in the energy splitting, $\d E_{00} = E_{NN} - E_1 - E_2$, suggesting that using the continuum dispersion relation to reconstruct the single nucleon energies $E_i= \sqrt{M_N^2 + n_i (2\pi/L)^2}$ is more appropriate for the QC2 as its derivation assumed such a dispersion relation.  This expectation was recently formally derived in Ref.~\cite{Hansen:2024cai}.

\begin{figure}
\includegraphics[width=\columnwidth]{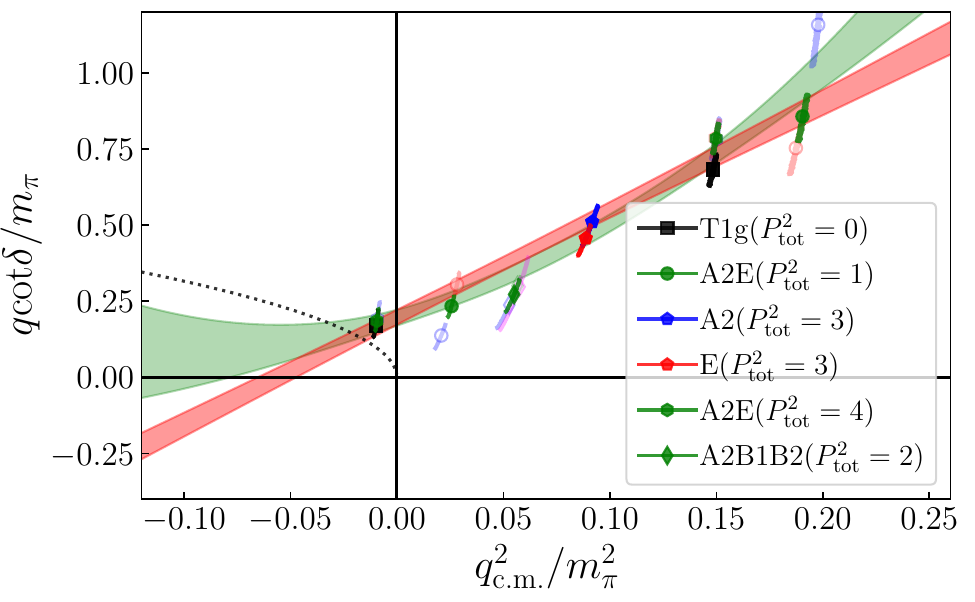}
\includegraphics[width=\columnwidth]{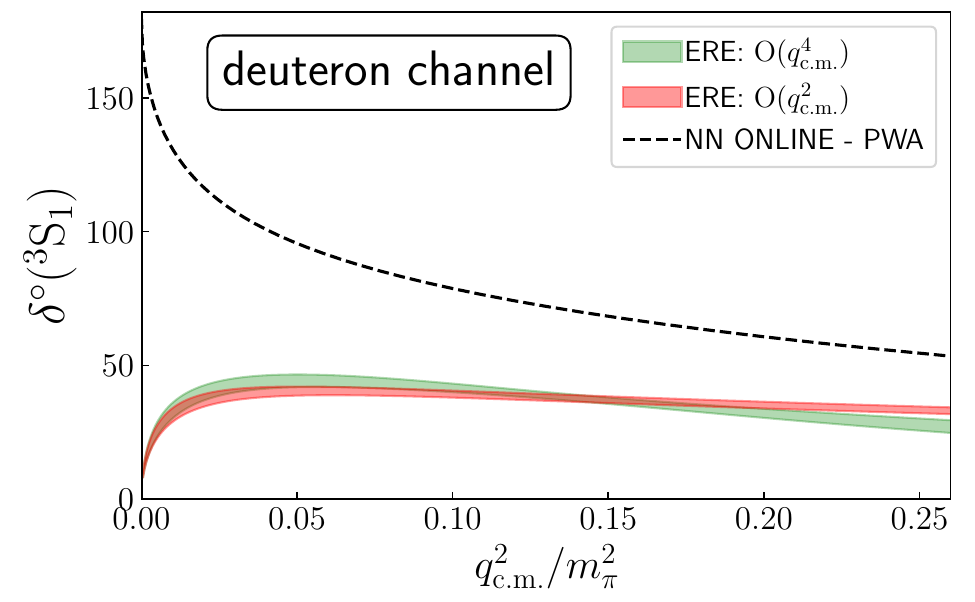}
\caption{\label{fig:deuteron_qcotd} 
Resulting phase shift expressed as  $q\cot\d$ (top) and $\d$ in degrees (bottom) from the single-channel QC2 analysis after irrep averaging.  The solid points are used in the analysis.  The open data are the original results before irrep averaging.  The dotted line in the $q\cot\d$ plot is $\sqrt{-q^2}$, i.e. the virtual bound state condition. A crossing between the $q\cot\delta$ and this curve implies the presence of a virtual bound state.
The NN ONLINE - PWA phase shift is obtained from \url{http://nn-online.org} using the Partial Wave Analysis of experimental data~\cite{Stoks:1993tb}.
}
\end{figure}

Regarding the $S-D$ partial wave mixing, Ref.~\cite{Briceno:2013bda} discussed in detail the expected mixing in these channels using the QC2 formalism with spin 1/2 particles~\cite{Briceno:2013lba,Briceno:2014oea}.  Depending upon the total momentum boost vector, different linear combinations of the spectrum results from different irreps can be used to remove the leading dependence upon the $S-D$ mixing, such as
\begin{align}
\mathbf{n} = (0,0,n):& E_{AE} = \frac{1}{3}E_{A_2} + \frac{2}{3}E_E\, ,
\nonumber\\
\mathbf{n} = (0,n,n):& E_{AB} = \frac{1}{3}E_{A_2} + \frac{1}{3}E_{B_1} + \frac{1}{3}E_{B_2}\, ,
\end{align}
with integer $n$.
In \figref{fig:deuteron_qcotd}, we plot the resulting determination of $q\cot\d$ in this channel from a single-channel QC2 analysis.  We plot the results in units of $m_\pi$ to more easily compare with our earlier results~\cite{Horz:2020zvv}.
Parameterizing $q\cot\d$ with an effective range expansion (ERE), 
\begin{equation}
\frac{q\cot\d}{m_\pi} = \frac{-1}{am_\pi} 
    +\frac{r m_\pi}{2}\left(\frac{q}{m_\pi}\right)^2
    +\l m_\pi^3 \left(\frac{q}{m_\pi}\right)^4
    +\cdots
\end{equation}
we determine the scattering length ($a$) and effective range ($r$).
The scattering length is stable under variations of the ERE order, while the effective range shows sensitivity to the truncation.  Some of the analysis results are presented in \tabref{tab:deuteron_results}.

%----------------------------------------
% Table: deuteron results
\begin{table}
\caption{\label{tab:deuteron_results}
Results of the ERE parameters in the deuteron channel at $\mathcal{O}(q_{\rm c.m.}^2)$ and $\mathcal{O}(q_{\rm c.m.}^4)$ in the ERE expansion.
}
\begin{ruledtabular}
\begin{tabular}{c|cccc}
% ERE ORDER & ampi & rmpi& chis/dof
ERE Order& $am_\pi$& $rm_\pi$& $\l m_\pi^3$& $\chi^2$/dof[dof]\\
\hline
$q_{\rm c.m.}^2$& -5.06(63)& 7.05(45)& --& 1.88[7]\\
$q_{\rm c.m.}^4$& -5.10(68)& 4.2(1.4)& 9.9(4.4)& 1.39[6]
\end{tabular}
\end{ruledtabular}
\end{table}
%----------------------------------------

The presence of a bound state requires a negative value of $q\cot\d|_{q=0}$, corresponding to a positive value of the scattering length with the above parameterization.  While our results leave a significant systematic uncertainty on the determination of the effective range parameter, depending upon the order used in the ERE, the determination of the scattering length is observed to be robust, and rules out a bound state on this ensemble by more than $5\s$. Instead, a virtual bound state is found, given by the condition $q\cot\d(q^2) = +\sqrt{-q^2}$ for negative $q^2$. This is illustrated in Fig.~\ref{fig:deuteron_qcotd} by the crossing between the fits and the ``virtual bound state condition''.

%------------------------------------------------------------------------------
%------------------------------------------------------------------------------

%------------------------------------------------------------------------------
\subsection{Di-neutron channel results\label{sec:results_dineuteron}}
%------------------------------------------------------------------------------

\begin{figure}
\includegraphics[width=\columnwidth]{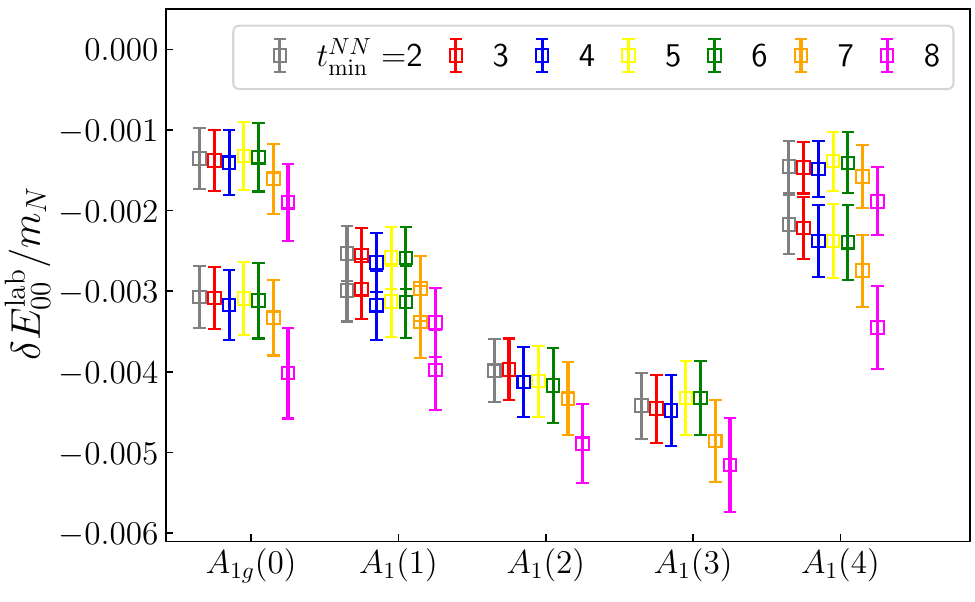}
\includegraphics[width=\columnwidth]{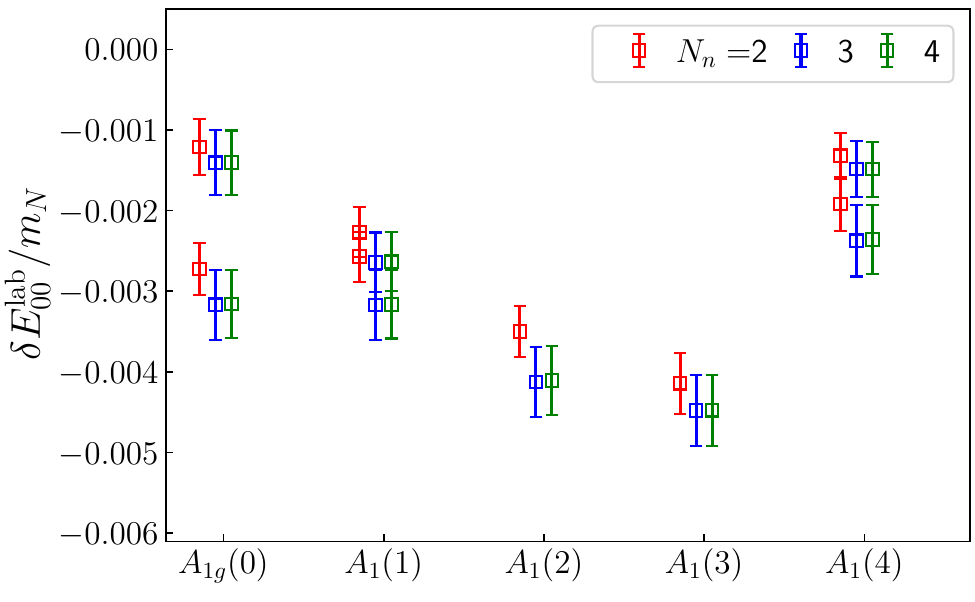}
\caption{\label{fig:dineutron_dE}
Stability of the ground state interaction energy , $\d E_{00}$ for all 8 irreps/levels used in this work in the di-neutron channel.
}
\end{figure}

\begin{figure}
\includegraphics[width=\columnwidth]{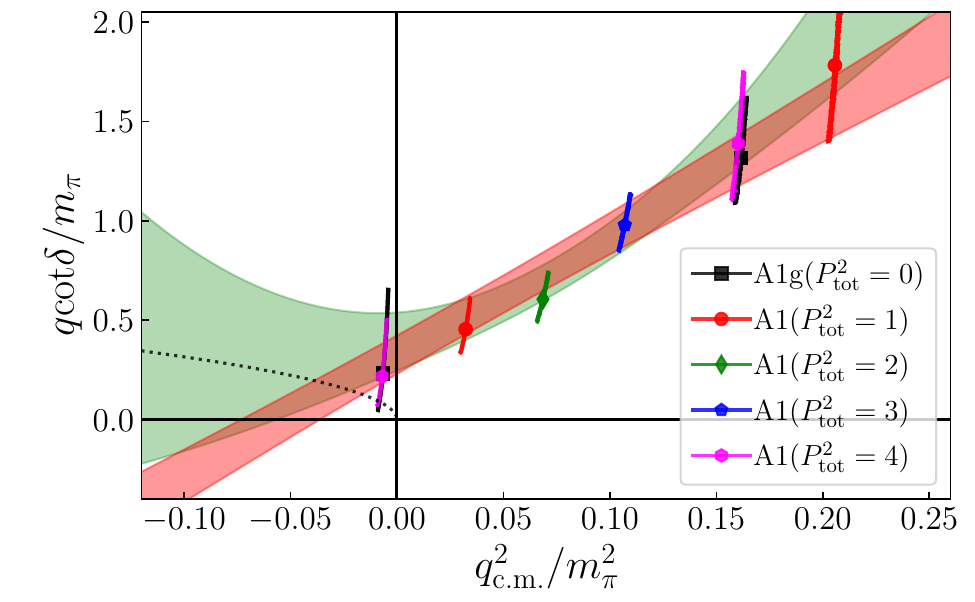}
\includegraphics[width=\columnwidth]{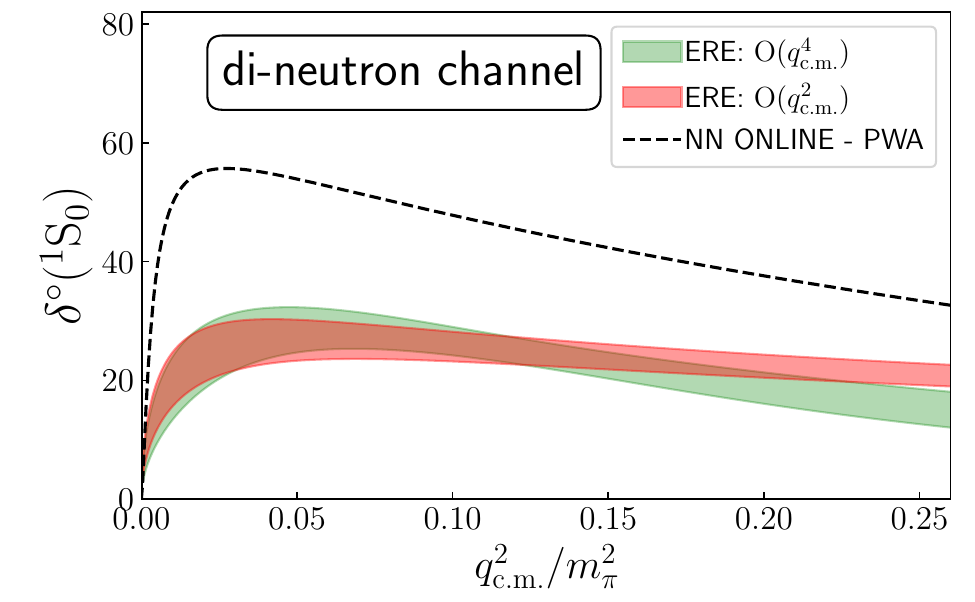}
\caption{\label{fig:dineutron_qcotd}
Resulting phase shift plots for the dineutron channel from the single-channel QC2. All notation as in Fig.~\ref{fig:deuteron_qcotd}
}
\end{figure}

We perform a similar stability analysis for the di-neutron scattering channel, finding the same optimal model used for the deuteron channel works well.
In \figref{fig:dineutron_dE}, we show the summary stability plots for the lab-frame energy shifts, $\d E_{00}$ for the eight irreps/levels used in this work.
Performing an amplitude analysis as with the deuteron, we determine the scattering length and effective range parameters at different orders in the ERE, listed in \tabref{tab:dineutron_results}.
This analysis also rules out a bound state in the di-neutron channel, and reveals the existence of a virtual bound state.
For the di-neutron, the issue of partial wave mixing does not affect the analysis, as it does for the deuteron channel.
\figref{fig:dineutron_qcotd} shows the resulting $q\cot\d$ and phase shift.
One interesting observation is that the effective range, $r_0$, in the di-neutron system is larger than the deuteron.

%------------------------------------------------------------------------------
%------------------------------------------------------------------------------

%----------------------------------------
% Table: deuteron results
\begin{table}[ht]
\caption{\label{tab:dineutron_results}
Results of the ERE parameters in the dineutron channel at $\mathcal{O}(q_{\rm c.m.}^2)$ and $\mathcal{O}(q_{\rm c.m.}^4)$ in the ERE expansion.
}
\begin{ruledtabular}
\begin{tabular}{c|cccc}
% ERE ORDER & ampi & rmpi& chis/dof
ERE Order& $am_\pi$& $rm_\pi$& $\l m_\pi^3$& $\chi^2$/dof[dof]\\
\hline
$q_{\rm c.m.}^2$& -3.06(91)& 12.2(1.4)& --& 0.78[6]\\
$q_{\rm c.m.}^4$& -2.55(95)& 5.4(5.0)& 24(15)& 0.21[5]
\end{tabular}
\end{ruledtabular}
\end{table}
%----------------------------------------

%------------------------------------------------------------------------------
%------------------------------------------------------------------------------

%-------------------------------------------------------------------------------

%-------------------------------------------------------------------------------
%  Hexaquarks
\section{Hexaquarks and the NN spectrum\label{sec:hexaquarks}}

\subsection{Adding HX operators to the basis\label{sec:add_HX}}
The results in the previous section did not include HX operators in the basis when constructing $C_{ij}(t)$.  The authors of Ref.~\cite{Amarasinghe:2021lqa} suggested such operators may more efficiently couple to bound states than those where the nucleons are individually projected in momentum space.  
Therefore, in this section, we explore the impact of including HX operators in the resulting identification of the NN spectrum.

The local HX operator we use is constructed in a similar manner and is designed to match that used in Refs.~\cite{NPLQCD:2012mex,Wagman:2017tmp,Amarasinghe:2021lqa} as closely as possible.  For this study, we focus on $\bm{P}_{\rm tot}=0$ for both the deuteron in the the $T_{1g}$ irrep and the di-neutron in the $A_{1g}$ irrep.
In the $T_{1g}$ symmetry channel, nine operators are used: $1$ HX and $8$ nucleon-nucleon, all of which are single-site operators.  The correlator matrix is averaged over the three rows of the $T_{1g}$ irrep.  In the $A_{1g}$ symmetry channel, five operators are used: $1$ HX and $4$ nucleon-nucleon.  Again, the HX and individual nucleons are single-site operators.  These operators are summarized in \tabref{tab:slaphhexaNNops}.

%------------------------------------------------------------------------------
\begin{table}
\caption{Example $NN$ operators used in the $I=0\ T_{1g}$
and the $I=1\ A_{1g}$ channels for total momentum $\bm{P}=0$.  The operator notation and the integers in the 2nd and 3rd columns are described in the text.
\label{tab:slaphhexaNNops}}
\begin{ruledtabular}
\begin{tabular}{lcc}
        $NN$ Operator &  $I=0\ T_{1g}$ & $I=1\ A_{1g}$ \\ \hline
        $N[G_{1g}(0)]_0\ N[G_{1g}(0)]_0$ & 1 & 1\\
        $N[G_1(1)]_0\ N[G_1(1)]_0$ &  2& 1\\
        $N[G(2)]_0\ N[G(2)]_0$ & 3 & 1\\
    $N[G(3)]_0\ N[G(3)]_0$ & 2& 1\\
\end{tabular}
\end{ruledtabular}
\end{table}
%------------------------------------------------------------------------------

In Table~\ref{tab:slaphhexaNNops}, each single-nucleon operator $N$ is specified by, in square brackets, the irrep of its little group, with the squared spatial momentum, in units of $(2\pi/L)^2$, shown in parentheses, and a subscript indicating a spatial identification number.  The spin and orbital structure associated with each identification number can be obtained from the authors upon request.  The notation for the irreps follows the conventions in Ref.~\cite{Morningstar:2013bda}. The subscripts $g/u$ denote even/odd parity.
The Clebsch-Gordan coefficients that fully define each operator are not given, but are available upon request.
The numbers of independent Clebsch-Gordan combinations for each operator type 
are shown in the 2nd column of Table~\ref{tab:slaphhexaNNops} for the isosinglet channel, and in the 3rd column for the isotriplet channel.

\begin{figure}
\includegraphics[width=\linewidth]{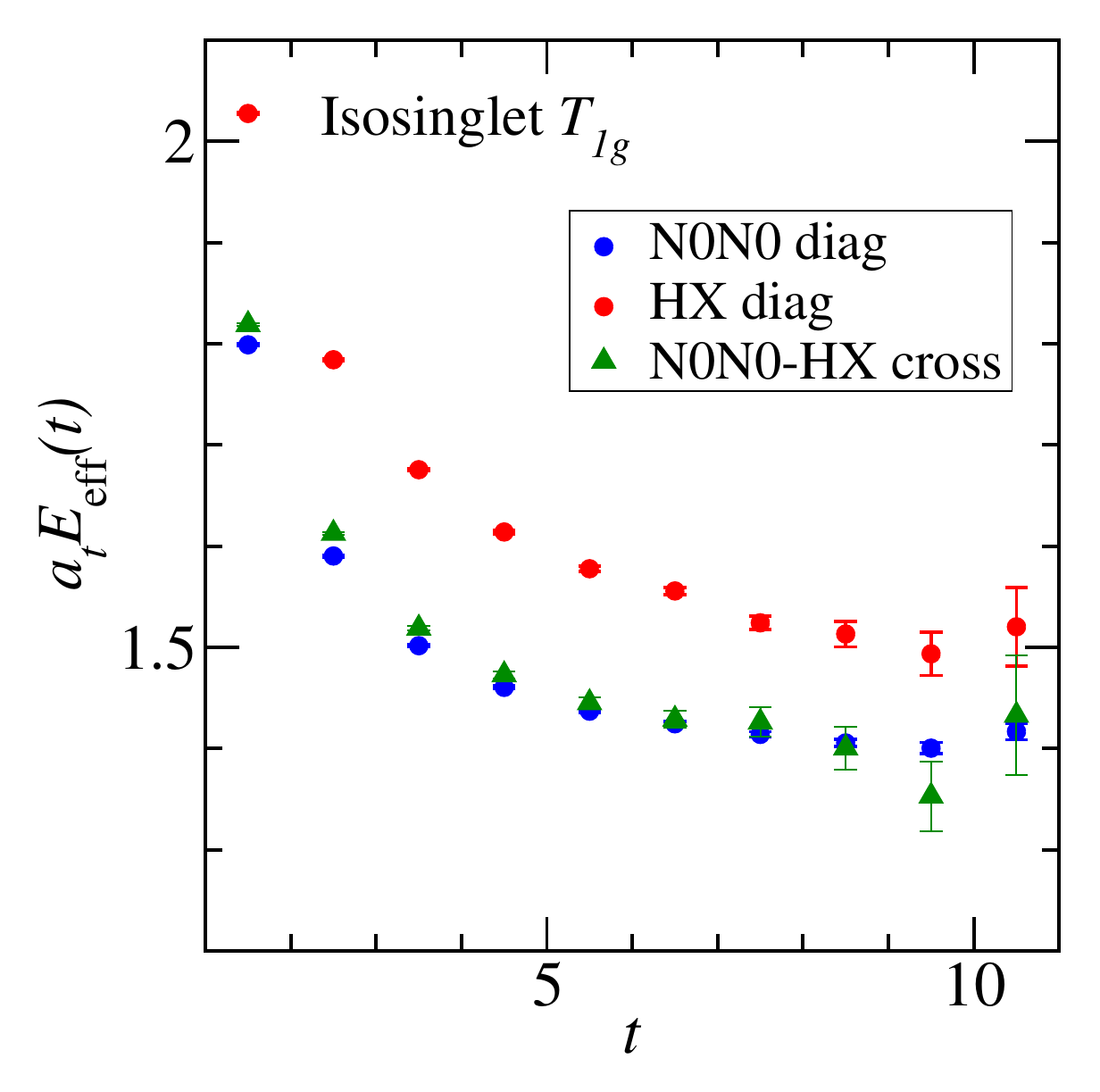}
\caption{
Comparison of the effective energy associated with the hexaquark operator (red points) with that of the two-nucleons at rest operator (blue points) and the off-diagonal correlator with an HX source (green) for the isosinglet $T_{1g}$ channel.  Excited-state contamination is significantly less for the two-nucleon operator. 
\label{fig:hexaeffsinglet}}
\end{figure}

\begin{figure}
\includegraphics[width=\linewidth]{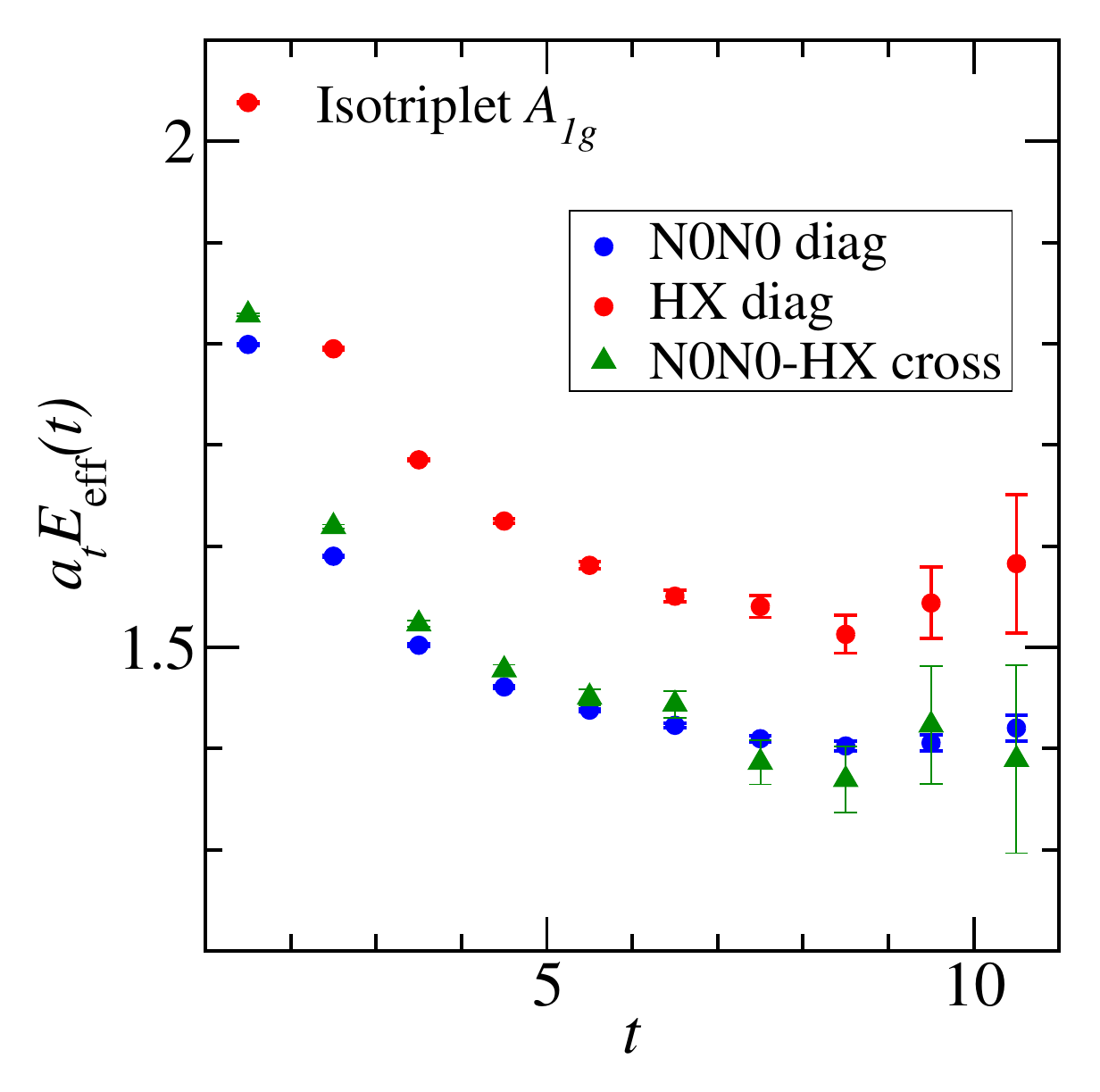}
\caption{Same as \figref{fig:hexaeffsinglet} except for the di-neutron channel.
\label{fig:hexaefftriplet}}
\end{figure}

The study is performed on a subset of the 1490 C103 configurations used in \secref{sec:slaphnn}, that being 400 of the first replica.  On each configuration, 8 randomized source times are  used, and the number of noise combinations used at the sink is $8$ for the HX operator and $14$ for each baryon operator.
As a first step, we compare the effective energies associated with the HX operators with those associated with the two-nucleons at rest (N0-N0) operators, as well as the effective energy associated with the N0N0-HX cross correlator, see \figref{fig:hexaeffsinglet} for the deuteron and \figref{fig:hexaefftriplet} for the di-neutron.

One sees that, for both isospin cases, the excited-state contamination is significantly less for the two-nucleon operator than the local HX operator.  One also sees that the statistical errors in the HX correlator are significantly larger than those from the nucleon-nucleon N0-N0 correlator.  The cross correlator in both isospin cases appears to produce an effective energy quite similar to that of the N0-N0 correlator, but with much larger statistical errors.  These observations suggest that HX operators are not necessary to extract the lowest-lying energy in each channel.

%------------------------------------------------------------------------------
%------------------------------------------------------------------------------

%------------------------------------------------------------------------------
%------------------------------------------------------------------------------    

%------------------------------------------------------------------------------    
\begin{figure}
\includegraphics[width=\linewidth]{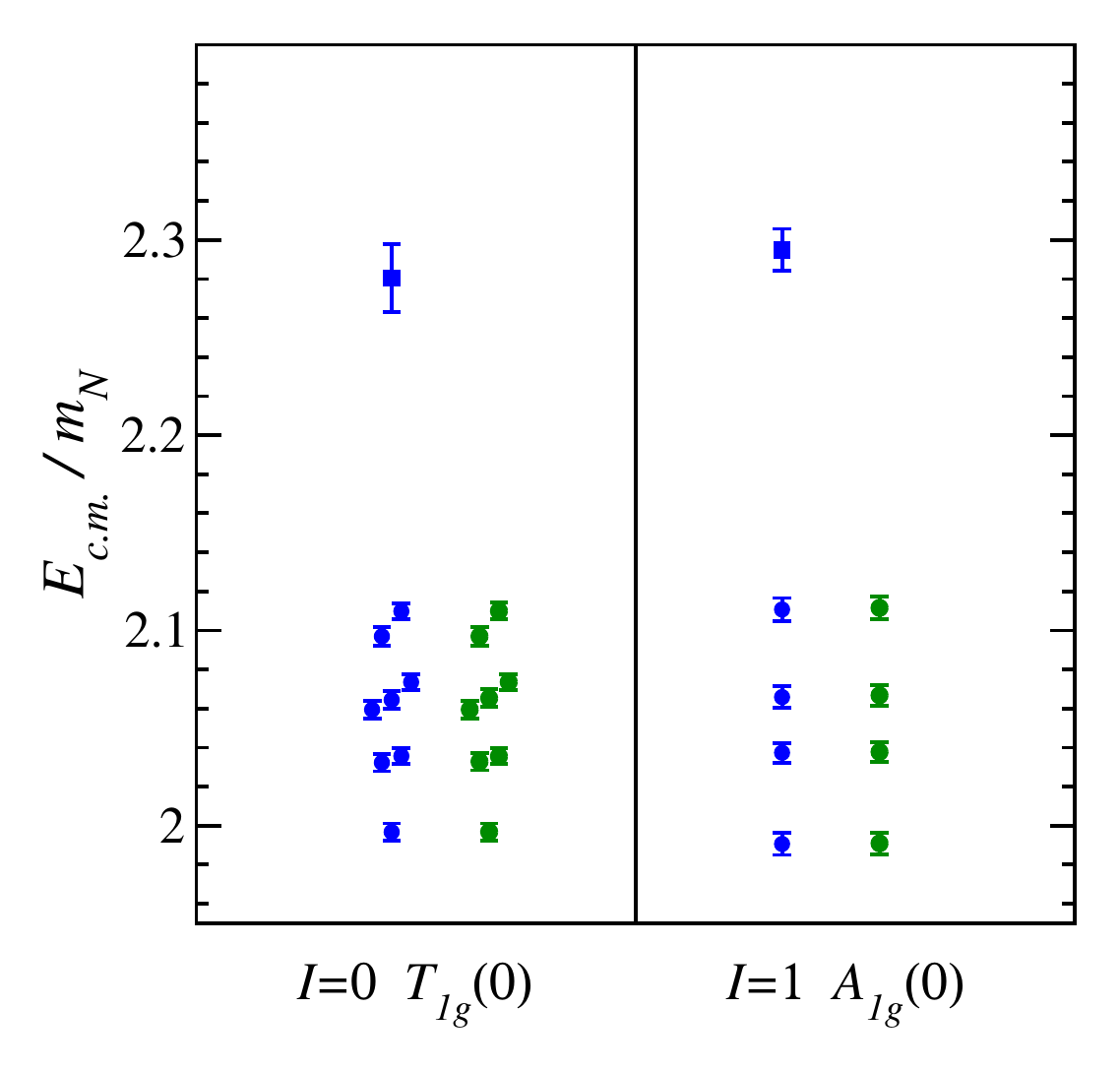}
\caption{Energy spectrum as ratios over nucleon mass for each isospin channel with zero total momentum.  The blue circles and square indicate results obtained using the entire correlation matrices, including the hexaquark operator.  Green circles show the energies obtained using the correlation matrices excluding the hexaquark operators.  The blue square in each channel indicates the energy corresponding to a hexaquark-dominated level.
\label{fig:hexaspectrum}}
\end{figure}
%------------------------------------------------------------------------------    

\begin{figure*}
\includegraphics[width=0.32\linewidth]{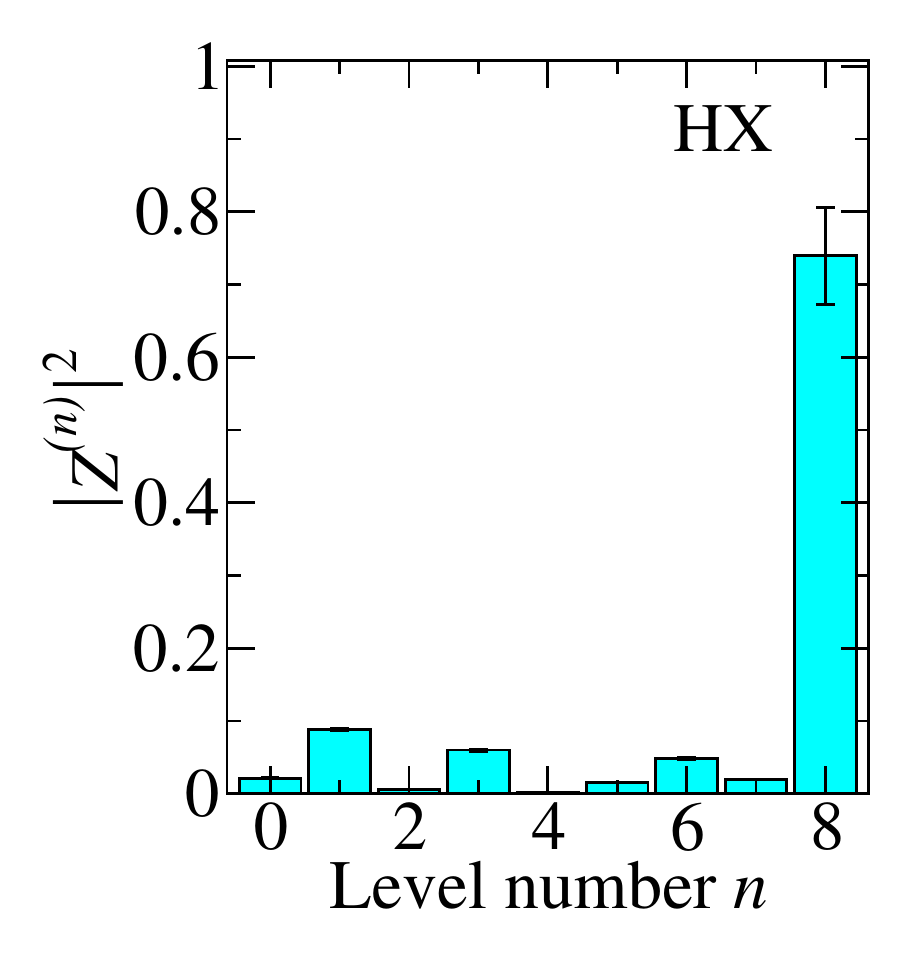}
\includegraphics[width=0.32\linewidth]{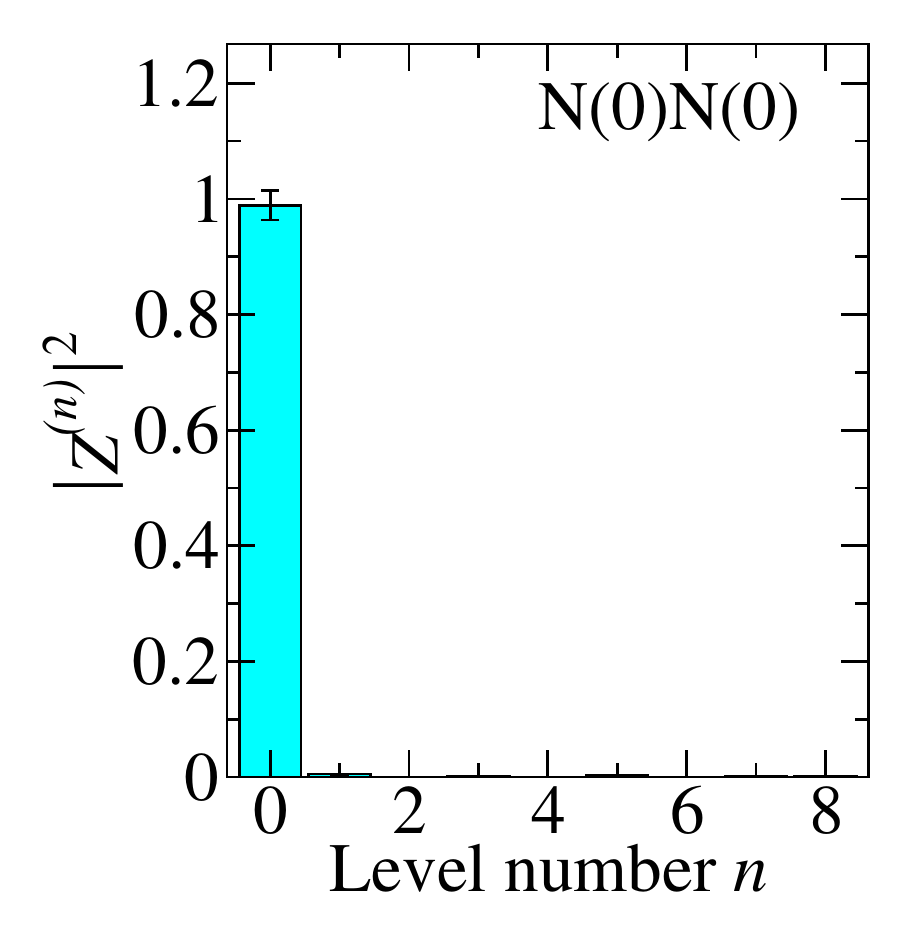}
\includegraphics[width=0.32\linewidth]{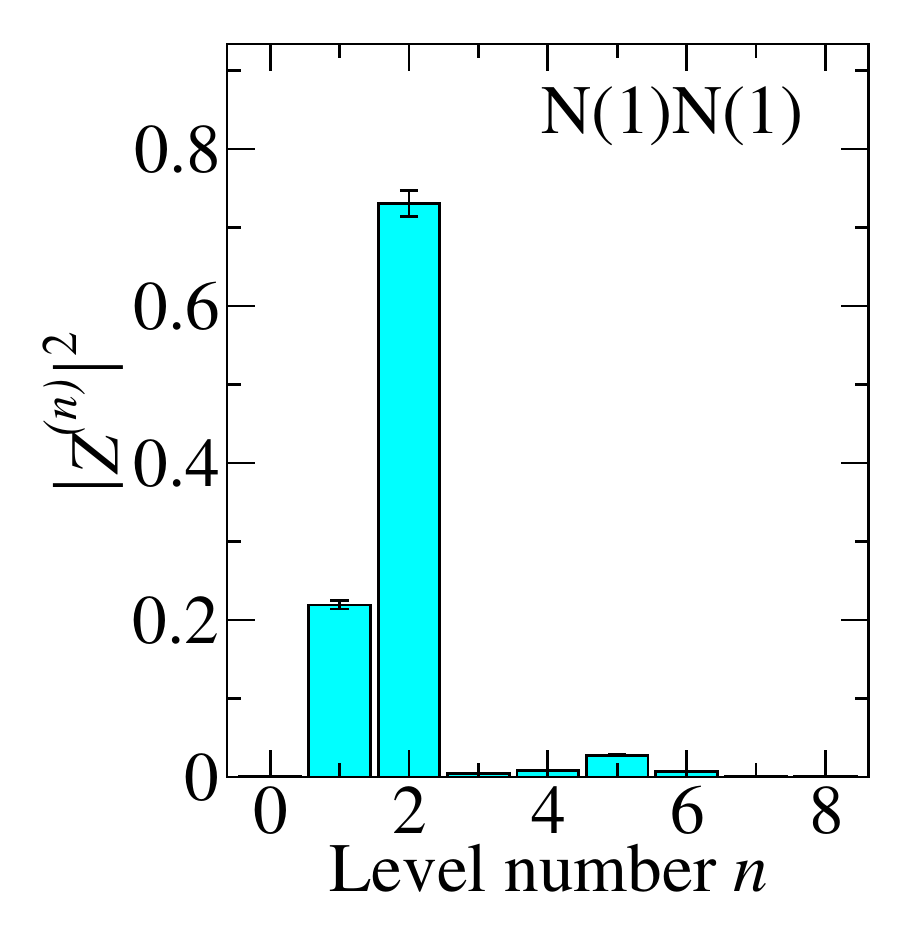}\\[-2mm]
\caption{Overlap factors for the hexaquark and two momentum based operators in the isosinglet $T_{1g}$ channel.   $N(\bm{d}^2)N(\bm{d}^2)$ refers to an operator expected to predominantly create or annihilate a two-nucleon state in which one nucleon has momentum $\bm{p}=(2\pi/L)\bm{d}$, where $\bm{d}$ is a three-vector of integers, and the other nucleon has momentum $-\bm{p}$.
\label{fig:hexaovlapI0}}
\end{figure*}

\begin{figure*}
\begin{center}
\includegraphics[width=0.32\linewidth]{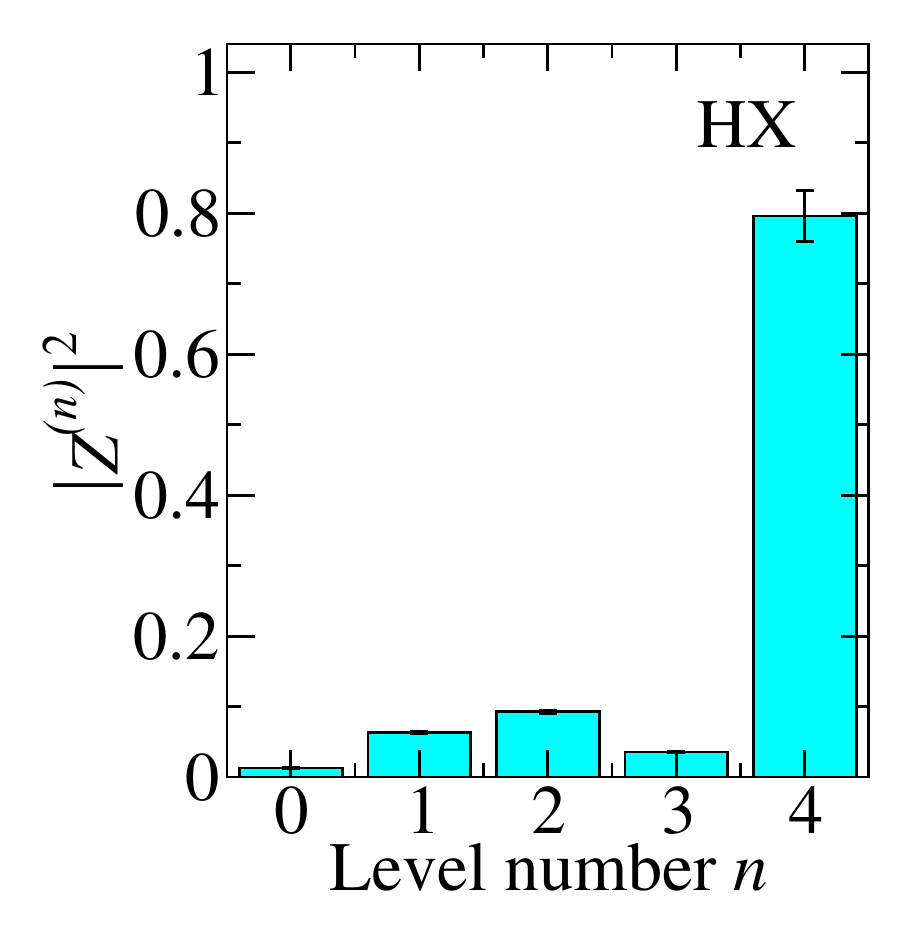}
\includegraphics[width=0.32\linewidth]{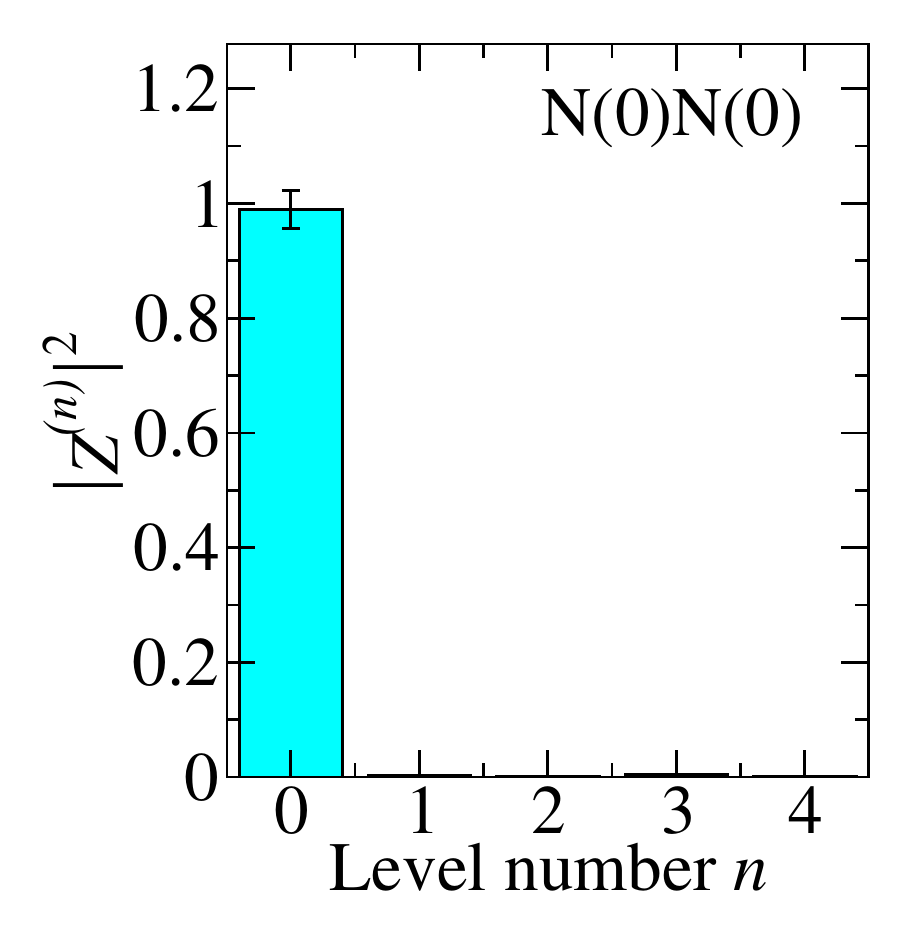}
\includegraphics[width=0.32\linewidth]{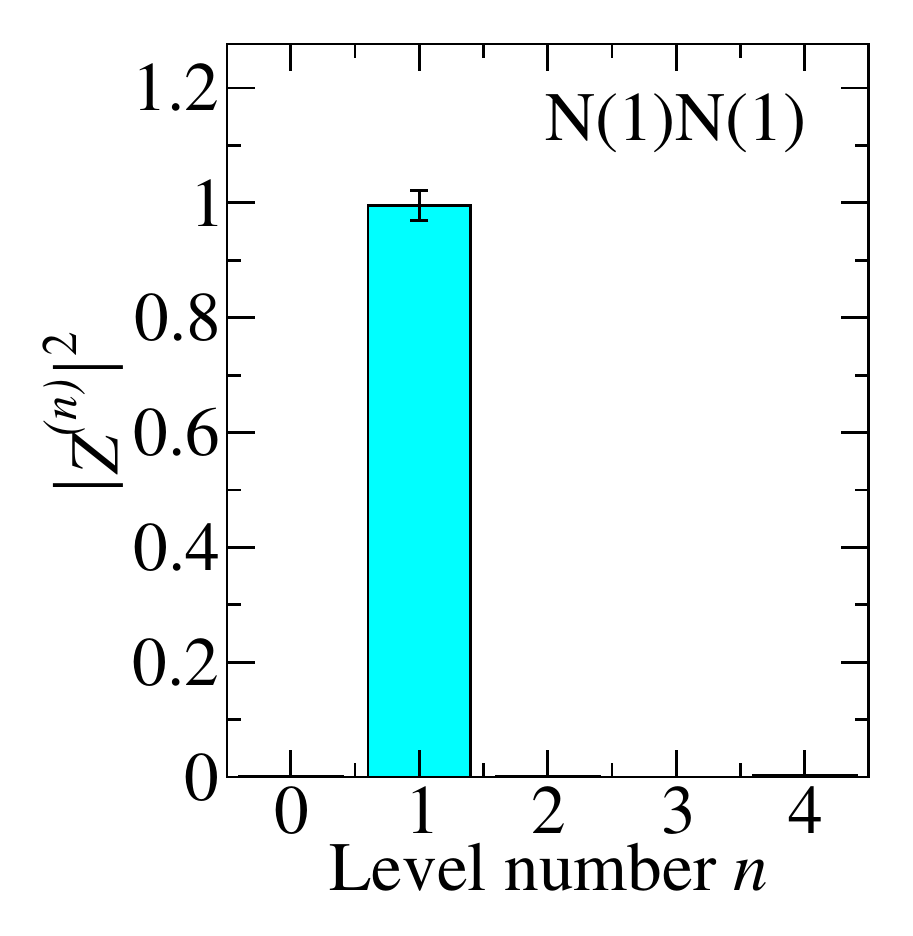}\\
\end{center}
\caption{Overlap factors for the hexaquark and two momentum-based $NN$ operators in the isotriplet $A_{1g}$ channel (similar to Fig.~\ref{fig:hexaovlapI0}).
\label{fig:hexaovlapI1}}
\end{figure*}

In \figref{fig:hexaspectrum}, we show the extracted NN spectrum for both the $T_{1g}$ (left) and $A_{1g}$ (right) symmetry channels with (left/blue) and without (right/green) the HX operators in the basis.  One observes that
\begin{enumerate}
\item The addition of the HX operators leads to the identification of a new state, but it is high up in the spectrum;

\item The determination of the low-lying spectrum is not affected by the inclusion of the HX operator.
\end{enumerate}
Using \eqnref{eq:Z_jn}, we can determine how well each operator couples to the various energy levels.  In \figref{fig:hexaovlapI0} and \figref{fig:hexaovlapI1}, we show the overlap of the HX operator and the two lowest momentum NN operators on the various energy levels for the deuteron and di-neutron respectively.
In both cases, the HX operator clearly couples strongly to the highest energy level and relatively poorly to all other levels in the system, consistent with the observations above.

From this study, we conclude that the HX operators are not needed to correctly identify the low-lying NN spectrum, nor do they couple well to these states of interest for extracting the low-energy NN amplitudes.

%------------------------------------------------------------------------------    
%------------------------------------------------------------------------------    

%------------------------------------------------------------------------------    
%------------------------------------------------------------------------------    

\subsection{Off diagonal correlators with HX operators\label{sec:hx_off_diagonal}}

\begin{figure}
\begin{center}
\includegraphics[width=1.65in]{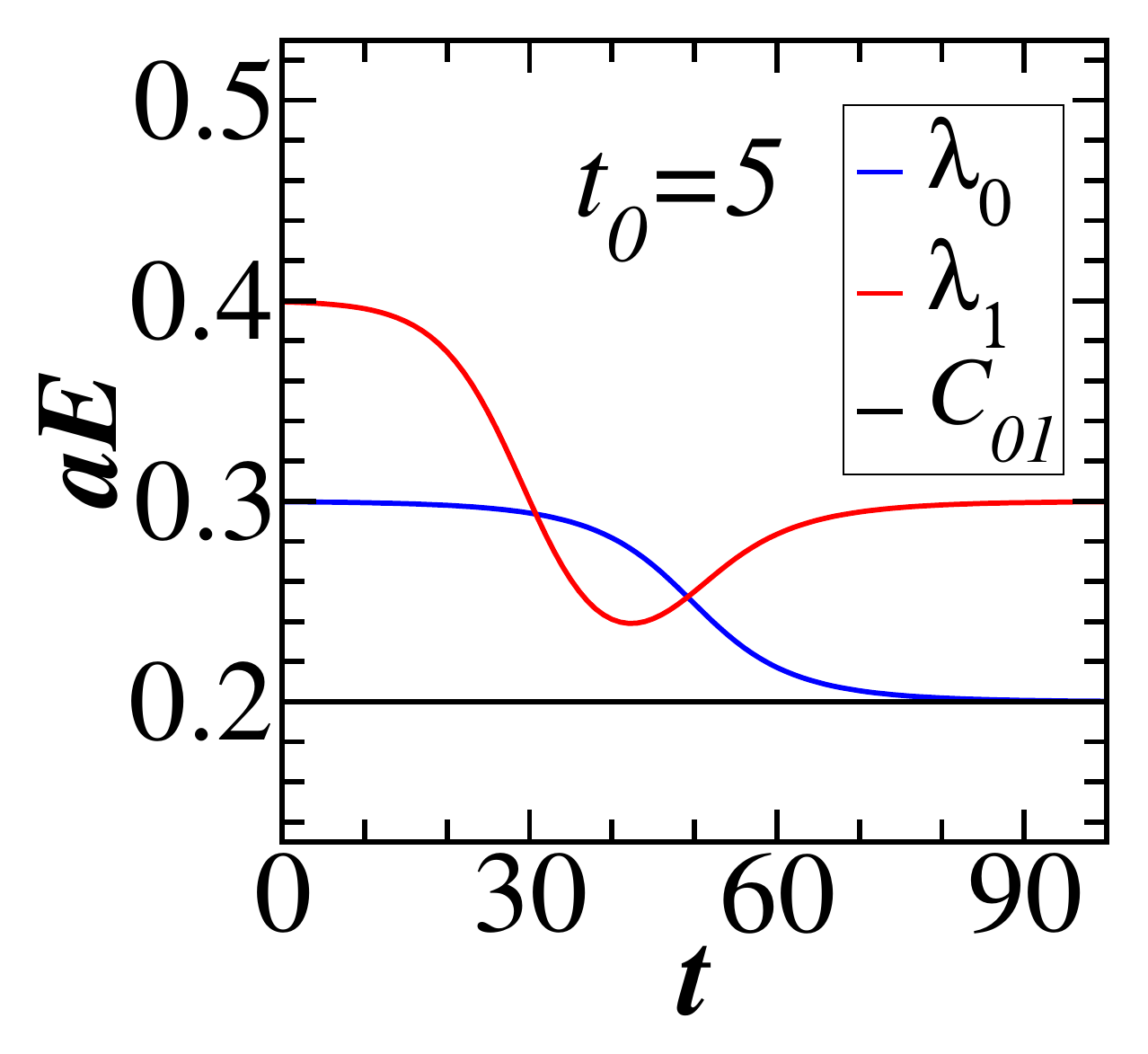}
\includegraphics[width=1.65in]{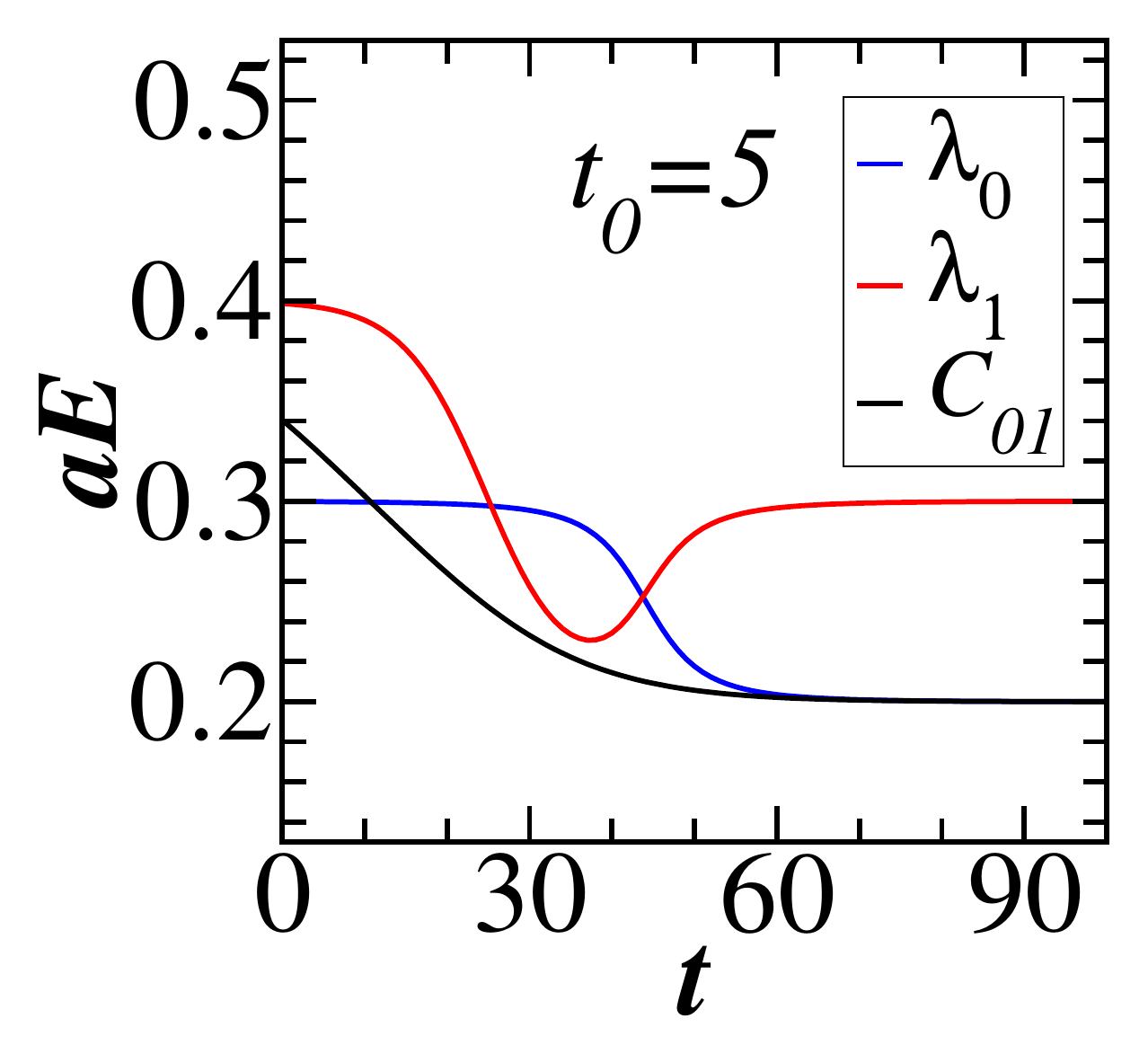}
\end{center}
\caption{
(Left) Effective masses corresponding to the two principal correlators $\lambda_0(t),\lambda_1(t)$,
shown as blue and red curves, respectively, and the effective mass corresponding to the off-diagonal
correlator $C_{01}(t)$, shown as a black curve, for the unrealistic toy model from 
Refs.~\protect\cite{Tews:2022yfb,Amarasinghe:2021lqa} with parameters values given in 
Eq.~(\ref{eq:toy1}).  The parameter values have been cleverly engineered so that the off-diagonal 
correlator leads to a perfect extraction of energy $E_0$.
(Right) The same quantities as in the left plot, but using the more realistic parameters
values of Eq.~(\ref{eq:toy2}). When parameter values are not chosen to produce a perfect
off-diagonal correlator, the black and blue curves tend to $E_0$ around the
same temporal separation $t\approx 50$.
\label{fig:nplqcdtoy}}
\end{figure}

In Chapter 15 of Ref.~\cite{Tews:2022yfb} and in Ref.~\cite{Amarasinghe:2021lqa}, the authors suggested that including local HX-type operators may be important for 
revealing their elusive bound state in the spectrum, arguing that the off-diagonal correlators may provide
a more effective means to find them.  These authors use a toy model to illustrate
their point.  They introduce a three-state system with energies $E_0<E_1<E_2$ and
two operators $A,B$, where the energies and overlap factors are given by
\begin{eqnarray}
    & E_0 = \eta-\Delta,\qquad E_1=\eta,\qquad E_2=\eta+\delta, &\\
   &   Z_A=(\epsilon,\sqrt{1-\epsilon^2},0),\quad 
  Z_B = (\epsilon,0,\sqrt{1-\epsilon^2}).&
   \label{eq:toy1}
\end{eqnarray}
Both operators $A,B$ create states which have a small but identical overlap $\epsilon$ with
the lowest-lying state of energy $E_0$, the state created by operator $A$ predominantly overlaps
the first excited state of energy $E_1$ and has an exact zero overlap with the $E_2$ state, and the state
created by operator $B$ predominantly creates the second excited state of energy $E_2$ and has an exact 
zero overlap with the $E_1$ state. The off-diagonal correlation function constructed from the two operators $A$ and $B$ in this case goes as,
 \begin{equation}
 C(t) = \epsilon^2 e^{-E_0 t} \ .
 \end{equation}
 Effective mass plots using values $\epsilon=0.05,\ \eta=0.3,\ \Delta=0.1,\ \delta=0.1$ are shown in the left plot of
Fig.~\ref{fig:nplqcdtoy}.  The effective masses for the so-called principal correlators
$\lambda_0,\lambda_1$ are shown, as well as the effective mass corresponding to the
off-diagonal correlator $C_{01}$. The principal correlators are the eigenvalues of
the matrix $C(t_0)^{-1/2}\,C(t)\, C(t_0)^{-1/2}$.  Here, we used $t_0=5$ with
eigenvector pinning to identify the levels.  The black curve shows the perfect extraction
of $E_0$ starting at zero temporal separation from the off-diagonal
correlator.  The blue curve associated with principal correlator $\lambda_0(t)$ eventually
plateaus at $E_0$, but not until large temporal separations are reached due to the small
coupling to the $E_0$ state.  The red curve is associated with the principal correlator
$\lambda_1(t)$ which plateaus eventually to $E_1$.

This is a contrived and unrealistic (if not impossible to attain) toy model since it has been engineered to produce 
a perfect energy extraction from the off-diagonal correlator due to the exact zeros in the
$Z_A,Z_B$ factors. To achieve this situation in general would require an exact diagonalization of the full volume of lattice operators, not the use of a single operator. Making only a single change, allowing an arbitrarily small, non-zero overlap with these states,
\begin{equation}
    Z_A=(\epsilon_1,\sqrt{1-\epsilon_1^2-\epsilon_2^2},\epsilon_2),\quad 
    Z_B = (\epsilon_1,\epsilon_2,\sqrt{1-\epsilon_1^2-\epsilon_2^2}).
\end{equation} 
immediately makes the effect go away, with the off-diagonal correlation function now scaling as,
\begin{equation} 
 C(t) = \epsilon_2 e^{-E_1 t} + \epsilon_2 e^{-E_2 t}  + \epsilon_1^2 e^{-E_0 t} + \cdots \ .
\end{equation}
Here we see that, in fact, it is the higher excited states that dominate at intermediate times, as the overlap onto the ground state is $\mathcal{O}(\epsilon^2)$.
 
An even more realistic model is studied in the right-hand plot of Fig.~\ref{fig:nplqcdtoy}.  For this plot, two simple changes are
made from the original toy model~\cite{Amarasinghe:2021lqa,Tews:2022yfb}: the overlaps with the ground state are not set to be the same for the states 
created by operators $A$ and $B$, and the exact zeros in the $Z_A,Z_B$ factors are replaced
by small but non-zero values:
\begin{equation}
   Z_A=(\epsilon_1,\sqrt{1-\epsilon_1^2},\epsilon_3),\quad 
  Z_B = (\epsilon_2,\epsilon_3,\sqrt{1-\epsilon_2^2}).
   \label{eq:toy2}
\end{equation} 
Effective mass plots using values $\epsilon_1=0.04$, $\epsilon_2=0.08$, $\epsilon_3=0.03$
with the same values of $\eta=0.3,\ \Delta=0.1,\ \delta=0.1$ are shown in the right plot of
Fig.~\ref{fig:nplqcdtoy}.  One sees that when parameter values are not cleverly engineered
to produce a perfect off-diagonal correlator, the black and blue curves tend to $E_0$ around
the same temporal separation $t\approx 50$, completely eliminating the advantage of the
off-diagonal correlator.  

\begin{figure}
\begin{center}
\includegraphics[width=1.65in]{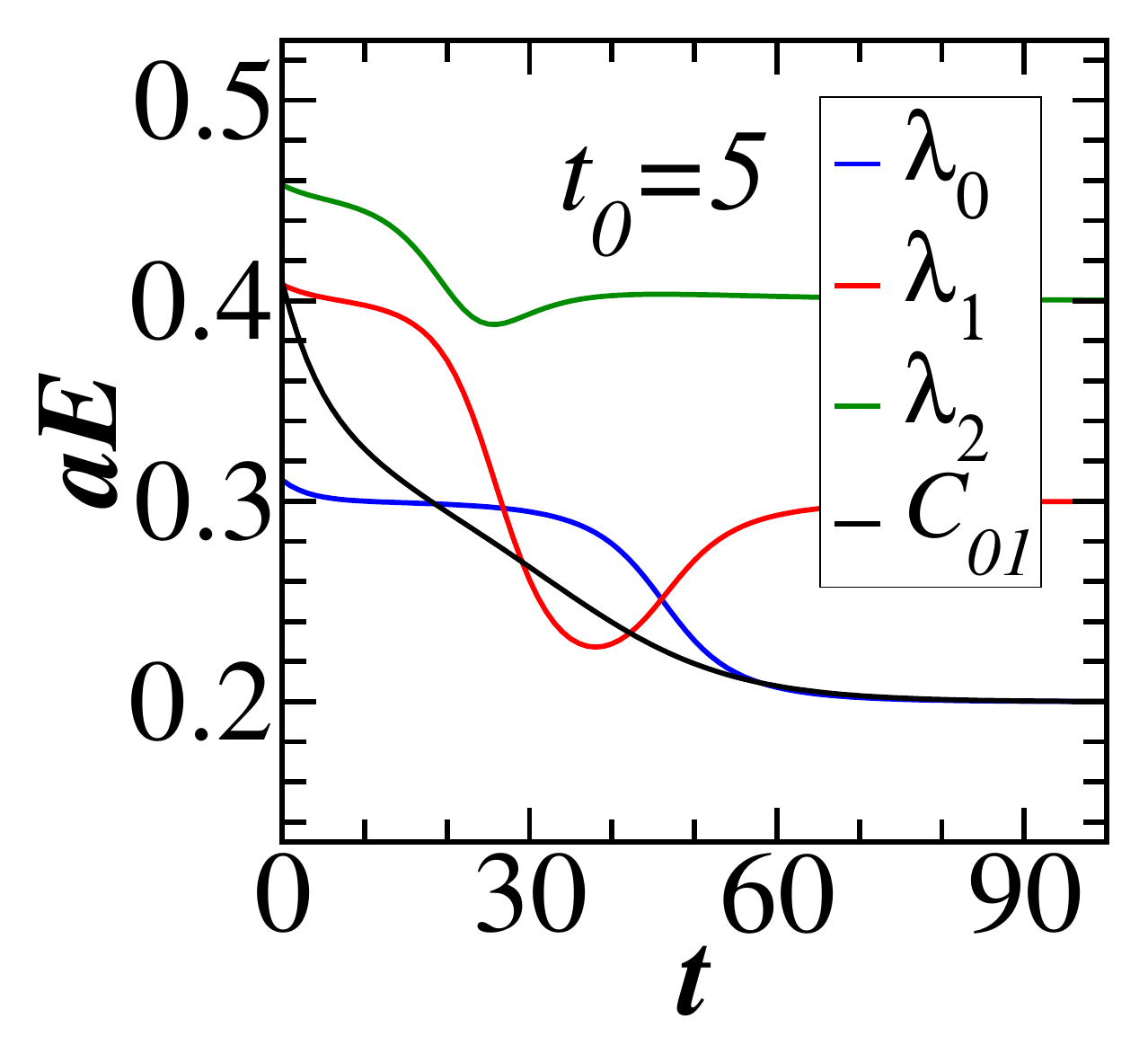}
\includegraphics[width=1.65in]{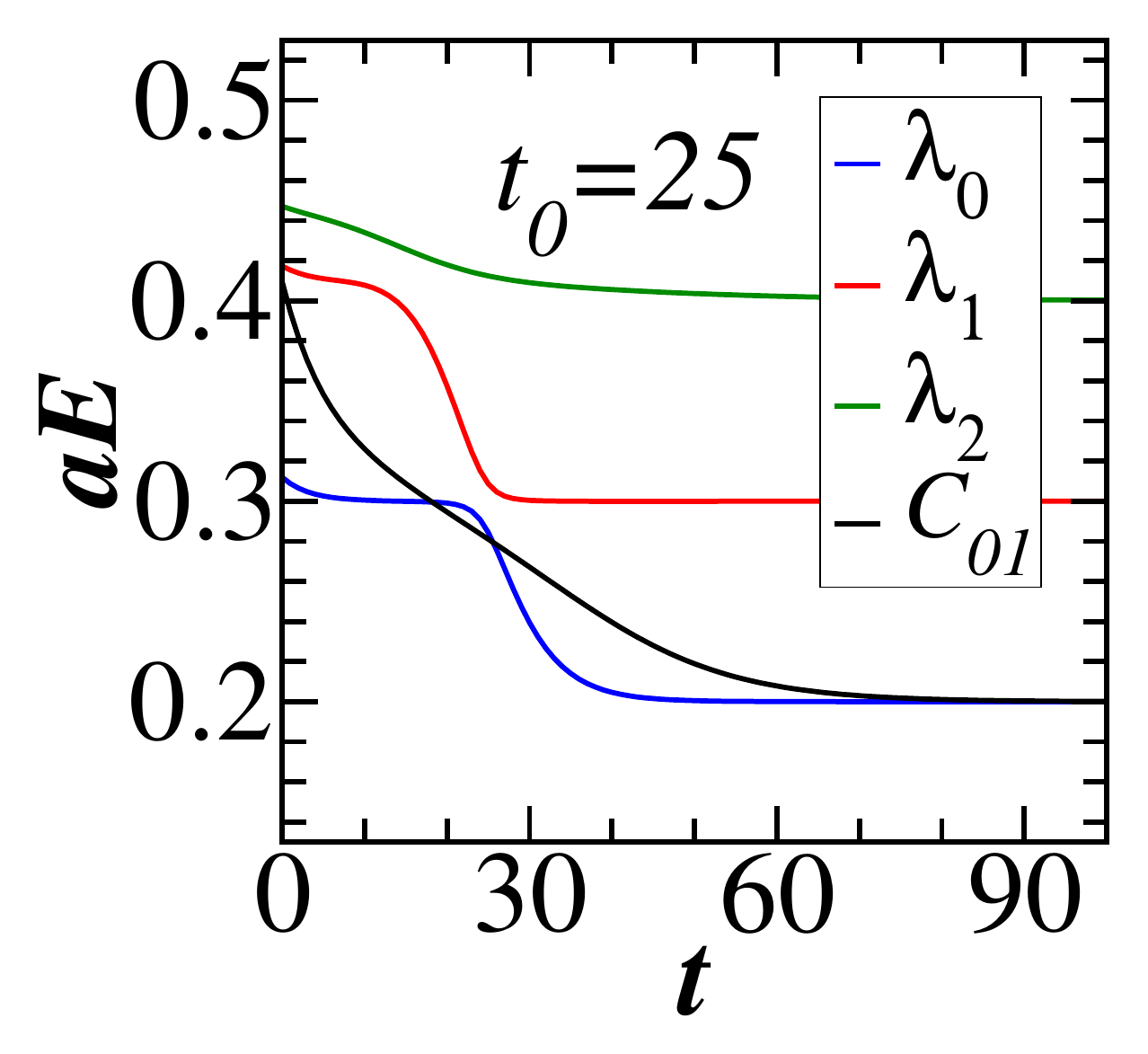}
\includegraphics[width=1.65in]{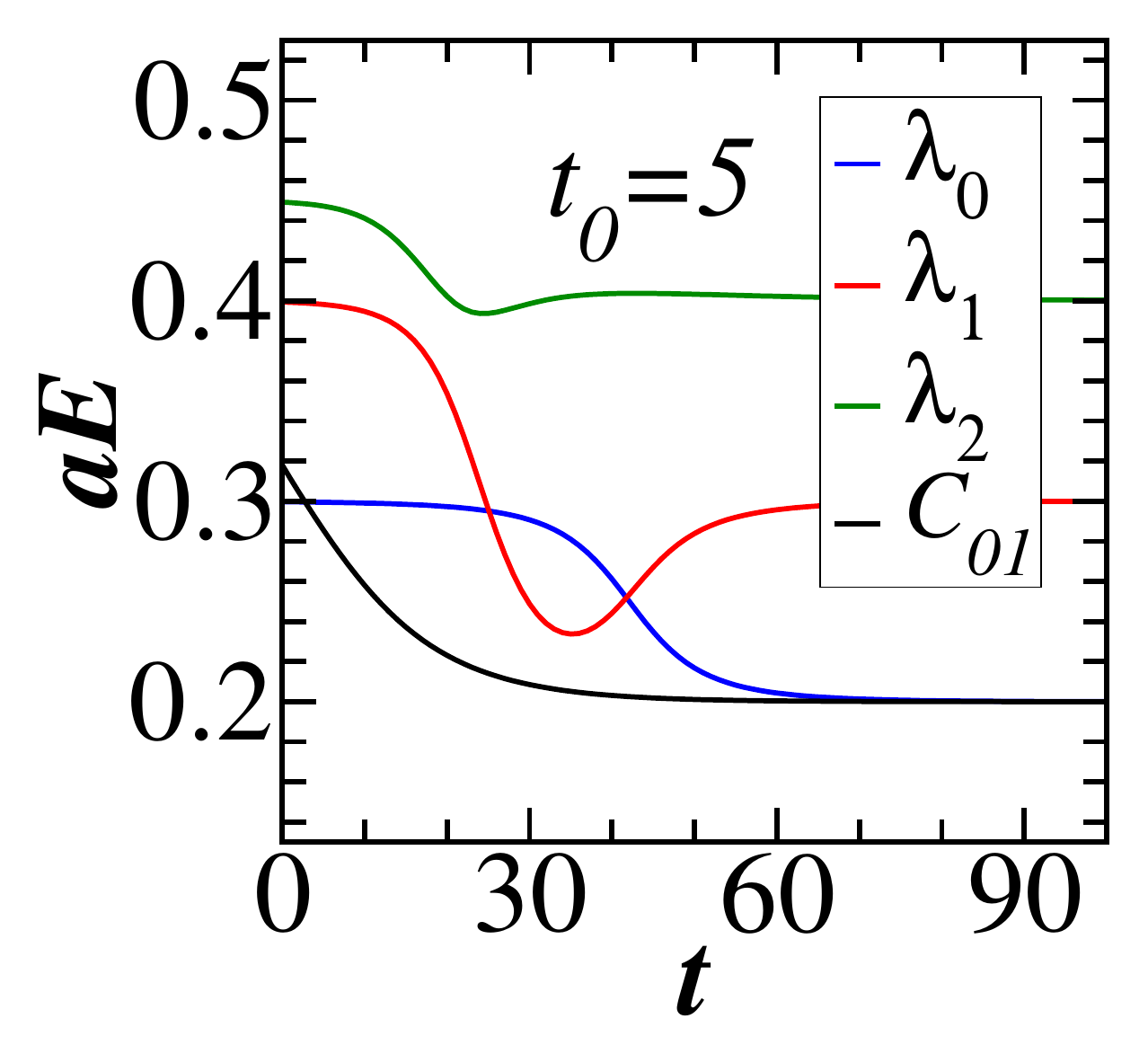}
\includegraphics[width=1.65in]{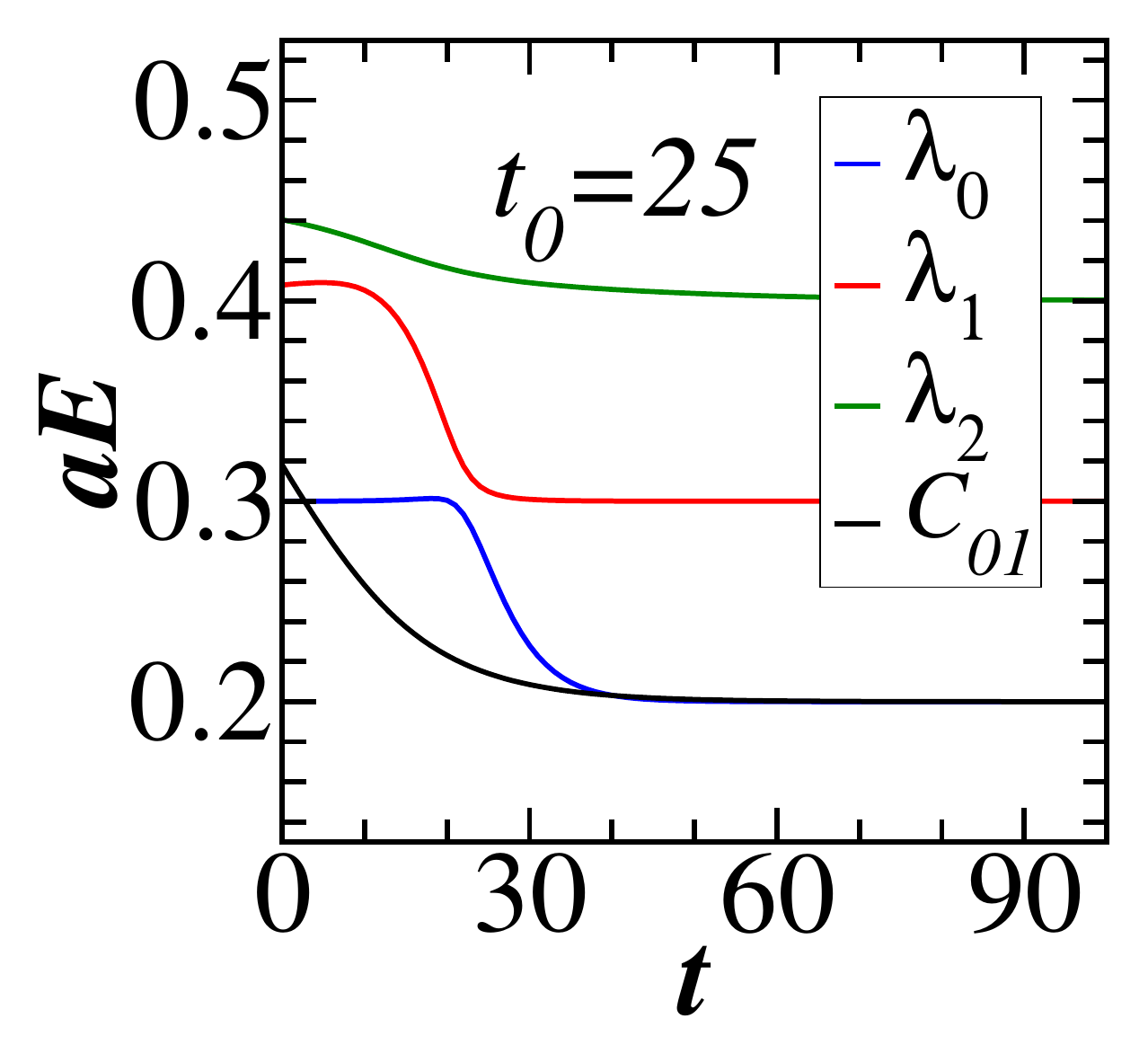}
\end{center}
\caption{
(Top left) Effective masses corresponding to the three principal correlators $\lambda_0(t),\lambda_1(t),\lambda_2(t)$,
shown as blue, red and green curves, respectively, and the effective mass corresponding to the off-diagonal
correlator $C_{01}(t)$, shown as a black curve, for the 10-level, 3-operator toy model described
in the text, with parameter values given in Eq.~(\ref{eq:bettertoy}).  Principal correlators have
been computed using $t_0=5$.  (Top right) Same as the plot on the top left, except the principal correlators
have been determined using $t_0=25$.  No advantage is observed for the off-diagonal correlator. (Bottom left and right) Same as top row, except $\delta=0.005$ is used.  On the left, an advantage is seen for the off-diagonal correlator, but this goes away as $t_0$ is increased (see right).
\label{fig:bettertoy}}
\end{figure}

This model is still unrealistic since there are no couplings to the many higher levels
as there would be in actual lattice QCD computations.
Lastly, consider a more realistic toy model with 10 levels and 3 operators $A,B,C$
and parameters chosen simply to be
\begin{eqnarray}
   & \mbox{Energies}:  (0.2, 0.3, 0.4, 0.45, 0.50, 0.55, 0.60, \nonumber\\
   &  \qquad\qquad 0.65, 0.70, 0.75), & \nonumber\\
   &   Z_A=(\epsilon, x, \delta, \delta,\delta,\delta,\delta,\delta,\delta,\delta),\quad x=0.90,&\nonumber\\
   &   Z_B=(\epsilon, \delta, x, \delta, \delta,\delta,\delta,\delta,\delta,\delta ),\quad \delta=0.10,&\nonumber\\
   &   Z_C=(\epsilon, \delta, \delta, x, \delta, \delta,\delta,\delta,\delta,\delta ),\quad \epsilon=0.05.& 
 \label{eq:bettertoy}
\end{eqnarray}
The effective masses corresponding to the three principal correlators  
$\lambda_0(t),\lambda_1(t),\lambda_2(t)$, as well as the effective mass corresponding to
the off-diagonal correlator $C_{01}(t)$, are shown in the top row of Fig.~\ref{fig:bettertoy}.  
The principal correlators have been determined using $t_0=5$ in the left plot, and
$t_0=25$ in the right plot.  The off-diagonal correlator shows no advantage.  In the bottom row, a small $\delta=0.005$ is assumed.  An advantage is observed in the bottom left plot for $t_0=5$, but this goes away as $t_0$ is increased, as shown in the bottom right plot.

\begin{figure}
\begin{tabular}{cc}
\includegraphics[width=0.49\columnwidth]{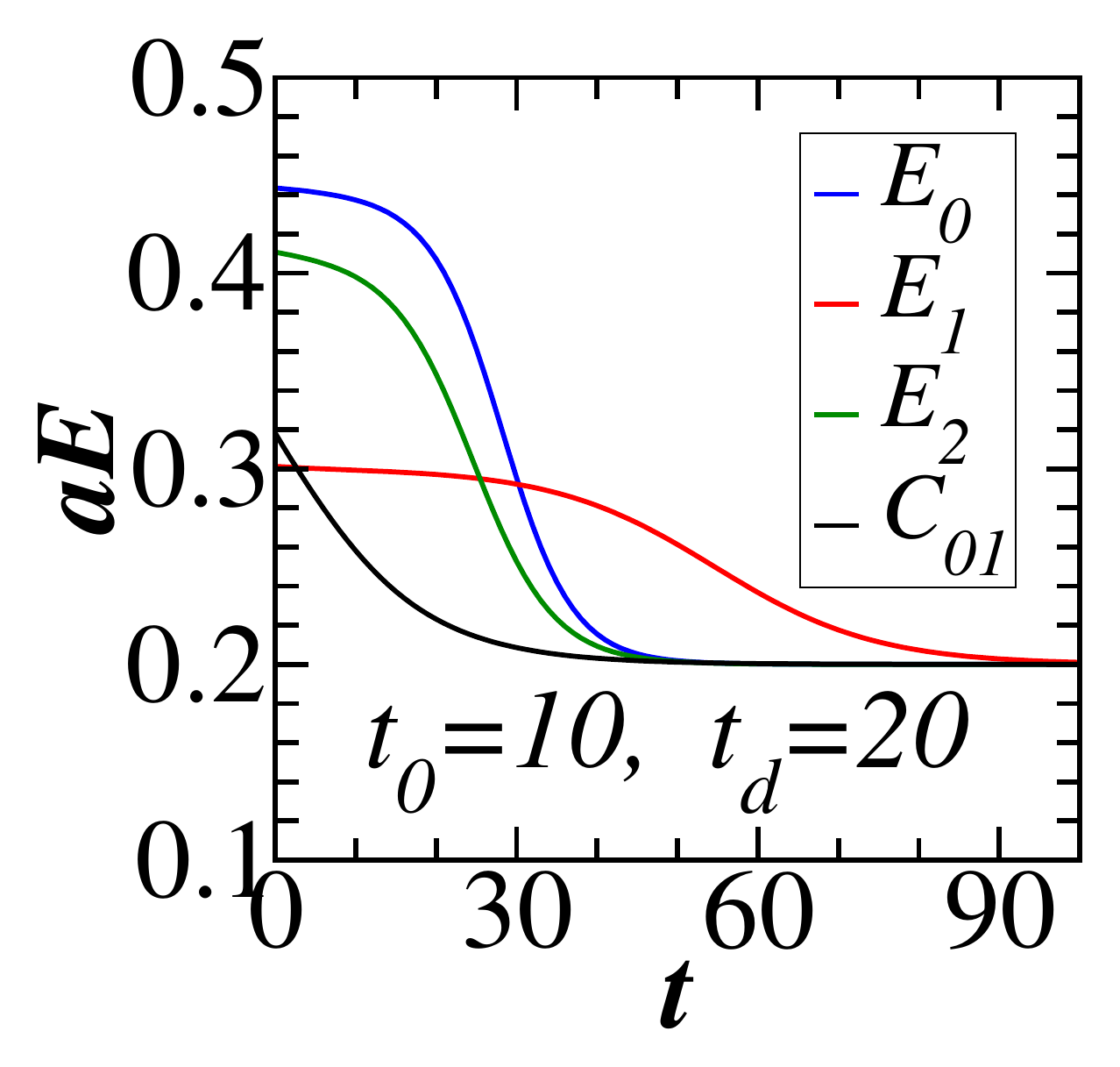}
&
\includegraphics[width=0.49\columnwidth]{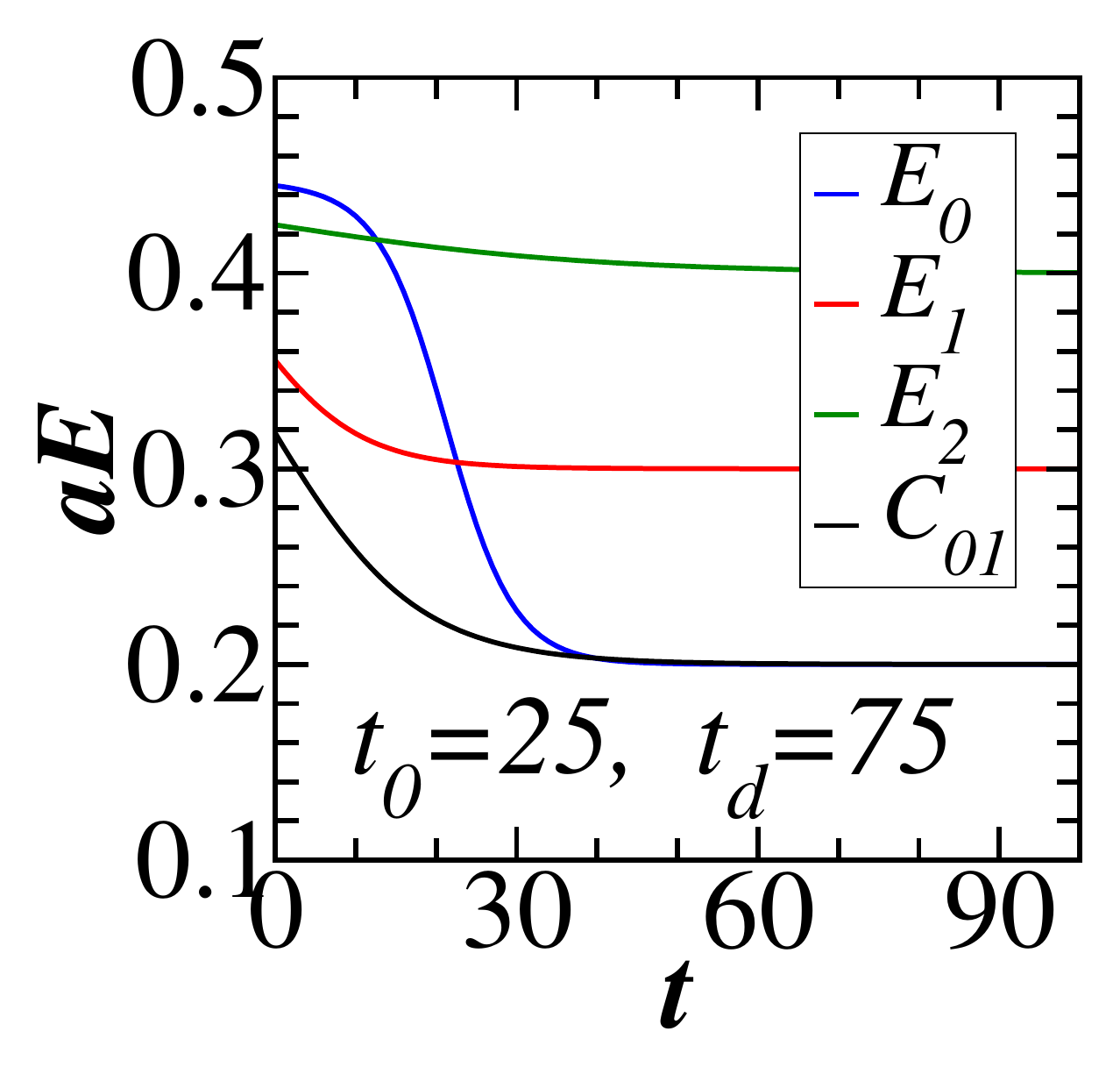}
%\\
%$t_0=10$, $t_d=20$&
%$t_0=25$, $t_d=75$
\end{tabular}
\caption{\label{fig:bettertoy_singlepivot}
Effective energies associated with diagonal elements of the rotated correlator matrix of \eqnref{eq:spivC} for the model given in \eqnref{eq:bettertoy} but with $\delta=0.005$ for two choices of $t_0, t_d$.}
\end{figure}

These toy-models also highlight the importance of carefully choosing the times used in constructing the rotation matrices through the GEVP, particularly $t_0$.
In the single-pivot method described in \secref{sec:slaphnn}, the effective masses associated with the diagonal elements of the rotated correlator matrix given in Eq.~(\ref{eq:spivC}) are guaranteed to monotonically decrease to the lowest-lying energy whose eigenstate has nonzero overlap with the state created by the rotated operator.  When the principal correlator method is used (described in \secref{sec:slaphnn}), as was done to generate \figref{fig:nplqcdtoy} and \figref{fig:bettertoy}, this monotonicity is no longer guaranteed since the overlap factors can acquire time dependence.  The dramatic variations observed in the effective masses shown \figref{fig:nplqcdtoy} and \figref{fig:bettertoy} for small temporal separations are a symptom of the fact that the operators are poorly coupled to the lowest energy state in the system.  In such a situation, the importance of studying the dependence on the extracted spectrum as a function of $t_0$ and $t_d$ is amplified.  The effective energies for two different choices of $t_0, t_d$ using the single-pivot method applied to the toy model of \eqnref{eq:bettertoy} are shown in \figref{fig:bettertoy_singlepivot}.  On the left, one sees that $t_0,t_d$ have been chosen much too small, producing a rotated correlator matrix that does not remain diagonal for times larger than $t_d$.  On the right, the rotated correlator remains diagonal for times larger than $t_d$, indicating a good choice of $t_0, t_d$.  This great sensitivity to $t_0, t_d$ in this model is again a symptom of the poor operator choices.
For the LQCD results presented in this work, we find no evidence of such behavior.

Our conclusion is that it is nearly impossible for an off-diagonal correlator to reveal
a level hidden from the correlator matrix method, especially for a hexaquark operator
that has been shown to contain copious amounts of excited contamination.
In fact, negative weights in the spectral representation of an off-diagonal correlator with
a slow fall-off of contributions from higher-lying states are more 
likely to cause plateau misidentifications. 

A preponderance of the evidence suggests that the use of an off-diagonal correlator and 
plateau misidentification was the most likely culprit for the discrepancy between early 
lattice QCD studies of $NN$ scattering at the $SU(3)_F$ symmetric point which claimed 
bound state formation (see Refs.~\cite{NPLQCD:2012mex,NPLQCD:2013bqy,Berkowitz:2015eaa}, 
among others) and more recent studies which make use of Hermitian correlation matrices
(see, for example, Refs.~\cite{Francis:2018qch,Horz:2020zvv,Amarasinghe:2021lqa,Green:2021qol,
Green:2022rjj,Geng:2024dpk} and this work) which find no bound state formation.

\begin{figure*}
\begin{center}
\includegraphics[width=0.46\textwidth]{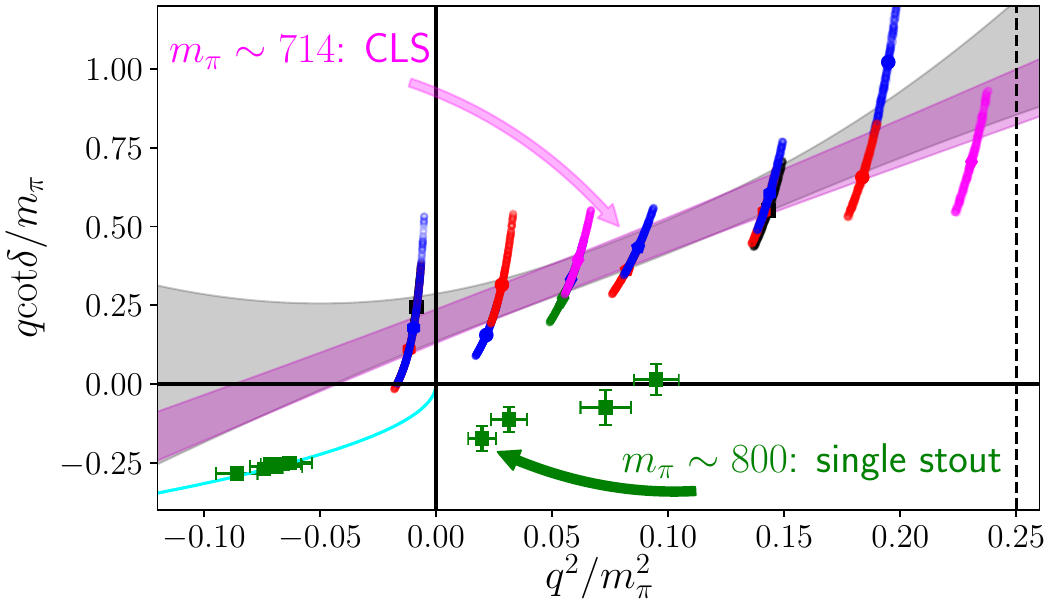}
\includegraphics[width=0.26\textwidth]{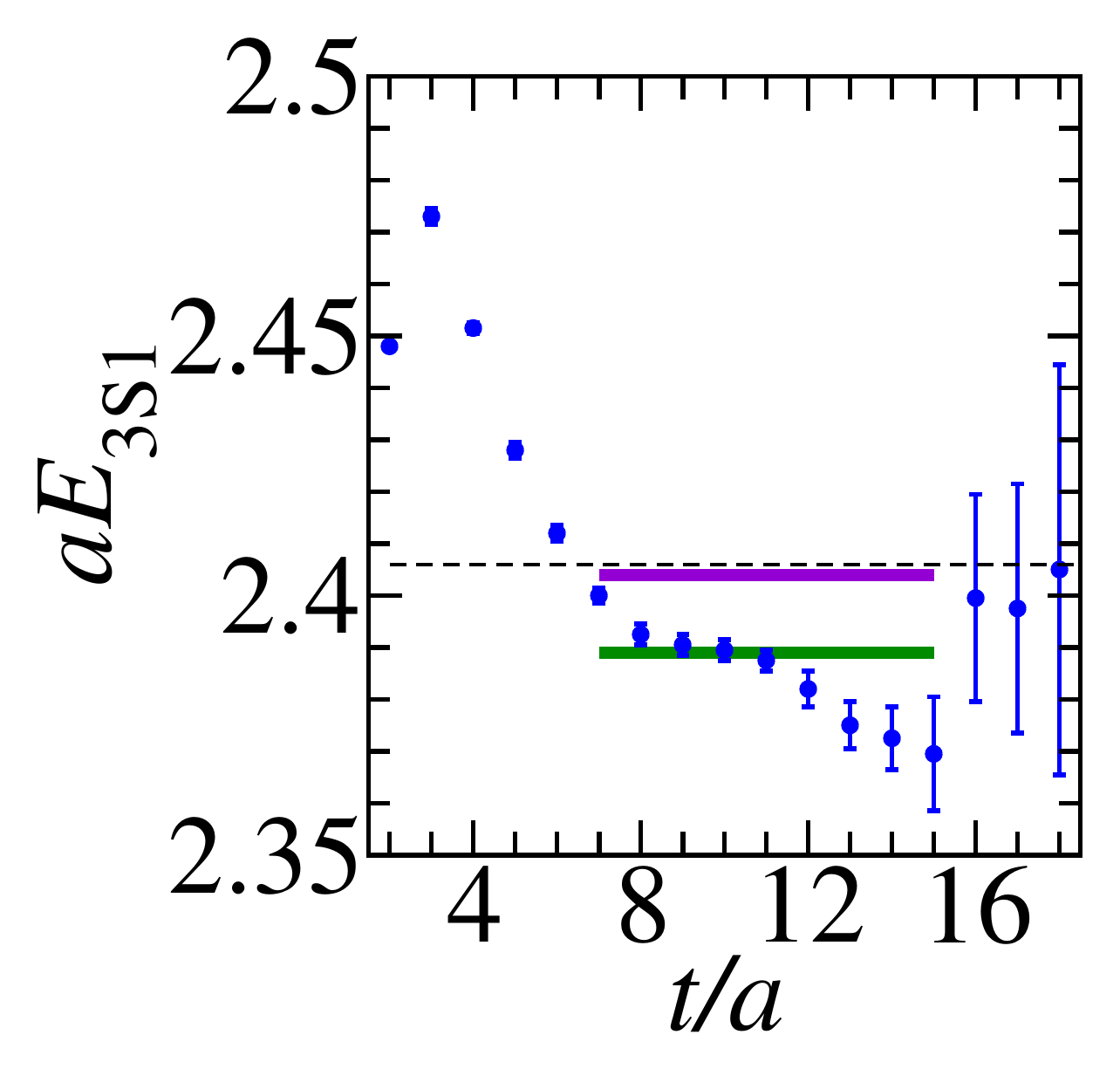}
\includegraphics[width=0.26\textwidth]{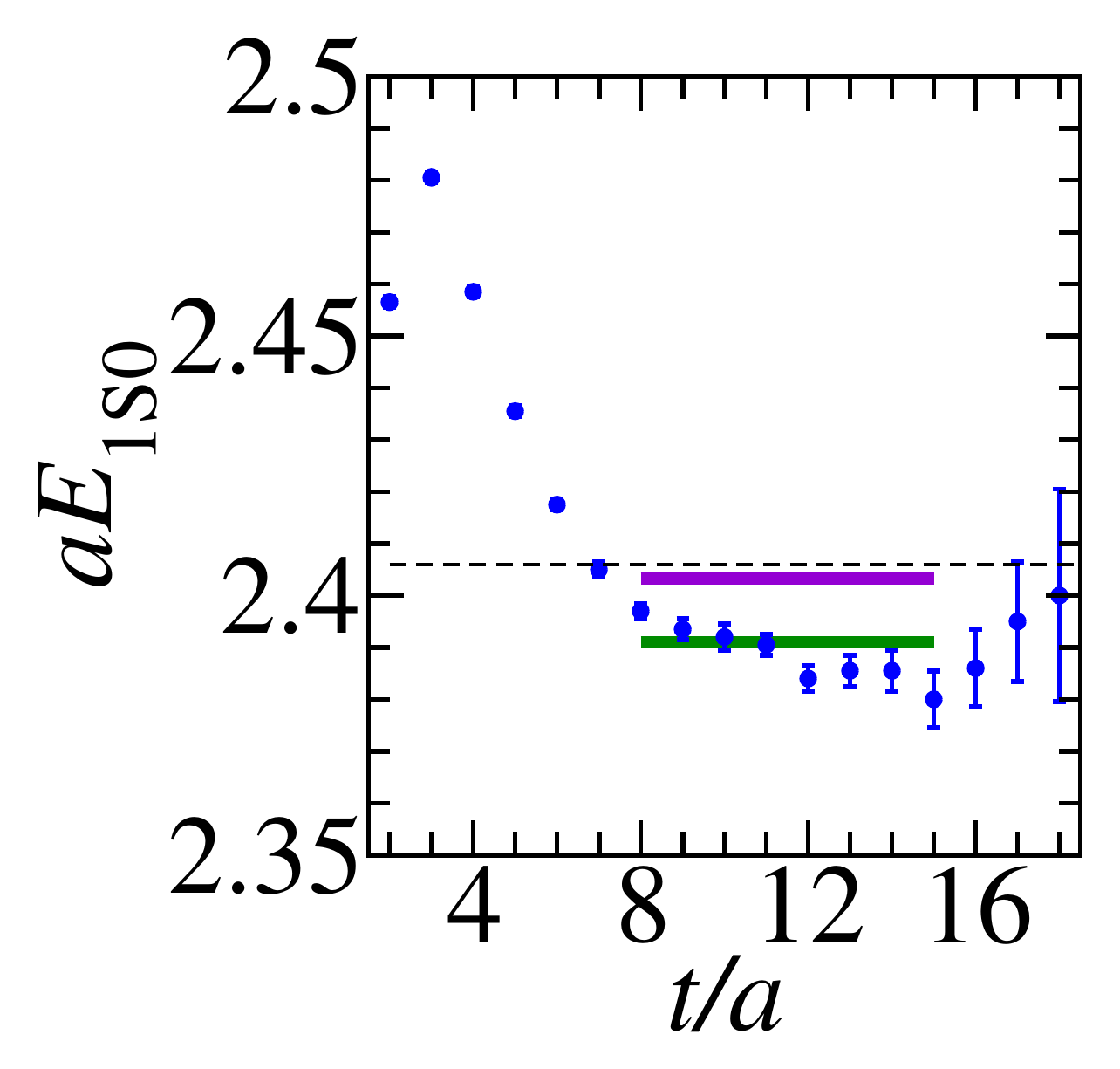}
\end{center}
\caption{(Left) Comparison of scattering phase shifts in the deuteron channel 
determined in Ref.~\cite{Horz:2020zvv}, shown as red, blue, magenta points and magenta
and gray bands, with phase shifts determined in Refs.~\cite{NPLQCD:2012mex,NPLQCD:2013bqy},
shown as green points.  (Middle/Right) Effective mass, shown as blue points, from 
Fig.~2 of Ref.~\cite{Beane:2017edf}
for the deuteron on a $48^3$ spatial lattice from an off-diagonal correlator involving a
local hexaquark operator at the source and a nucleon-nucleon operator of zero momentum 
at the sink. The horizontal green line indicates the approximate energy extraction from Fig.~4
of Ref.~\cite{NPLQCD:2012mex}, with the violet line indicating 
approximately where the energy extraction from Ref.~\cite{Horz:2020zvv} would be for 
this $48^3$ lattice. (Bottom right) Effective mass, similar to that in the center plot,
but for the dineutron.
\label{fig:NNdiscrepancy}}
\end{figure*}

The discrepancy can be seen in Fig.~\ref{fig:NNdiscrepancy}. The
scattering phase shift $q\cot(\delta)/m_\pi$ for the deuteron is shown in the
upper plot of Fig.~\ref{fig:NNdiscrepancy}.  Early results from 
Refs.~\cite{NPLQCD:2012mex,NPLQCD:2013bqy}, which used an off-diagonal correlator are shown 
in green, suggesting a bound state, whereas more recent results from Ref.~\cite{Horz:2020zvv},
which used a Hermitian correlation matrix, are shown by the red, blue, and magenta points with
gray and magenta bands, suggesting no bound state.  Note that 
Refs.~\cite{NPLQCD:2012mex,NPLQCD:2013bqy} used a 
tadpole-improved L\"uscher-Weisz gauge action and a stout-smeared clover fermion action
with lattice spacing 0.145~fm, while Ref.~\cite{Horz:2020zvv} used a tree-level improved
L\"uscher-Weisz gauge action and a non-perturbatively $O(a)$-improved clover Wilson fermion
action from CLS with lattice spacing 0.086~fm.  The quark masses
also differ, leading to a pion mass of 800~MeV in Refs.~\cite{NPLQCD:2012mex,NPLQCD:2013bqy} 
and about 710~MeV in Ref.~\cite{Horz:2020zvv}.  These lattice actions, spacings, and
volumes are similar enough that they are unlikely to be the cause of the large discrepancy
seen in the upper plot of Fig.~\ref{fig:NNdiscrepancy}.  

The discrepancy boils down to a difference in energy extractions.  Effective masses
from Ref.~\cite{Beane:2017edf} for the deuteron (lower left plot) and dineutron
(lower right plot) are shown in Fig.~\ref{fig:NNdiscrepancy} for a $48^3$ spatial lattice.
A single off-diagonal correlator involving a local hexaquark operator at the source and
a nucleon-nucleon operator of zero momentum at the sink was used to determine these
effective masses, which are shown as blue circles with errors.   The horizontal green
lines indicate the approximate energy extractions from Ref.~\cite{NPLQCD:2012mex}
with the vertical thickness indicating the statistical uncertainties.
The horizontal dashed lines indicate the energies of two non-interacting nucleons at rest
in the deuteron and dineutron cases.  The occurrences of the horizontal green lines well 
below the horizontal dashed lines, along with similar differences in other energy
extractions, essentially lead to the bound states observed, for example,
in the upper plot.  The horizontal violet boxes indicate approximately where the energy 
extractions from Ref.~\cite{Horz:2020zvv} would be for this $48^3$ lattice.  The lowest-lying
energies in Ref.~\cite{Horz:2020zvv} occur  slightly below the non-interacting energies.
Hence, the discrepancy essentially arises from the differences between the green and 
violet lines, along with similar differences from other energy determinations.

\begin{figure}
\begin{center}
\includegraphics[width=1.65in]{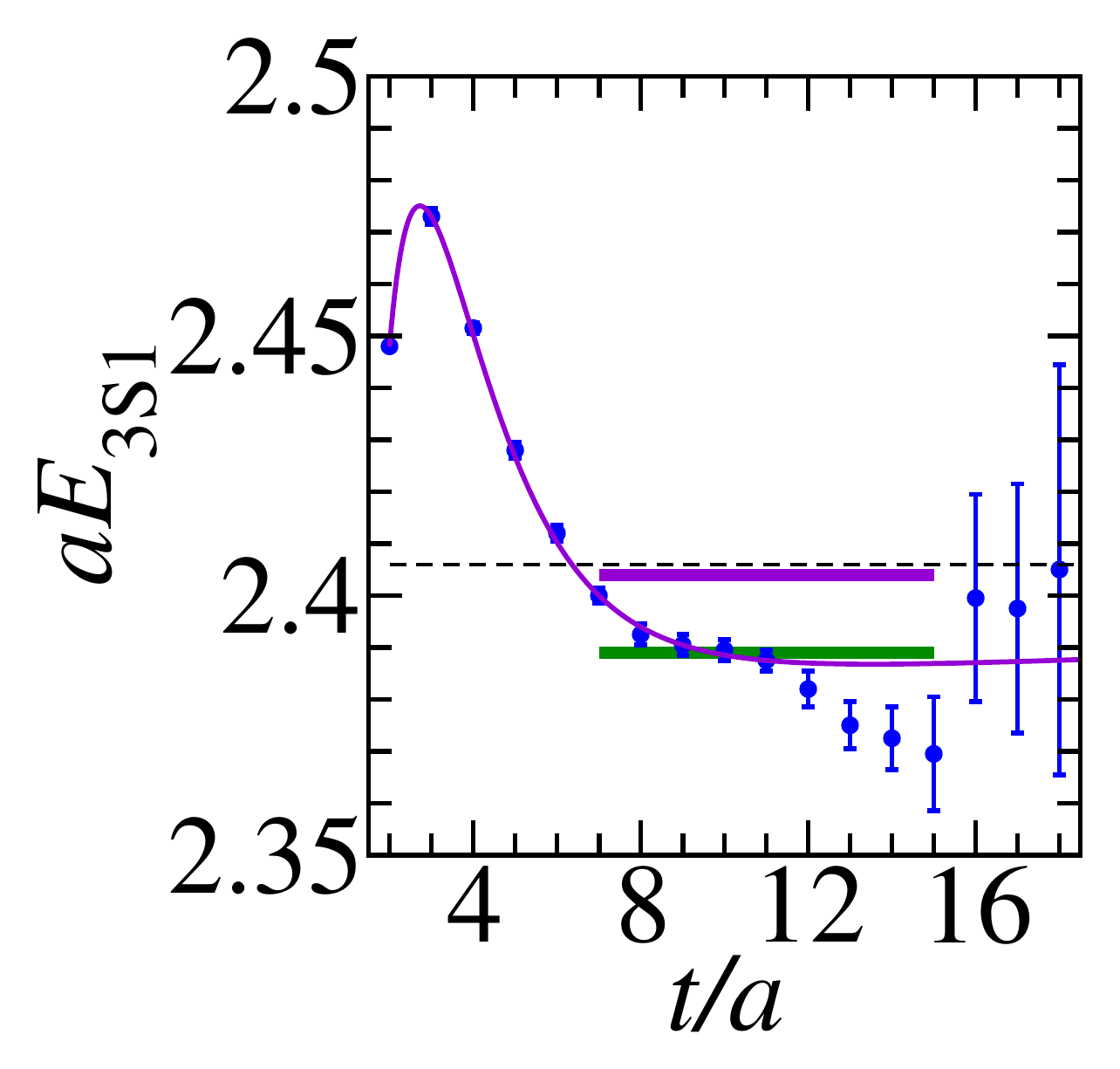}
\includegraphics[width=1.65in]{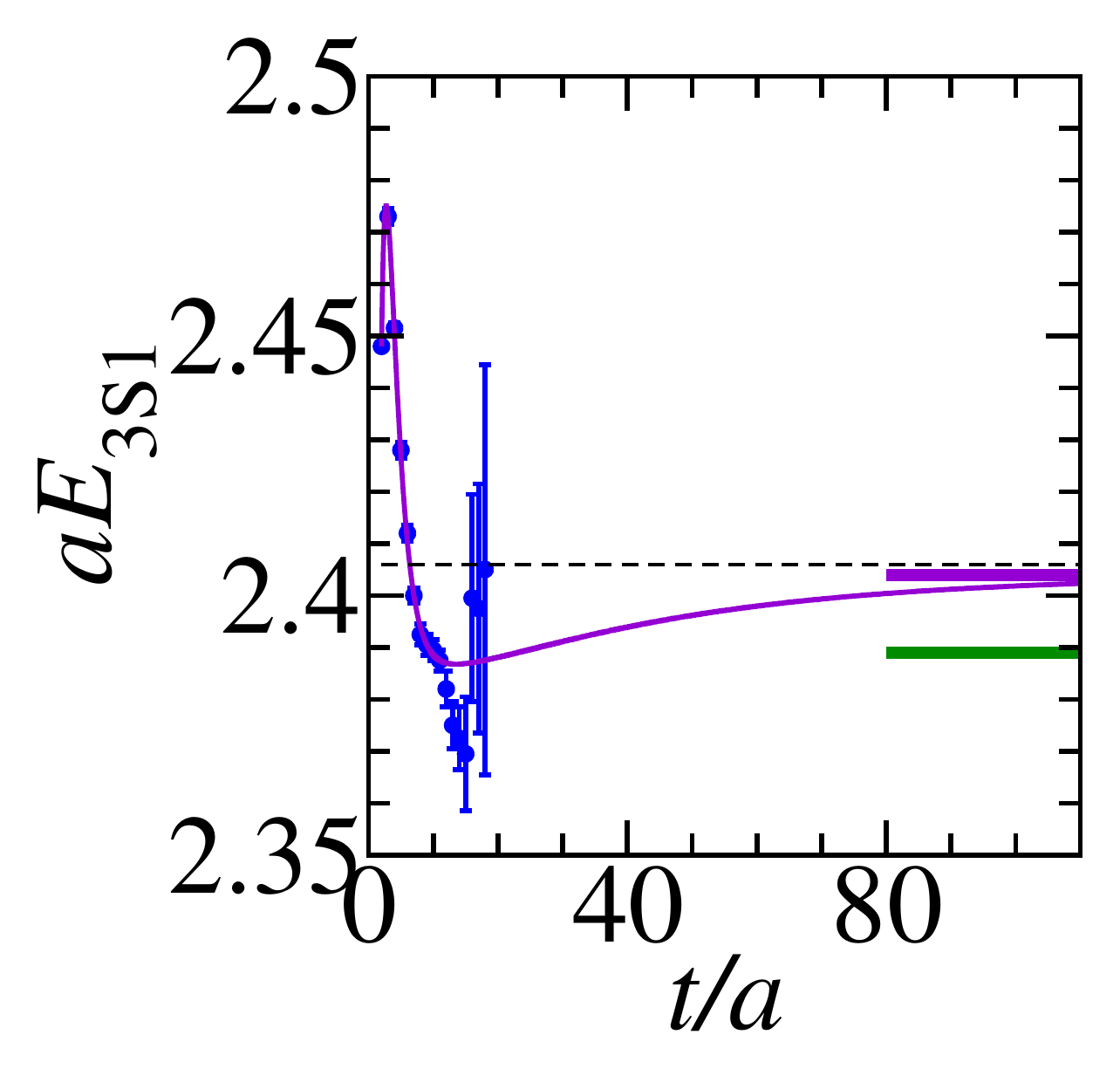}\\
\includegraphics[width=1.65in]{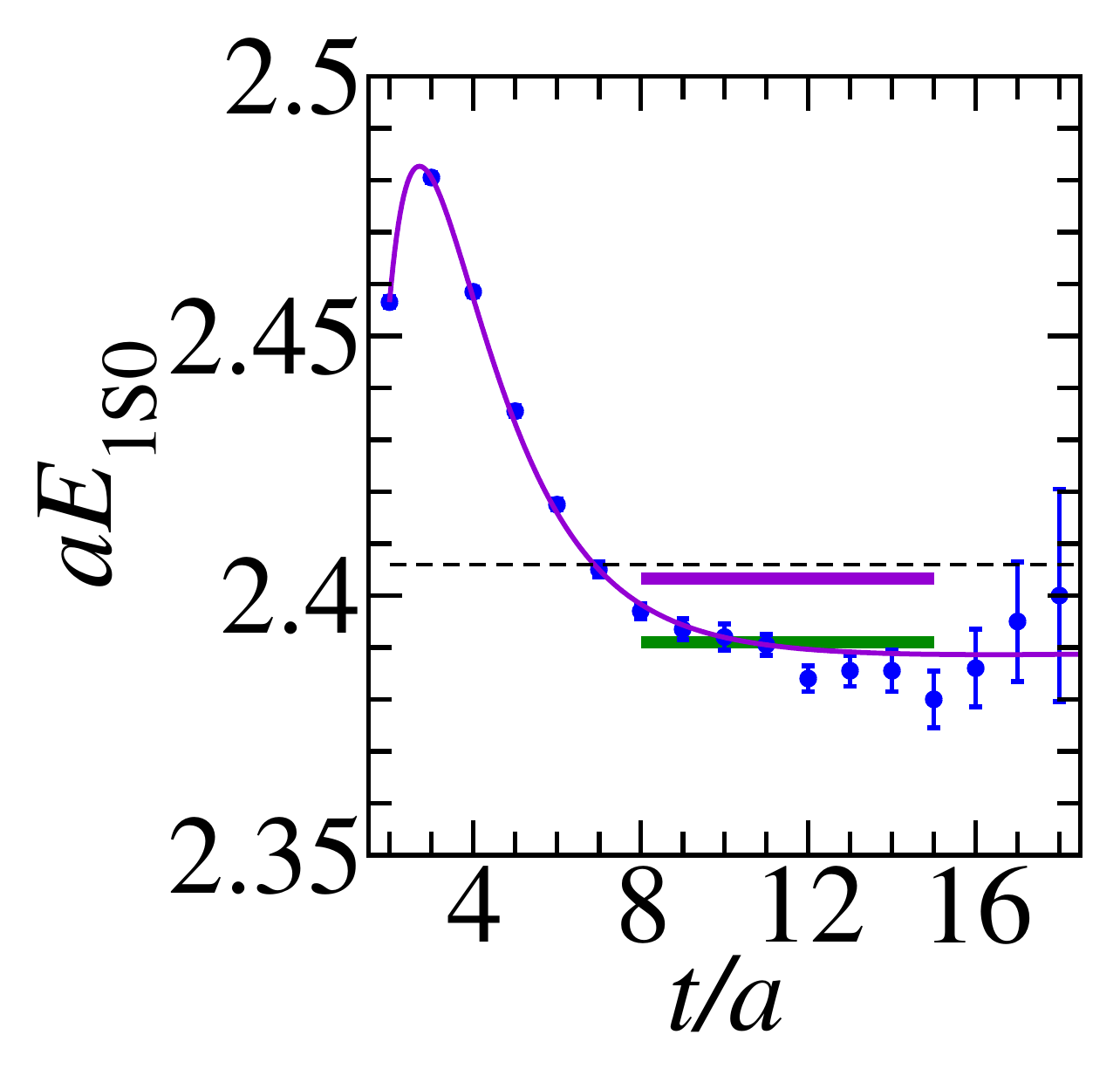}
\includegraphics[width=1.65in]{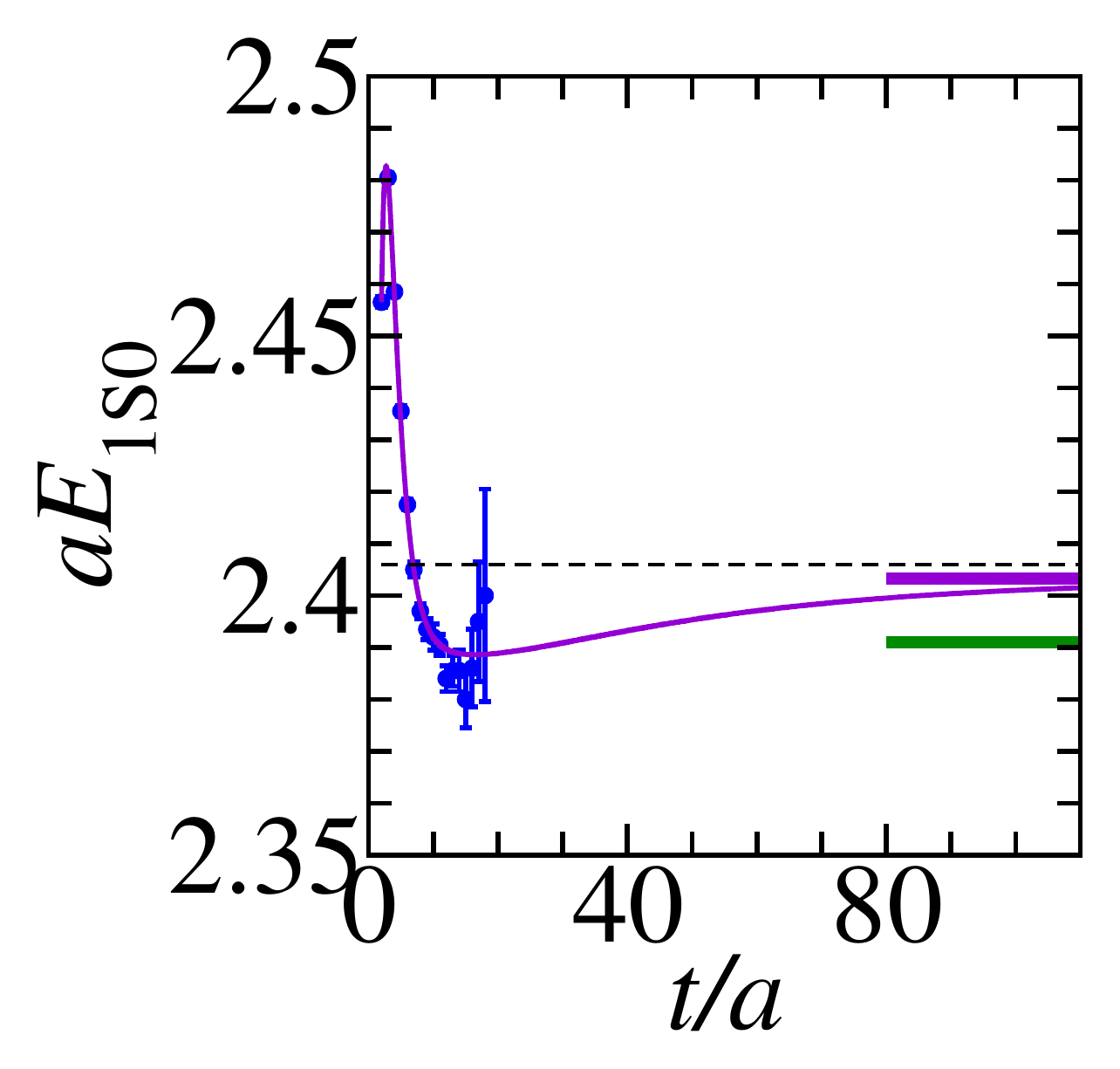}
\end{center}
\caption{(Upper left) Same as the bottom left plot of Fig.~\ref{fig:NNdiscrepancy},
but also includes the fit from Eqs.~(\ref{eq:NNfalsefit}), (\ref{eq:NNfalsegaps}), 
(\ref{eq:NNfalseAvalstriplet})
as a violet curve. For temporal range $t/a=10-20$, the violet curve shows a
remarkable but false plateau.  (Upper right) Same as the upper left plot in this figure, 
but continued to much larger temporal separations, showing the  slow approach to the
asymptotic limit.  (Lower left) Same as the bottom right plot of Fig.~\ref{fig:NNdiscrepancy},
but also includes the fit from Eqs.~(\ref{eq:NNfalsefit}), (\ref{eq:NNfalsegaps}), 
(\ref{eq:NNfalseAvalssinglet})
as a violet curve. (Lower right) Same as the lower left plot in this figure, but continued
to much larger temporal separations. 
\label{fig:falseplateaux}}
\end{figure}

Temporal correlators in lattice QCD admit a spectral representation
of the form shown in Eq.~(\ref{eq:Cij}),
which ignores negligible temporal wrap-around contributions.  For a diagonal 
$i=j$ correlator, the weights of the exponentials in the above spectral
representation are guaranteed to all be \textit{positive}
 \begin{equation}
     C_{ii}(t) = \sum_{n=0}^\infty \left\vert Z_i^{(n)}\right\vert^2 e^{-E_n t},
 \end{equation}
which greatly restricts the behavior of such correlators.  Off-diagonal 
$i\neq j$ correlators are not subject to such restrictions and can have 
\textit{negative} weights.  The initial rise of the blue points at small temporal
separation in Fig.~\ref{fig:NNdiscrepancy} certainly indicates
the unwelcome presence of large negative weights in the spectral representation. 
Furthermore, excited-state contamination in a
simple single off-diagonal correlator decays slowly as $e^{-(E_1-E_0)t}$,
where $E_0$ is the energy of the lowest-lying state and $E_1$ is the
energy of the second lowest-lying state.  Contamination in a diagonal
correlator obtained from a generalized eigenvalue problem optimization
decays much more quickly as $e^{-(E_N-E_0)t}$ for an $N\times N$ correlator 
matrix. Given the possibility of negative weights and the slow decay of 
excited-state contamination in a single off-diagonal correlator, the 
likelihood of plateau misidentification is uncomfortably high.

To illustrate how plateau misidentification could have occurred, consider a five-exponential form for the off-diagonal correlator
 \begin{eqnarray}
    C(t)&=&e^{-E_0 t}\Bigl(1+A_1e^{-\Delta_1 t}+A_2e^{-\Delta_2 t} \nonumber\\
        &&\qquad\qquad +A_3e^{-\Delta_3 t}+A_4e^{-\Delta_4 t}\Bigr).
  \label{eq:NNfalsefit}
 \end{eqnarray}
For the two lowest gaps, we take values that are expected for a $48^3$ spatial
lattice.  The other 2 gaps are set to be large enough to handle the observed
short-time behavior of the effective mass.  In particular, we take
\begin{eqnarray}
   \Delta_1&=&0.025,\quad\Delta_2=\Delta_1+0.025,\nonumber\\ 
   \Delta_3&=&\Delta_2+0.5,\quad\Delta_4=\Delta_3+1.0,
   \label{eq:NNfalsegaps}
\end{eqnarray}
then using the $E_0$ values shown by the violet boxes, we can solve for the weights
$A_1,A_2,A_3,A_4$ using correlations at times  $t=2,3,7,11$.
For the deuteron $(I=0,\ ^3S_1)$, we find
\begin{eqnarray}
    A_1 &=& -1.0483,\ A_2 = 0.4133,\nonumber\\
    A_3 &=& 0.6495,\ A_4 = -1.7750,
\label{eq:NNfalseAvalstriplet}
\end{eqnarray}
and for the dineutron $(I=1,\ ^1S_0)$, we obtain
\begin{eqnarray}
     A_1 &=& -1.0986,\ A_2 = 0.4993,\nonumber\\
     A_3 &=& 0.7127,\ A_4 = -1.9065.
\label{eq:NNfalseAvalssinglet}
\end{eqnarray}
The resulting effective masses are shown as the violet curves in 
Fig.~\ref{fig:falseplateaux}.  The violet curves reproduce the observed behaviors
at small temporal separations and show amazingly flat plateaux-like behavior for
a temporal range from about $t/a=10$ to 20, as illustrated in the left-hand plots of
Fig.~\ref{fig:falseplateaux}.  However, the right-hand plots display the behaviors
for larger temporal separations, showing how the approaches to the asymptotic limits
given by the violet boxes are exceedingly slow.  These plots are for illustrative 
purposes only to show how this could have happened; they do not prove that this did happen. This illustration is similar to that presented in Ref.~\cite{Iritani:2018vfn}.

%-------------------------------------------------------------------------------

%-------------------------------------------------------------------------------
%  HAL QCD Potential
\section{HAL QCD Potential\label{sec:potential}}

Finally, we compare with the HAL QCD potential method for extracting the NN phase shifts~\cite{Ishii:2006ec,Ishii:2012ssm,Aoki:2020bew}.  Prior to this work, HAL QCD was the only collaboration to have used this method.

This formalism uses as input the following off-diagonal two-nucleon correlation function, 
\begin{equation}\label{eq:CNN_r}
C_{NN}(t,\mathbf{r}) = \sum_{\mathbf{x}} 
    \langle N(t,\mathbf{x+r})N(t,\mathbf{x})
    N^\dagger(0)N^\dagger(0)\rangle\, ,
\end{equation}
where the annihilation fields are spatially local and the sum over $\mathbf{x}$ projects the total system to zero momentum.
A sum over $\mathbf{r}$ yields an off-diagonal correlator that can be used in the QC2 formalism.
Rather than following this approach, the HAL QCD Potential method instead focuses on utilizing the spatial distribution of the correlation function to extract a Nambu-Bethe-Salpeter (NBS) wavefunction rather than spectra. This wavefunction may then be used to solve for the potential, and in turn, scattering phase shifts, using the NBS equation. 

In the large $t$ limit, one has,
\begin{equation}
C_{NN}(t,\mathbf{r}) = \sum_n \psi^n(\mathbf{r}) Z_n e^{-E_n t} \ ,
\end{equation}
where the sum is over elastic two-nucleon scattering states, $Z_n$ denotes the overlap of the creation operator with the $n^{\rm th}$ state, $E_n = 2\sqrt{M_N^2 + q_n^2}$ denotes the energy of this state, and $\psi^n(\mathbf{r})$ denotes the NBS wavefunction.

Note that this assumes that $t$ is sufficiently large that inelastic states do not contribute and this must be carefully checked - while it is clear that contributions from inelastic excited states should die off exponentially more quickly than low-lying elastic scattering states, studies of single-nucleon correlation functions have shown that in practice it may be notoriously difficult to suppress these contributions within the time ranges typically available for two-nucleon correlation functions. This, of course, is also an issue within the \luscher method. In both cases, the situation can be ameliorated through the use of the ratio of correlation functions,
\begin{equation}\label{eq:R_t_r}
R(t,\mathbf{r}) = \frac{C_{NN}(t,\mathbf{r})}{C_N(t)^2}\, ,
\end{equation}
where near-cancellation between inelastic single-nucleon excited state contributions leads to significant suppression (though in this work we choose instead to adopt the ``conspiracy" fits detailed in the previous Section for the \luscher method). This situation must be carefully checked within both the \luscher formalism and the HAL QCD potential method, as contributions from excited states enter differently into the spectrum versus the potential, as detailed below, and as the interpolating operators are typically different for the two setups. 

The ratio $R(t,\mathbf{r})$ then satisfies a time-dependent version of the NBS equation in a given scattering channel,
\begin{equation}\label{eq:time_dependent_HAL}
\left[ \frac{\partial_t^2}{4M_N} - \partial_t -H_0 \right] R(t,\mathbf{r})
=
\int d^3 r'\, U(\mathbf{r},\mathbf{r'}) R(t,\mathbf{r'})\, ,
\end{equation}
where $H_0 \equiv \nabla^2/M_N$. 
This modified, time-dependent, Schr\"odinger-like equation, referred to as the {\it time-dependent NBS} equation in Ref.~\cite{Ishii:2012ssm}, is valid for a correlator $R(t,\mathbf{r})$ that has contributions from excited elastic NN scattering states, as well the ground state, though it still assumes the inelastic single nucleon and two-nucleon excited states do not contribute.

Numerical determination of the potential, $U(\mathbf{r},\mathbf{r'})$ in general, is challenging.  In practice, in both the literature and in this work, the leading local approximation of this non-local potential is most commonly used:
\begin{equation}
\label{eq:gradExpansion}
U(\mathbf{r},\mathbf{r'}) = \left[
    V_C(r) +\cdots
    \right]\d^3(\mathbf{r-r'})\, ,
\end{equation}
where additional terms include gradient terms as well as spin-orbit mixing.  HAL QCD has investigated both of these extra contributions~\cite{Murano:2013xxa,HALQCD:2018gyl}.  In this work we  neglect higher order terms in the gradient expansion.   Further, we restrict our attention to the $T_{1g}$ ($A_{1g}$) irreps for the deuteron (di-neutron) which are dominated by the central potential $V_C(r)$ at low energy. As discussed in \appref{sec:make_hal_potential}, we verified that remaining $\ell\geq4$ contributions are negligible by performing an $\ell=0$ projection of the NN correlator data and comparing the resulting potential and phase shifts.  This is similar to the use of the Misner method~\cite{Misner:1999ab} by the HAL QCD collaboration~\cite{Miyamoto:2019jjc}.

While such a gradient expansion may suggest some sort of low-energy approximation, there is no known relevant energy scale to properly cut off such an expansion, which is, in general, uncontrolled. Possibilities for testing the relative sizes of such corrections have been proposed by the HAL QCD collaboration, and involve the use of two-nucleon correlation functions having different source operators~\cite{HALQCD:2018gyl}. In such cases it is imperative to fully control the large-time behavior of the calculated potential for each source operator. We will explore the time-dependence in detail in this work.

In this work, we investigate the effects of different analytic forms used to fit the potential, including incorporating regulated one- and multiple-pion exchange potentials, and exploring the use of harmonic oscillator (H.O.) basis functions as well as Gaussians for the short range contribution $V_s(r)$.  The HAL QCD collaboration has primarily used a Gaussian basis to describe the potential~\cite{Ishii:2012ssm}, but they have also explored the possibility of adding a one- and two-pion exchange terms~\cite{HALQCD:2018gyl}.  The use of harmonic oscillator basis functions is new in this work.  The various potentials we fit to the results include
\begin{equation}
    V(r) = V_{s}(r) + V_\pi(r)\; \left[\; +\; V_{2\pi}(r)\;\right]\, ,
\end{equation}
where the $[+\; V_{2\pi}(r)]$ indicates it is optionally included.

For the one- and two-pion exchange potentials, we utilize a regulated Yukawa potential of the form
\begin{eqnarray}\label{eq:pot_lr}
    V_{[2]\pi}^{(p)}(r) = -g^2 \frac{e^{-m r}}{r} \left(1 - e^{-c r^2}\right)^p\ , 
\end{eqnarray}
where $m,c,p$ are free parameters.
The coupling is modeled as $g^2=e^a$ and we have absorbed the $m^{-1}$ that usually accompanies the $r^{-1}$.
Parameter $m$ models the mass of the exchange particle.  
Parameters $c$ and $p$ control a regulator that removes the non-physical short-range divergence that comes from treating nucleons as point particles. The regulator is borrowed from the Argonne $v_{18}$ one pion exchange potential implementation~\cite{Wiringa1994wb}, equations 17 through 19, with $c$ (related to the nucleon size) controlling the range of regulator and $p$ its strength.  
We performed fits with $p=1$ and $p=2$, with results favoring the more rapid cutoff with $p=2$.
We find no sensitivity to the number of terms (representing multi-meson fits) included in the fit function, and also find that the fit is fairly insensitive to any of these fit parameters.

For the short-range terms, we explore three different functional forms
\begin{equation}
\label{eq:V_sr}
V_s(r) = \left\{\begin{matrix}
    \sum_i v_i e^{-(r/b_i)^2}& \textrm{``Gaussian"}\\
        \sum_i v_i e^{-1/2(r/b)^2} L^{1/2}_{i-1}\left(\frac{r^2}{b^2}\right)& \textrm{``harmosc"}\\
    \sum_i v_i e^{-(r/b_i)^2} H_i(r/b_i)& \textrm{``1Dosc"}
\end{matrix}\right.\, ,
\end{equation}
where $v_i$ are strengths for the various terms.
For all three functions, the $b_i$ are length scales associated with different order terms.  In the case of the 3D harmonic oscillator, the $b$ parameter is the same for all terms to preserve the orthogonality of the basis functions.  $L_i$ and $H_i$ are Laguerre and Hermite polynomials respectively.
The Gaussian basis is what is often used by HAL QCD.
The 3D Harmonic oscillator choice is a complete orthogonal basis commonly used on calculations of nuclei with nucleon degrees of freedom.
The 1D oscillator functions are used to demonstrate that the functional form used to describe the potential has  little impact on the resulting phase shift, provided the basis is able to describe the numerical potential data in the relevant region of $r$ that effects the phase shift at low energies.

Given a parameterization of the potential, with parameters constrained from the results, we investigate extracting the $S$-matrix in two different ways, the first of which is the standard method of numerically solving the Schr\"odinger equation to determine the asymptotic wavefunction. The second uses the variable phase method~\cite{calogero1967variable}, see Chapter 19, equation 19.   A simple first-order differential equation for $S(R)$ involving $V(r) \Theta(R-r)$ is given, where the Heaviside function forces the potential to 0 outside of range $R$.    The initial condition is $S(0) = I$, and $S(\infty) = S$, the desired $S$-matrix.   One advantage with this method for future work is that coupled channels are directly supported.  Also see~\cite{Das1981} for an extension to non-local potentials.

In addition to the aforementioned tests of systematics of the HAL QCD potential, we also explore the effects of the finite-volume on the extracted potential, and the effects of discretization of the NBS equation. The results of these studies are incorporated into an overall systematic error on the extracted scattering phase shifts through a bootstrap analysis. Details of these systematics are discussed below.
\\
\subsection{Details of the calculation}

Zero momentum wall sources without any 3D smearing were used to compute the quark propagators on the gauge configurations after they were Coulomb gauge fixed.  These quark propagators were used to compute both the N and NN correlation functions.  We performed the calculation on 16 equally spaced time-sources on all 1490 configurations, with a random starting time on each configuration for a total of 23,840 sources.  The annihilation interpolating fields utilzed point-sinks, projected to zero-momentum for the single nucleons.  The full relative momentum of the NN correlation functions were computed at zero total momentum, utilizing the FFTW library~\cite{FFTW.jl-2005}, allowing for a reconstruction of the full position-space correlator at the sink.
These sources are the most commonly used by the HAL QCD collaboration based on their studies of wall versus smeared correlation functions~\cite{HALQCD:2018gyl}.

Details of constructing the potential from Eqs.~\eqref{eq:time_dependent_HAL} and \eqref{eq:gradExpansion} are provided in \appref{sec:make_hal_potential}.

Below, we detail the four main sources of systematic error associated with the potential method, as well as our methods for mitigation.

\subsubsection{Excited state contamination}

%-----------------------------------------
% HAL potential vs time
\begin{figure}
\includegraphics[width=\linewidth]{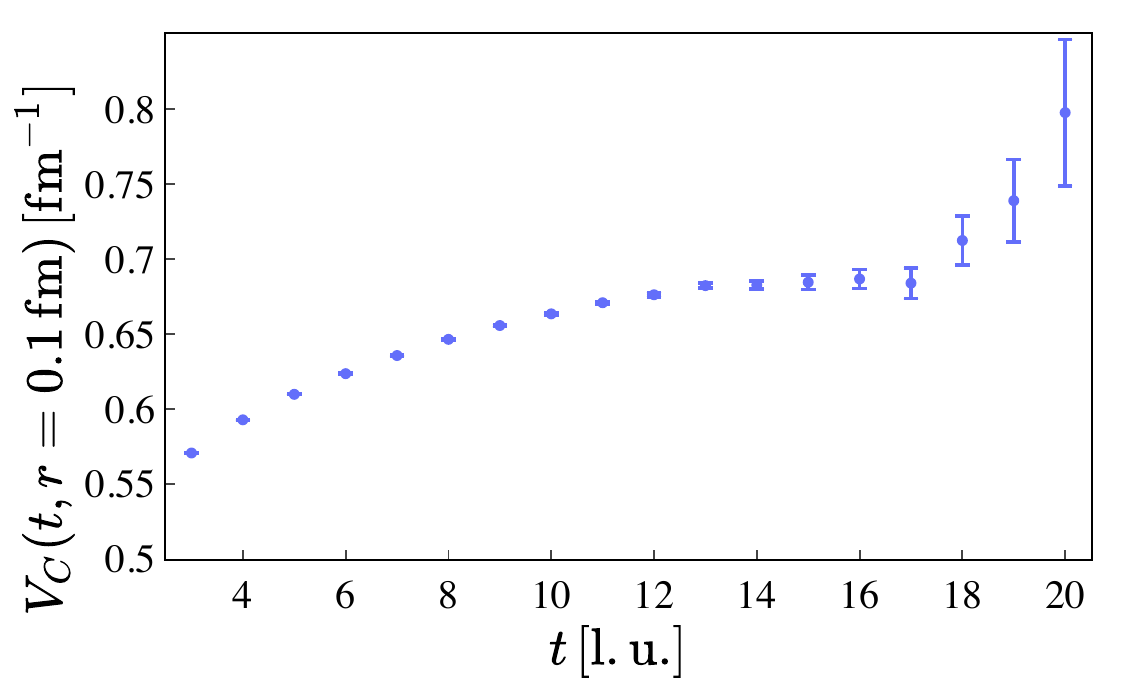}
\includegraphics[width=\linewidth]{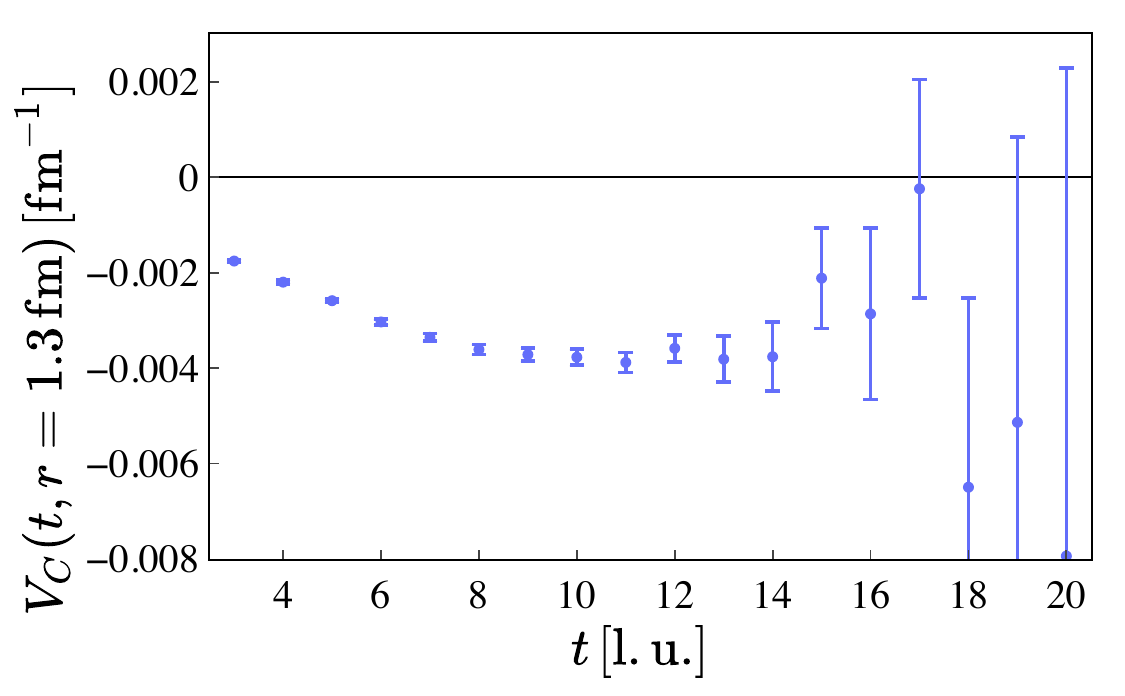}
\caption{Example plots of the extracted potential $V_C(t,\mathbf{r})$ as a function of Euclidean time, $t$, in the deuteron channel for small (top) and large (bottom) values of $r$. Both axes are in units of fm. 
\label{fig:eff_mass_t}}
\end{figure}
%-----------------------------------------

Inserting \eqnref{eq:gradExpansion} into \eqnref{eq:time_dependent_HAL} and solving for the local potential leads to the following relation,
\begin{eqnarray}\label{eq:V_central}
V_C(t,\mathbf{r}) = \frac{1}{R(t,\mathbf{r})} \left[ \frac{\partial_t^2}{4M_N} - \partial_t -H_0 \right] R(t,\mathbf{r}) \, .
\end{eqnarray}
If $R(t,\mathbf{r})$ is free from inelastic excited state contamination, then $V_C(t,\mathbf{r})=V_C(\mathbf{r})$.
At finite $t$, we can estimate contamination to this relation from inelastic excited states by using the functional form of the correlation functions and inserting them into the ratio correlator, \eqnref{eq:R_t_r}.
This yields
\begin{align}\label{eq:VC_excitedStates}
V_C(t,\mathbf{r}) &= V_C(\mathbf{r})
    +\d V_C(\mathbf{r})
    \alpha^I_1(\mathbf{r}) 
    e^{-\D^{\rm NN,I}_1 t}
\nonumber\\&\phantom{=}
    -2\tilde{\alpha}_1\D_1^{\rm N} 
    e^{- \D^{\rm N}_1 t}\, ,
\end{align}
where $\D^{\rm NN,I}_1$ is the mass gap from the ground state NN energy to the first inelastic NN state and $\D^{\rm N}_1$ is the mass gap to the first single nucleon inelastic mass gap.
See Appendix~\ref{sec:HAL_inel_es} for a derivation of this formula.  This motivates a multi-exponential fit of the form
\begin{equation}
V_C(t,\mathbf{r}) = V_C(\mathbf{r})
    +\sum_{n=1}\alpha_n(\mathbf{r}) e^{-\D_n(\mathbf{r}) t}\, ,
\end{equation}
where $\alpha_n(\mathbf{r})$ and $\D_n(\mathbf{r})$ are independently fit for each $\mathbf{r}$.

\begin{figure}
\includegraphics[width=\linewidth]{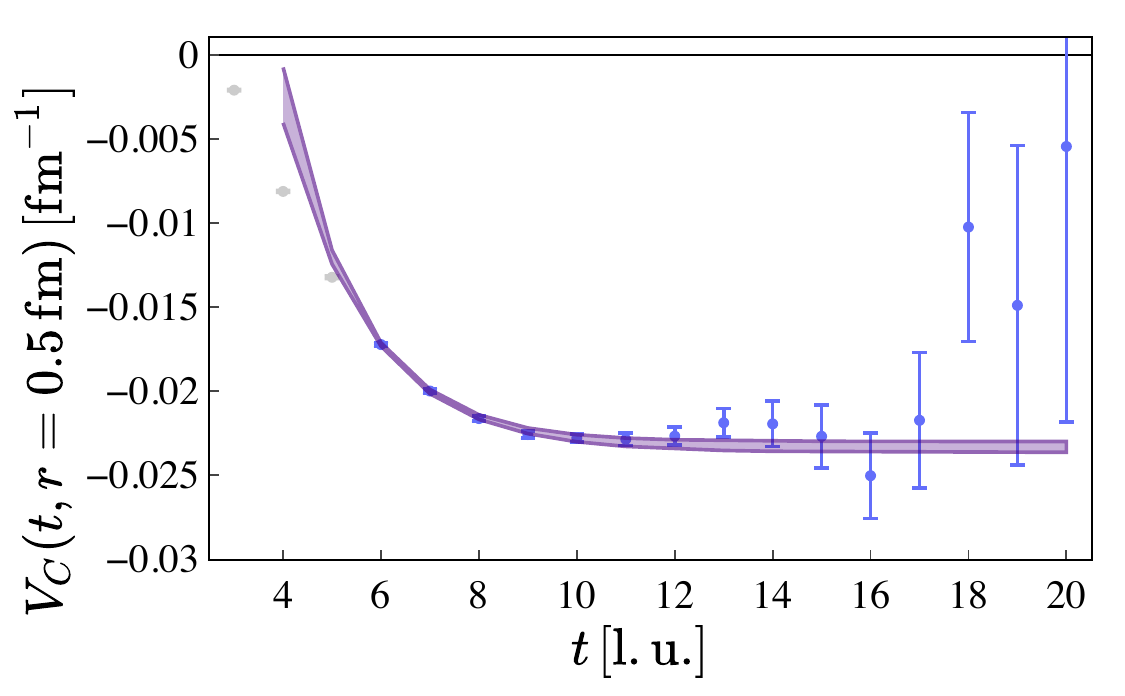}
\includegraphics[width=\linewidth]{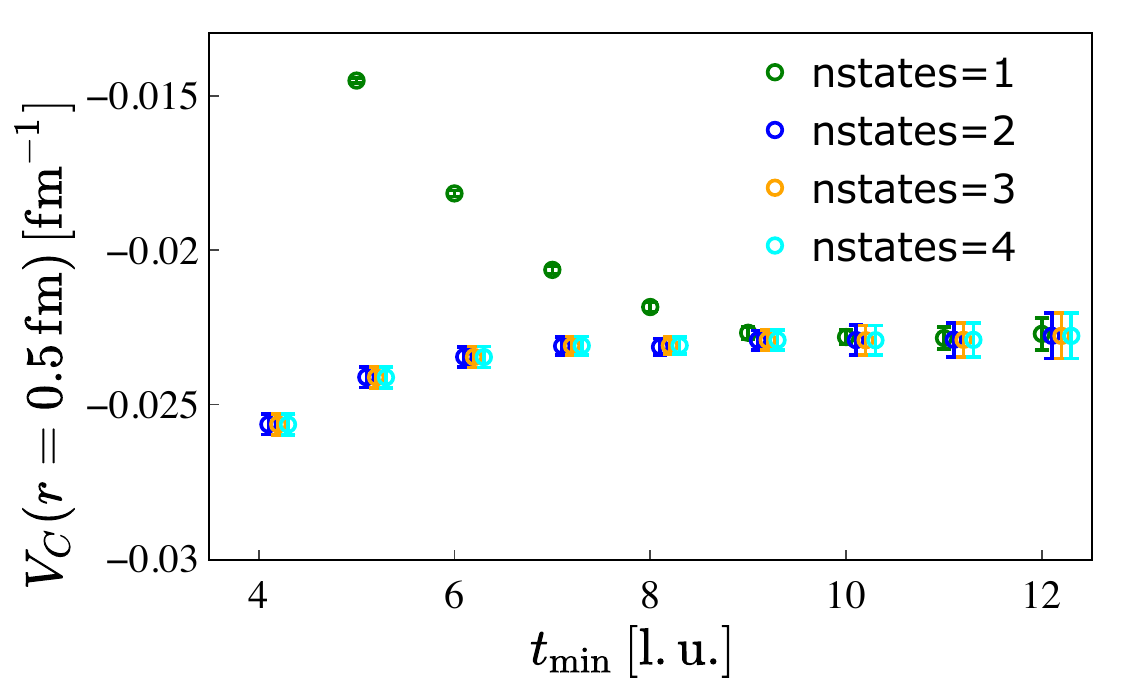}
\caption{(Top) Example plot in the deuteron channel for the extracted potential $V_C(t,\mathbf{r})$ as a function of Euclidean time, $t$, for intermediate $r$, showing a fit to a sum of exponentials, \eqnref{eq:VC_excitedStates}. Gray data points are not included in the fit. (Bottom) The extracted infinite time potential, $V_C(\mathbf{r})$, from fits to the data in the top panel, as a function of the minimum time of data included in the fit ($t_{\mathrm{min}}$), and for a given number of states included in the sum of exponentials (nstates). 
\label{fig:pot_stab_t}}
\end{figure}

In this work, we use this functional form to investigate extrapolating $V_C(t,\mathbf{r})$ to obtain $V_C(\mathbf{r})$.  Note that the overlap factors are a function of $\mathbf{r}$, and therefore, the contamination from excited states will differ for different values of $\mathbf{r}$. One may expect, for example, that as one reduces $\mathbf{r}$ to the point where the radii of the two nucleons begin to overlap, greater contamination from inelastic excited states intrinsic to the NN system will occur. We find that this is indeed the case, as shown in \figref{fig:eff_mass_t} for small (top) and large (bottom) reference values of $\mathbf{r}$. One also finds a drop in the signal-to-noise ratio as $\mathbf{r}$ is increased. This can presumably also be traced to the $\mathbf{r}$-dependence of the overlap factors, giving greater overlap of the noise onto a 3-pion state as $\mathbf{r}$ is increased~\cite{Parisi:1983ae,Lepage:1989hd}. For these reasons, the fit ranges and number of states in multi-exponential fits must be carefully chosen, and may in general depend on $\mathbf{r}$. \figref{fig:pot_stab_t} illustrates the results of such a study at a fixed intermediate value of $\mathbf{r}=0.5$~fm.

\begin{figure}
\includegraphics[width=\linewidth]{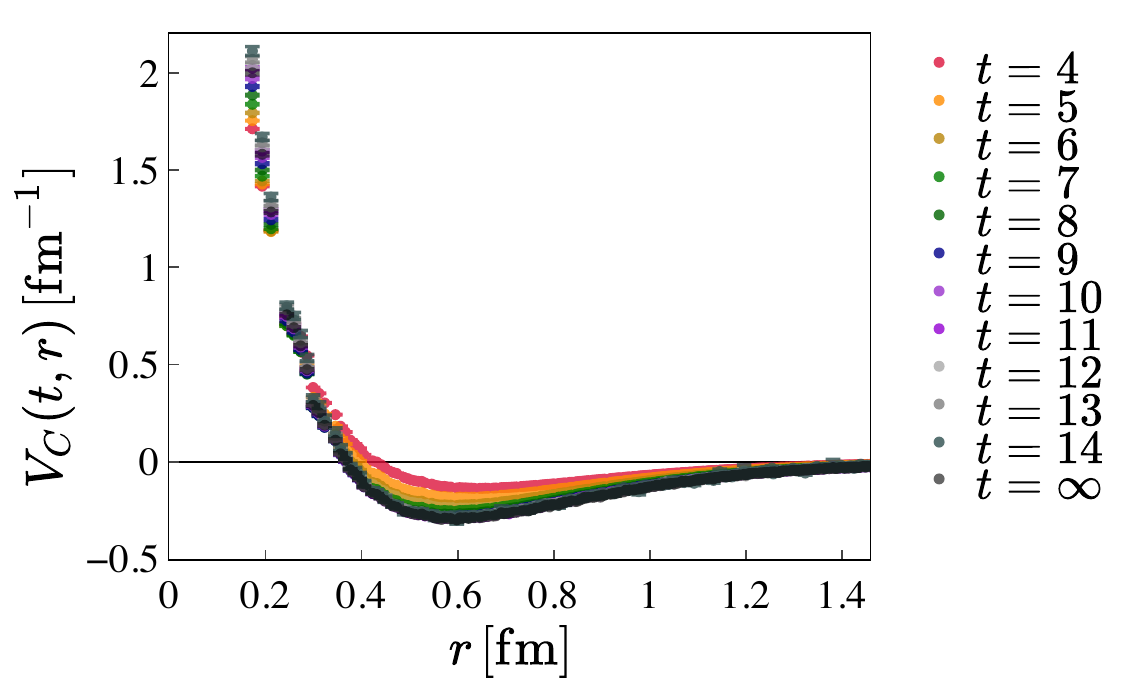}
\caption{Potential, $V_C(t,\mathbf{r})$, in the deuteron channel for various Euclidean time, $t$, as a function of $r$, in units of fm.
\label{fig:pot_t}}
\end{figure}

\begin{figure}
\includegraphics[width=\linewidth]{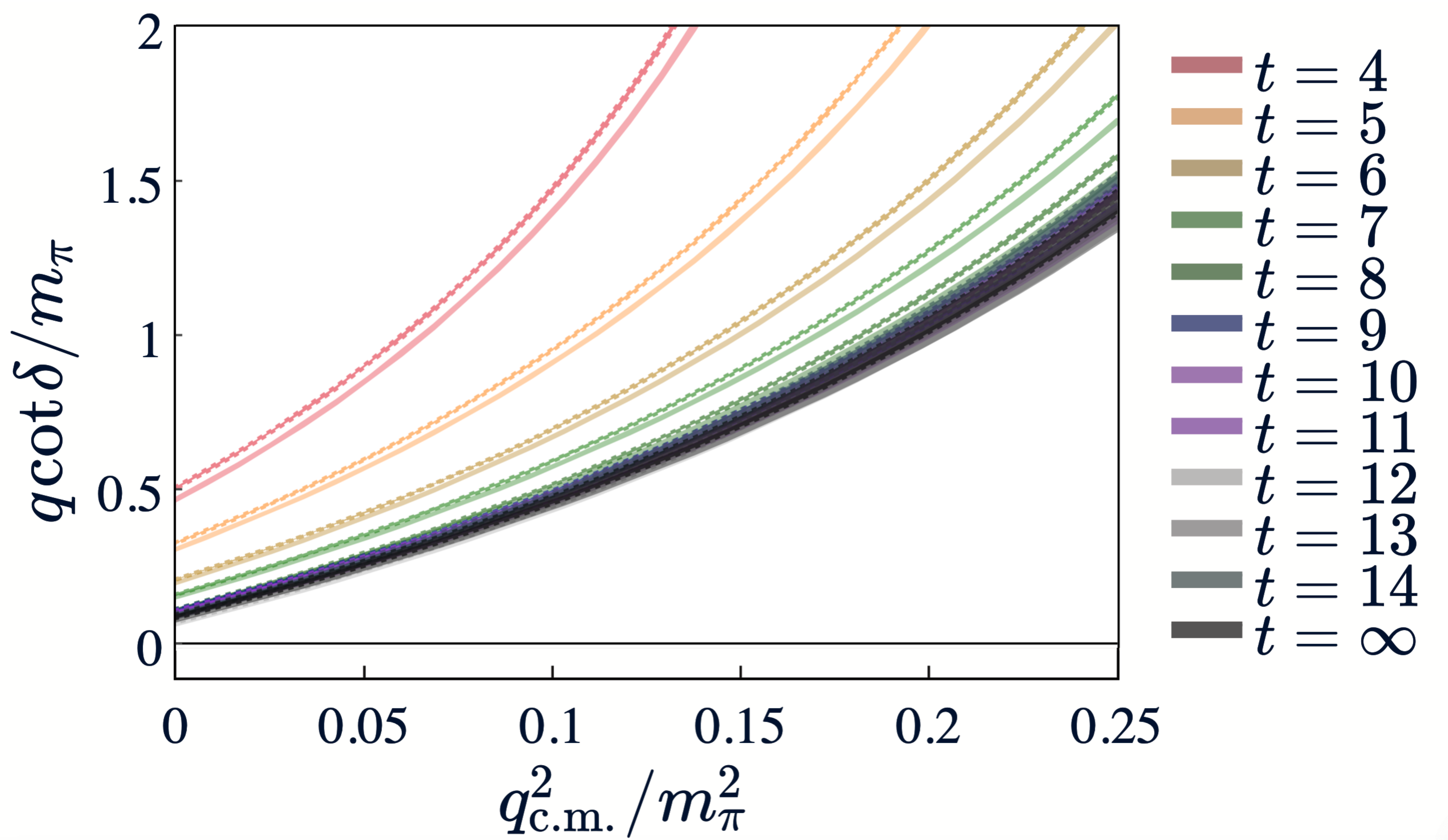}
\caption{Scattering phase shift in the deuteron channel for various Euclidean time, $t$, as a function of momentum transfer in the center of mass frame, $q_{\rm CM}$, in units of the pion mass, $m_{\pi}$. Phase shifts are determined using two examples of our fit functions from Eq.~\ref{eq:V_sr}, 1Dosc (solid) or gaussian (dashed), for the short-range portion of the potential.
\label{fig:phase_t}}
\end{figure}

In Fig.~\ref{fig:pot_t}, we show the potential in the deuteron channel for various Euclidean times, as well as the extrapolation of the potential to infinite time. 
The same plot for the di-neutron channel is presented in Appendix~\ref{seq:hal_nn}.
Performing such extrapolations is clearly merited for controlling systematics in precision calculations utilizing the potential method, as choosing a single fixed value of the time, $t$, may be biased due to correlated fluctuations. Furthermore, due to the $\mathbf{r}$-dependence of the overlap onto excited states discussed previously, it may not be possible to choose a single value of $t$ such that inelastic excited states are well-controlled for all values of $r$. In Fig.~\ref{fig:phase_t}, we show the resulting phase shift as a function of scattering momentum squared utilizing a given fixed value of $t$. Different forms of the short-distance portion of the fit function, denoted ``harmosc" (``gaussian") for the Harmonic Oscillator (Gaussian) basis. We find that for early and intermediate times, particularly in the deuteron channel, the different choices for the form of the short-distance potential give results that disagree significantly from each other, even for low energies (large $\mathbf{r}$), signaling combined discretization/excited state contaminations. For later times and for the extrapolated potential, we find no significant differences between the different fit forms and include both in our quoted systematic error band.

\begin{figure}
\includegraphics[width=\linewidth]{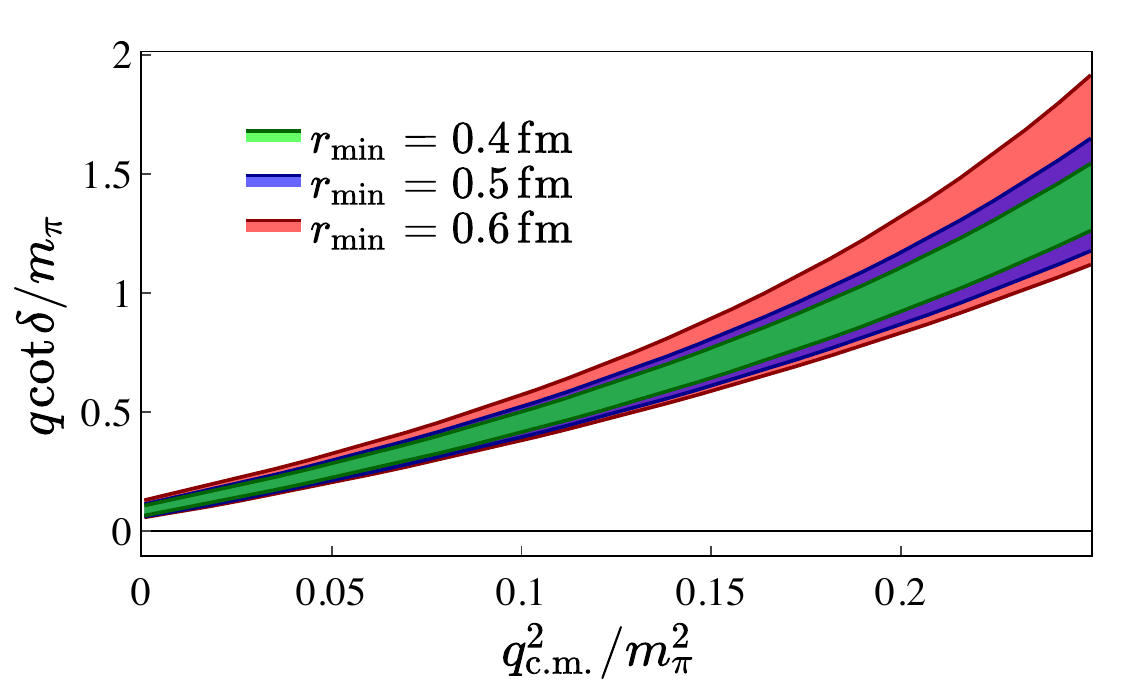}
\caption{Scattering phase shift, $q \cot \delta$, versus $q_{\mathrm{CM}}^2$ in units of $m_{\pi}$ in the deuteron channel using various cuts to the data, $r \geq r_{\mathrm{min}}$, in the fits of the potential. 
\label{fig:phase_rmin}}
\end{figure}

Due to the increased sensitivity to inelastic excited states for small $r$, we furthermore vary the minimum $\mathbf{r}$-value ($\mathbf{r}_{\mathrm{min}}$) included in fits to $V_C(\mathbf{r})$ to incorporate a systematic for any remaining excited state contributions for $\mathbf{r} \lesssim $~0.4 fm. \figref{fig:phase_rmin} shows the effects on the scattering phase shifts in the deuteron scattering channel as a function of the scattering momentum. 
We note here that varying $\mathbf{r}_{\mathrm{min}}$ within our fits simultaneously quantifies sensitivity to both inelastic excited state and discretization effects, which are each amplified for small $\mathbf{r}$. The results of this study will be discussed in the next Section. 

\subsubsection{Discretization effects\label{sec:hal_discretization}}

In addition to the discretization effects present in the correlation functions themselves, the potential method utilizes a discretized version of the Schr\"odinger equation. Furthermore, the assumption that all elastic excited states satisfy the same Schr\"odinger equation applies only to the continuum theory.  At finite lattice spacing, higher elastic excited states also contain larger discretization errors, leading to further possible time dependence even within the elastic state time-regime, and further combined finite Euclidean time/discretization effects. Similarly, as discussed in \secref{sec:results_deuteron} and \ref{sec:results_dineuteron}, the irreps with boosted nucleons had greater sensitivity to the lattice cutoff for these energies using the \luscher formalism. This leads to the expectation that on a given lattice ensemble, even if all other systematics have been well-controlled, the two formulations may disagree, and the size of any discrepancies may be indicative of the sizes of discretization effects.  Further studies at finer lattice spacings are necessary to test this.

Here we investigate contributions from discretization effects which are particular to the potential method. We take the ratio of correlation functions for times sufficiently large that only elastic excited states contribute, \eqnref{eq:R_HAL_elastic}.
A discretized version of Eq.~(\ref{eq:V_central}) leads to
\begin{align}\label{eq:V_a}
V_C(a,\mathbf{r}) &= \frac{1}{R(t,\mathbf{r})}  \sum_n R_n(t,\mathbf{r})\bigg[
    \frac{\cosh \D^{\rm NN,E}_n -1}{2M_N}
\nonumber\\&\qquad\qquad\qquad
    +\sinh \D^{\rm NN,E}_n
    -\frac{\nabla_a^2 \psi_n(\mathbf{r})}{M_N \psi_n(\mathbf{r})}
    \bigg]
\nonumber\\&= V_C(\mathbf{r})
    + \frac{1}{R(t,\mathbf{r})}\sum_n\bigg[
        \frac{(\D^{\rm NN,E}_n)^4}{48M_N}
\nonumber\\&\qquad
    +\frac{(\D^{\rm NN,E}_n)^3}{6}
    -\frac{\nabla_a^2+q_n^2}{M_N\psi_n(\mathbf{r})}
    \bigg] R_n(t,\mathbf{r})
\end{align}
where
\begin{equation}
R_n(t,\mathbf{r}) = \psi_n(\mathbf{r}) 
    Z_n^{\rm N} e^{-\D^{\rm NN,E}_n t}\, .
\end{equation}
and $\nabla_a$ is any discretized version of the Laplacian.  We have used the symmetric definition for the temporal derivative, such that the leading lattice spacing dependence from this term comes in beyond $\mathcal{O}(a^2)$.

Thus, we see that even within the regime dominated by elastic scattering state contributions, there is residual time dependence whose coefficients are of order $\mathcal{O}(a^2)$. 
In turn, discretization effects will be present in the extracted potential coming from the sum of all states whose contribution is non-negligible within a given time range. 
In contrast to \eqnref{eq:VC_excitedStates} for which the contamination comes from a relatively large inelastic energy gap, $\D^{\rm NN,I}_1$ and $\D^{\rm N}_1$, from this expression, we see that discretization effects lead to corrections that are only suppressed by elastic NN excitations, $\D^{\rm NN,E}\ll \D^{\rm NN,I}_1, \D^{\rm N}_1$.
These effects must be carefully considered for precision calculations, and when comparing results between the \luscher and potential methods.

\begin{figure}
\includegraphics[width=\linewidth]{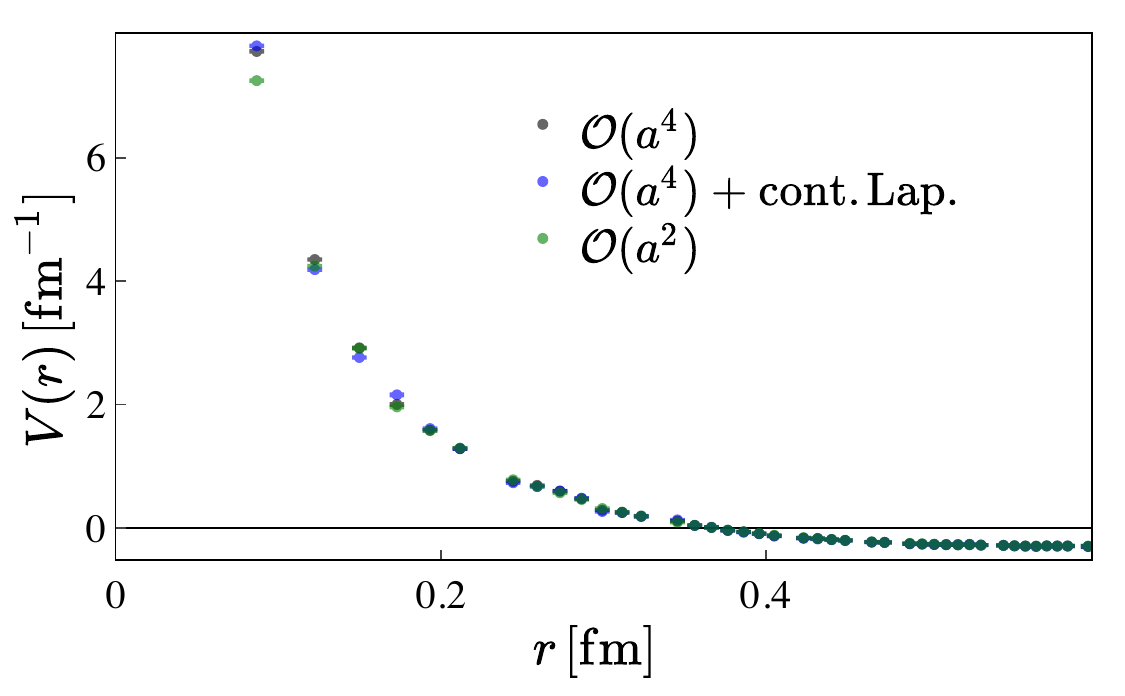}
\includegraphics[width=\linewidth]{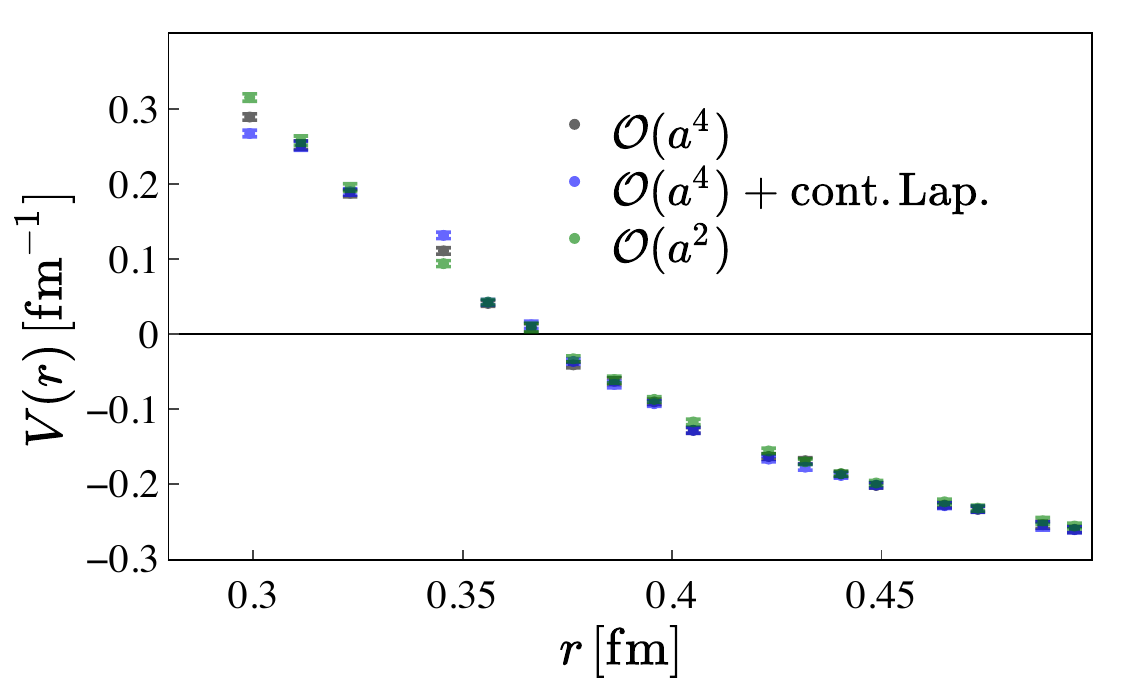}
\caption{Potential as a function of $r$ in the deuteron channel. Black data uses temporal and Laplacian lattice derivatives for the Schr\"odinger Equation which are improved to $\mathcal{O}(a^4)$, blue data uses the $\mathcal{O}(a^4)$ improved temporal derivative, with a continuum Laplacian up to a cutoff set by the edge of the Brillouin zone, and purple data uses unimproved ($\mathcal{O}(a^2)$) derivatives. The bottom plot shows a zoomed portion indicating the minimum $r \sim 0.35~$fm above which we no longer see significant discrepancies between the different levels of improvement. All units are in fm. 
\label{fig:pot_deriv}}
\end{figure}

Restoring explicit lattice spacing dependence to \eqnref{eq:V_a} shows each correction term scales as $\mathcal{O}(a^2)$ with respect to the leading term.  Thus, na\"ive discretization of the spatial and (symmetric) temporal derivatives leads to $\mathcal{O}(a^2)$ discretization effects. In this work, we attempt to reduce some of these discretization effects by using improved derivatives for the Schr\"odinger equation. For example, we can improve both the spatial Laplacian and temporal derivatives by including lattice points that are multiple lattice sites displaced from each other. Because we have the full position-space correlator, we are able to improve the derivative to all orders up to the lattice cutoff by computing the Laplacian directly in momentum space. The temporal derivative, however, mixes contributions at different Euclidean times, which will have different excited state dependence. Therefore, we limit the number of temporal hops to two when computing improved temporal derivatives. 
The results of these changes are shown in \figref{fig:pot_deriv}. We find that the potential only differs significantly to different levels of improvement for $r\lesssim 0.2$~fm. We were unable to find stable fits to $V_C(\mathbf{r})$ including data below $r\sim 0.4$~fm, limiting our study of the stability of the phase shift to values of $r_{\textrm{min}}$ greater than this value. However, for the low scattering momenta of interest, we find little sensitivity of the phase shift to $r_{\textrm{min}} \lesssim 0.5$~fm, therefore, we are insensitive to the effects of discretization of the Schr\"odinger Equation at $r\lesssim 0.2$~fm. We choose to adopt the $\mathcal{O}(a^4)$ improved Schr\"odinger Equation for our final results.

Finally, we test systematics coming from the short-range portion of the potential through various tests of the fit model. The HAL QCD collaboration has utilized a Gaussian basis to describe the potential as well as exploring the possibility of adding a one-pion exchange term for the long-range tail and rho-exchange for intermediate range contributions~\cite{Aoki:2009ji}. In this work, we thoroughly investigate the effects of different analytic forms used to fit the potential, including incorporating regulated one- and multiple-pion exchange potentials (see Sec.~\ref{sec:pot_fv}), and exploring the use of Harmonic Oscillator basis functions versus Gaussians. With differences in the results from different bases for the short-range potential, and variation in $r_{\rm min}$ included in the fit, we can probe discretization effects coming from short-range contributions. Fits utilizing $r_{\textrm{min}} \lesssim 0.5$~fm and all three forms for the short-distance fit function are included in our final quoted systematic.

Another discretization systematic arises from the modified dispersion relation, \eqnref{eq:nucleon_dispersion} for which we found $\xi=0.979(5)$, \eqnref{eq:xi}.
A proposal to account for this effect is to modify \eqnref{eq:time_dependent_HAL} by replacing $\nabla^2 \rightarrow \xi \nabla^2$, see App.~\ref{sec:hal_dispersion}.  As with the other discretization effects discussed in this section, we find that this does not have a statistically significant effect on the potential or subsequent value of $q\cot\d$. 
Since HAL QCD noted some impact of higher $\ell\geq4$ contributions that they removed with an $\ell=0$ projection using the Misner method~\cite{Misner:1999ab}, we also performed an $\ell=0$ projection, finding that the $\ell \geq4$ contributions are sub-percent variations that appear as a small noise contribution and do not affect our phase shift results, see \appref{sec:make_hal_potential}.

\subsubsection{\label{sec:pot_fv} Finite volume effects}

Similarly to the \luscher method, the potential method takes into account the finite volume by using a non-perturbative solution to the Schr\"odinger equation, subject to a periodic boundary condition. This boundary condition is then unfolded to deduce the infinite-volume phase shift. The \luscher method assumes a contact interaction, thus, corrections scale with the physical size of the interaction as $\mathcal{O}(e^{-m L})$, representing the probability for an exchanged scalar of mass $m$ to travel through the boundary. The potential extracted using the potential method is similarly polluted by the long-range tails of periodic copies of the potential occurring from the boundary condition. The overlap of these mirror images also scales as $\mathcal{O}(e^{-m L})$, assuming again that the long-range portion of the potential can be described through Yukawa exchange of a scalar particle. 

\begin{figure}
\includegraphics[width=\linewidth]{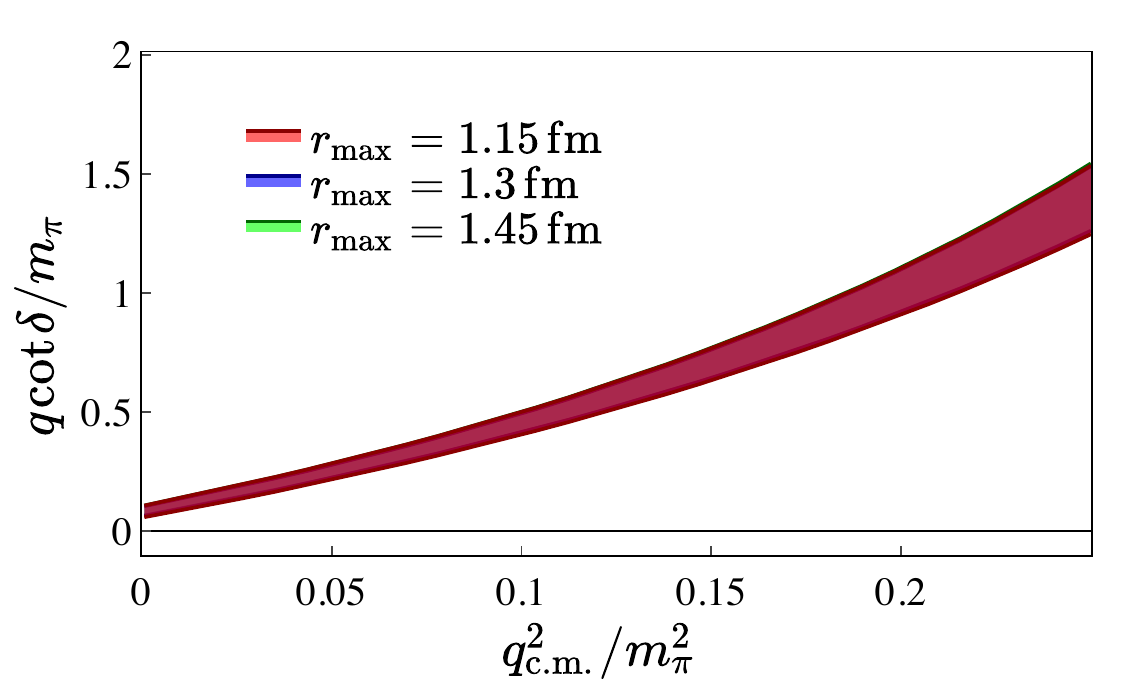}
\caption{Scattering phase shift, $q \cot \delta$, versus $q_{\mathrm{CM}}^2$ in units of $m_{\pi}$ in the deuteron channel using various cuts to the data, $r \leq r_{\mathrm{max}}$, in the fits of the potential. 
\label{fig:phase_rmax}}
\end{figure}

We test sensitivity to finite volume effects in multiple ways. First, we test different forms for the long-range terms in the fit function to the potential. For this, we use a sum of regulated Yukawa potentials, varying the number of included terms \eqnref{eq:pot_lr}. 
We furthermore test sensitivity of the fit to different cutoffs at large $r>r_{\mathrm{max}}$. The fits are again nearly completely insensitive to these changes, as shown in \figref{fig:phase_rmax}. We include fits using $r_{\mathrm{max}}\gtrsim1.3$~fm as part of our quoted systematic band.

\subsubsection{Derivative expansion}

While the above-mentioned systematics differ in certain respects between the \luscher method and the potential method, 
these differences are expected to vanish in the continuum and infinite-volume limits. However, a systematic within the potential method remains unresolved in these limits, namely, the use of a leading-order cutoff in the derivative expansion for the potential. Sensitivity to such effects can be probed using correlation functions having different spatial dependence at the source~\cite{HALQCD:2018gyl}. In such cases, care must be taken to disentangle effects from excited states, discretization, and derivative cutoff combined (see sections above). In this work, the wall source was the only source from which we obtained a signal, thus, we cannot probe the sizes of such effects.

\subsection{Final results}

\begin{figure*}
\includegraphics[width=0.49\textwidth]{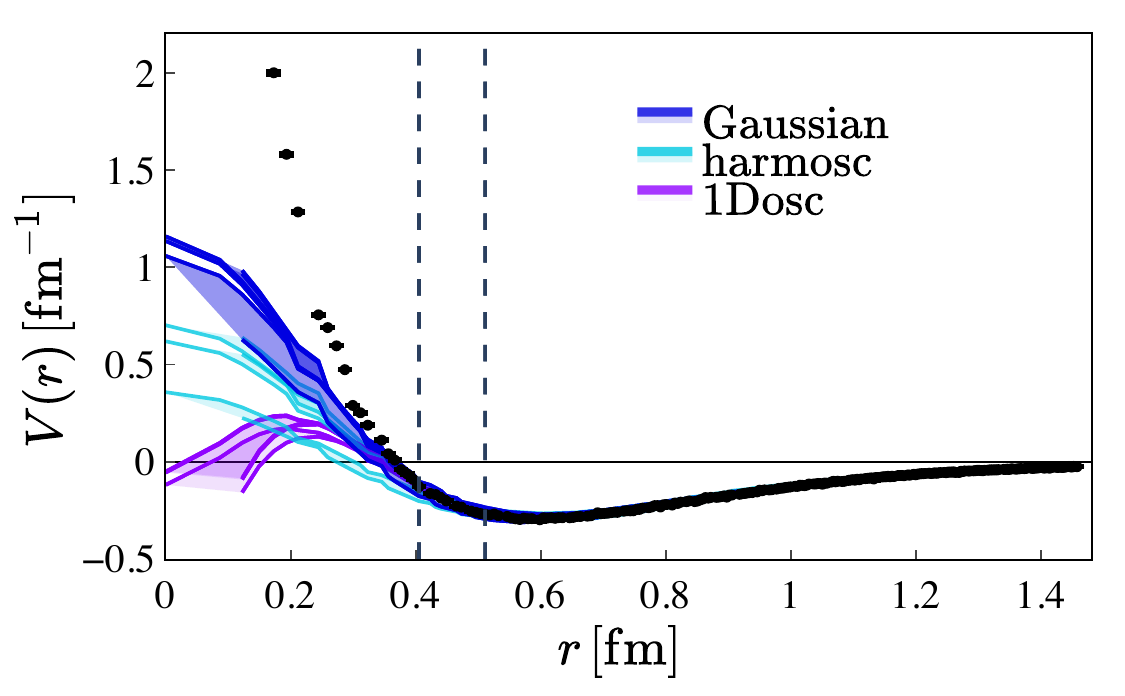}
\includegraphics[width=0.49\textwidth]{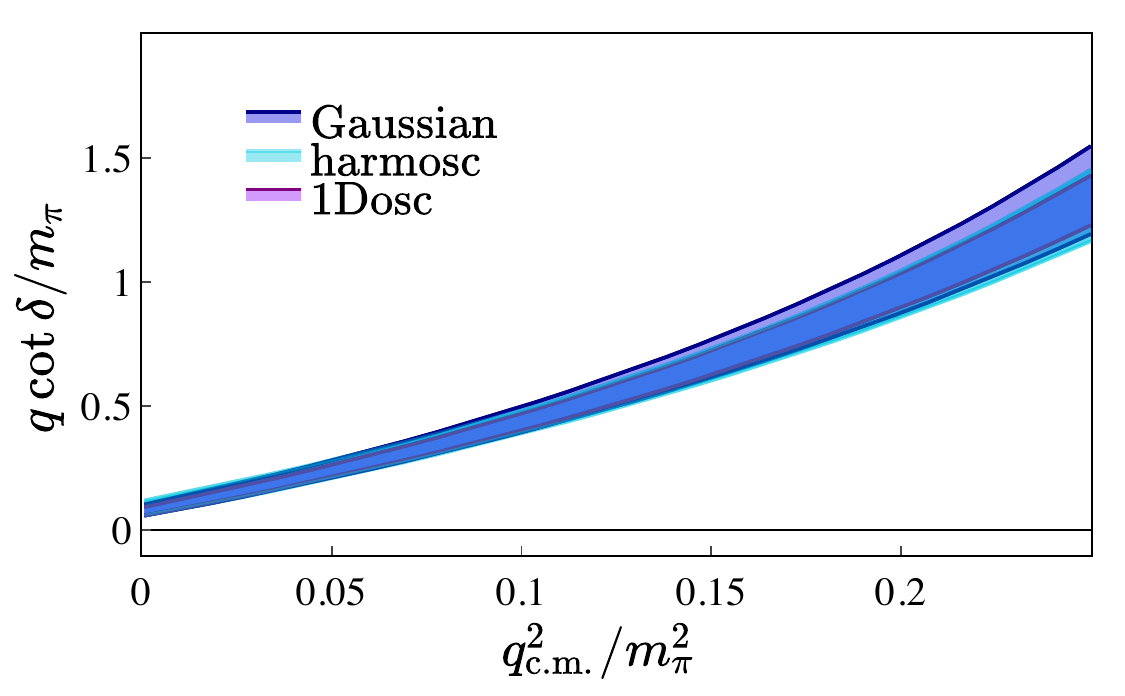}
\caption{Potential, $V_C(\mathbf{r})$ (left), in the deuteron channel as a function of $r$, in units of fm (black data). Colored bands represent all fits included in the final analysis, where the different colored bands utilize different functions as a basis for the short-range contributions (Eq.~\ref{eq:V_sr}). Vertical dashed lines represent the two possible minimum values of $r$ included in the fits.
The right figure shows the resulting $q\cot\d$ arising from these two different choices of the short-range potential over the full range of fits explored.  One can see that the choice of short-range potential and the choice of $r_{\rm min}$ do not have any significant impact on the determination of $q\cot\d$ in this low-energy regime.
\label{fig:pot_fits}}
\end{figure*}

In \figref{fig:pot_fits}, we show the potential extrapolated to $t\to\infty$ as a function of $r$ (black data). Fits to this data that are included in our final error band for the phase shift are shown as blue (purple) bands, representing fits utilizing both single- and double-meson exchange, \eqnref{eq:pot_lr}, and a sum of 2-3 Gaussians (Harmonic Oscillator functions) to represent the short-range portion of the potential. While the fits using Gaussians and Harmonic Oscillator functions show completely different behavior for $r$ smaller than the included fit range, we find no significant difference in the resulting phase shifts within the energy range of interest (\figref{fig:pot_fits} (right)). 
This also means the phase shift in this energy range is not sensitive to the divergent nature of the short range part of the potential~\cite{Aoki:2013zj}.
We therefore conclude that the physics of interest is insensitive to the behavior of the potential below this range. We also find no significant differences in our fitted potential having greater than 1 long-range term and greater than 2 short-range terms. The chosen fits furthermore include two different long-range cutoffs, $r_{\mathrm{max}} \sim 1.3, 1.45$~fm, and two different short-range cutoffs, $r_{\textrm{min}} \sim 0.4,0.5$~fm, for which we saw relative stability in the phase shift results (see previous Sections). 

Due to the large number of highly correlated data points, we could not accurately determine the covariance matrix for the entire data set of $V_C(\mathbf{r})$ on our given ensemble size. Therefore, we cut the data by only including every fifth point within $r_{\textrm{min}}\lesssim r \lesssim r_{\textrm{max}}$, then performed further fits with starting point varying between $r_{\textrm{min}} \to r_{\textrm{min}} + \delta r_{\textrm{min}}$, where $r_{\textrm{min}} +\delta r_{\textrm{min}}$ is the fourth point away from $r_{\textrm{min}}$ in the sequence. All such fits are included in our final quoted band for the phase shift. 

\begin{figure}
\includegraphics[width=\linewidth]{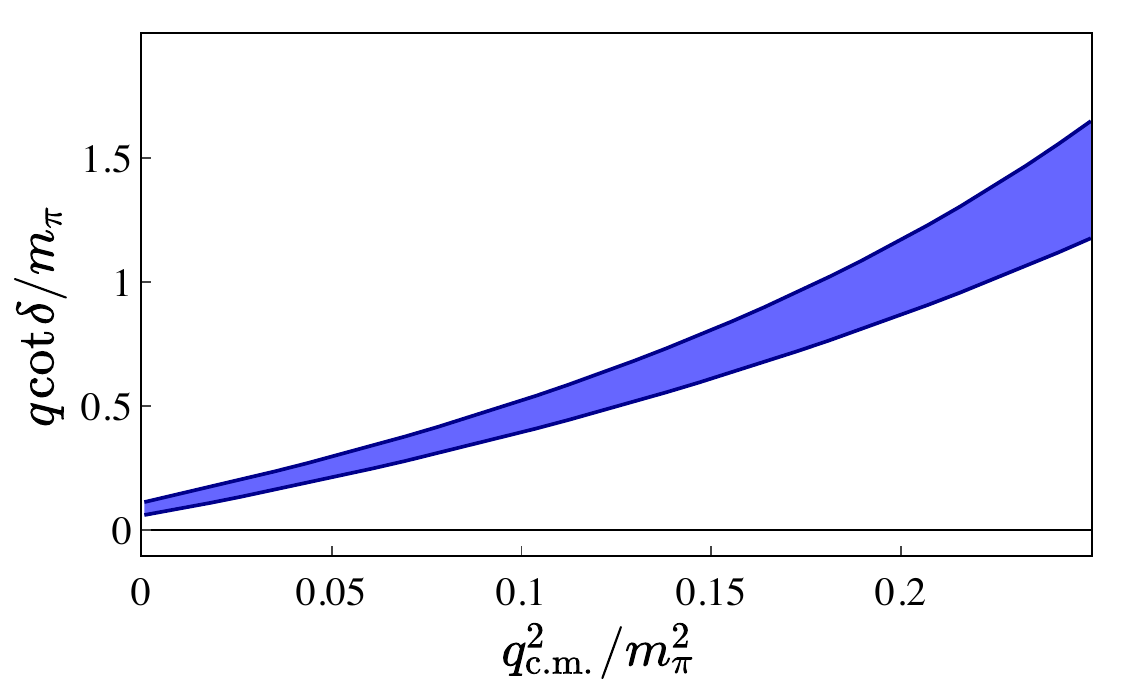}
\caption{Scattering phase shift, $q \cot \delta$, versus $q_{\mathrm{CM}}^2$ in units of $m_{\pi}$ in the deuteron channel. The band includes all fits to the potential as discussed in the text. 
\label{fig:phase_tot}}
\end{figure}

The scattering phase shift as a function of scattering momentum in the deuteron scattering channel is shown in \figref{fig:phase_tot} with the di-neutron presented in App.~\ref{seq:hal_nn}. The band represents the total contribution from all fits as discussed above, and includes estimates of excited state, discretization, and finite-volume effects.

%-------------------------------------------------------------------------------

%-------------------------------------------------------------------------------
%  Discussion and Conclusion
\section{Discussion and conclusion\label{sec:discussion}}

%----------------------------------
% FIGURE: Luscher and HAL
\begin{figure}
\includegraphics[width=\columnwidth]{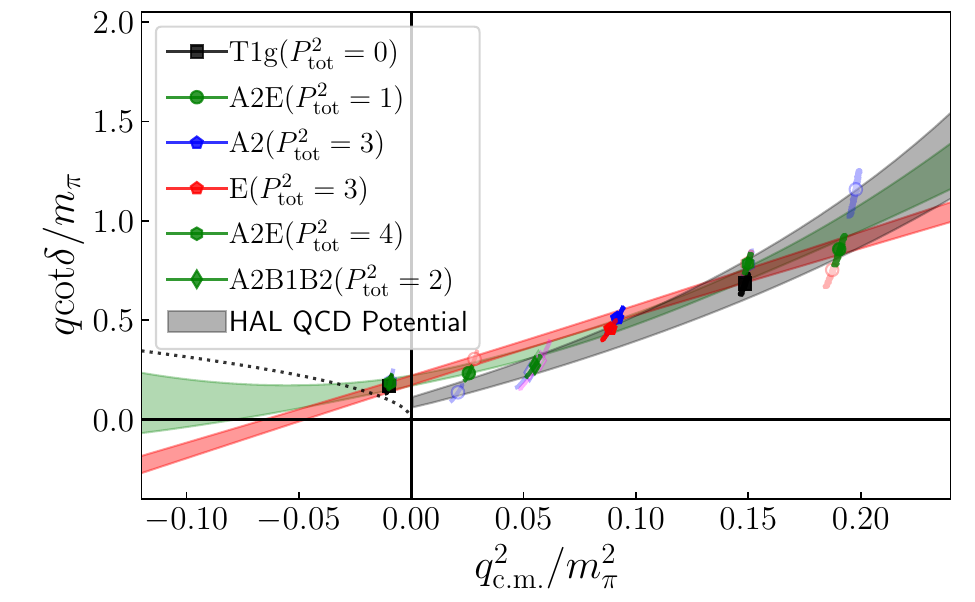}
\includegraphics[width=\columnwidth]{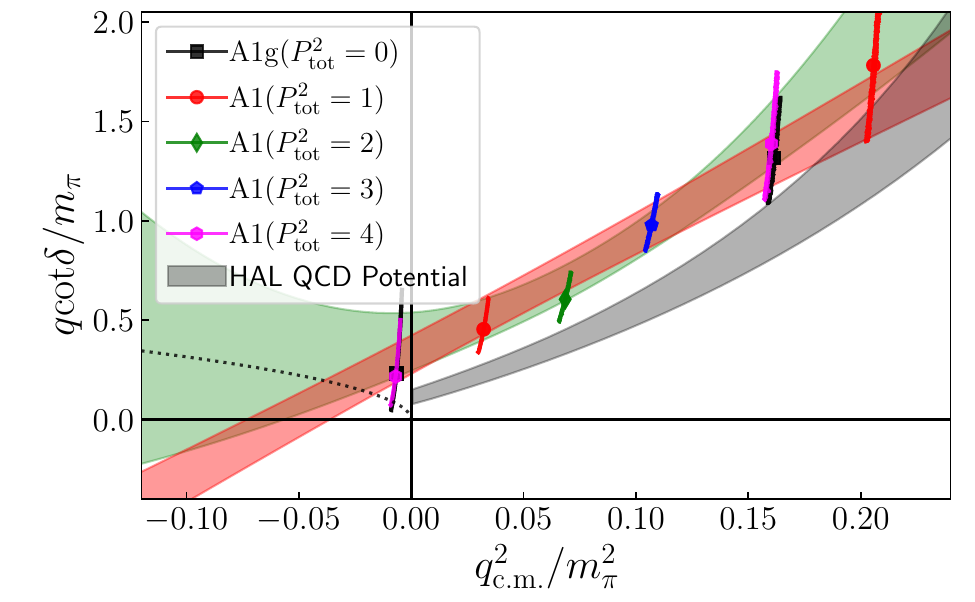}
\caption{Comparison of $q\cot\d$ between the Luscher analysis and the HAL QCD potential for the deutron (top) and di-neutron (bottom). 
\label{fig:luscher_hal}}
\end{figure}
%----------------------------------

We presented a high-statistics calculation of the NN scattering amplitudes in the SU(3) flavor limit with a heavy pion mass of $m_\pi\approx714$~MeV.
The calculation was designed to target a mass region where a discrepancy in the literature on the nature of the NN systems, whether or not they form bound states, has persisted for more than a decade, see for example the Lattice 2012 and 2013 reviews~\cite{Doi:2012ab,Walker-Loud:2014iea}.  In order to address this controversy, for the first time, we computed the NN amplitudes with several methods in the literature on the same set of underlying gauge ensembles, allowing us to isolate systematic uncertainties arising in the methods.

The stability of the extracted spectrum was investigated by varying all common aspects of the correlator analysis.  We introduced the {\it conspiracy model} to describe the remaining excited state contamination after the GEVP analysis removes the low-lying excited state contamination.  This model assumes the excited states in the resulting diagonal NN correlators are well described by the inelastic excitations of the single-nucleons used to build the NN operators.  We find this model is more probable than one that treats the NN excited states in an agnostic fashion, as determined through the Bayes evidence, see Secs.~\ref{sec:conspiracy} and \ref{sec:results_deuteron}.

A thorough analysis of various sources of systematic uncertainty that arise in the HAL QCD Potential method was conducted, identifying the most significant one being the infinite time extrapolation of the potential, which in principle as well as in practice with our results, requires a separate extrapolation at each range as the excited state contamination is $|\mathbf{r}|$-dependent.  After controlling this systematic uncertainty, all choices of parameterizing the potential with various analytic forms as well as cutting off the long or short-range parts of the potential, lead to consistent determinations of $q\cot\d$.

In \figref{fig:luscher_hal}, we plot the resulting $q\cot\d$ results from both the single channel finite-volume QC2 analysis as well as from the HAL QCD potential, which show qualitative agreement between the methods.
The source of small quantitative discrepancy between these methods is not currently resolved.  As discussed in the present work, the discretization errors between the methods are not the same so a continuum extrapolation is required for such a comparison.  If discretization effects are the source of discrepancy, those in \eqnref{eq:V_a} seem the most likely as they are only exponentially suppressed by energy gaps to the elastic NN scattering modes, and thus both difficult to resolve and more relevant for low-energy.  Another obvious candidate to check is the 
derivative expansion of \eqnref{eq:gradExpansion}, which we have not explored in this work.  
Finally, it would be interesting to use the HAL QCD method to construct optimized operators~\cite{Lyu:2025ncq} and then use the correlation functions built from them to see if improvements in the extraction of the energy and subsequent QC2 can be obtained.
Despite this small discrepancy, both methods exclude bound states.

In \secref{sec:add_HX}, we investigated the impact of including hexa-quark operators in the basis of NN operators used to construct the positive-definite matrices of correlation functions.  In both NN channels, we found that the hexa-quark operators did not have any statistically significant impact on the determination of the low-lying spectrum, indicating that such operators produce states with negligible overlap onto the scattering states of interest.

In \secref{sec:hx_off_diagonal},
we discussed the toy-model presented by in Ref.~\cite{Amarasinghe:2021lqa} and in Chapter 15 of Ref.~\cite{Tews:2022yfb} that argues the off-diagonal correlators may be able to capture a deeply-bound state that the other operators are not sensitive to.  Recognizing the unrealistic and contrived nature of their model, we made small, simple and physically reasonable changes to the model and showed that the off-diagonal hexa-quark correlator produced no advantage whatsoever.

The present work utilized a single lattice spacing and also a different clover-action from that in Refs.~\cite{NPLQCD:2012mex,Berkowitz:2015eaa,Wagman:2017tmp}.  Ref.~\cite{Green:2021qol} observed significant discretization effects for the H dibaryon.  We do not expect this to be the source of discrepancy, however.  The H dibaryon is a special case, and in the $SU(3)$ flavor limit, it is expected to have a purely attractive interaction~\cite{Inoue:2010es}.  Second, preliminary results for the di-nucleon on the same ensembles as the H dibaryon show milder discretization effects as well as a trend to move further from being a bound state as the continuum limit is approached~\cite{Green:2022rjj}.

We conclude that di-nucleons do not form bound states at heavy pion mass.
In the case of the deuteron, the QC2 analysis rules out a bound state on this ensemble by $>5\sigma$.  In the case of the di-neutron, a bound state is excluded at the $\lesssim3\s$ level.
Previous identification of bound and deeply-bound di-nucleons~\cite{NPLQCD:2012mex,Yamazaki:2012hi,Yamazaki:2015asa,Berkowitz:2015eaa,Wagman:2017tmp} most likely arose from a misidentification of a plateau in the effective energy arising from opposite sign contributions allowed in off-diagonal correlation functions, as first suggested by HAL QCD~\cite{Iritani:2016jie}.
This observation also invalidates all two- and higher-body LQCD matrix element calculations that utilized these off-diagonal correlation functions as those results relied upon the assumption of deeply dound di-nucleons to relate the finite and infinite volume matrix elements, which are in general non-trivial in magnitude, see for example~\cite{Briceno:2015dca}, thus leaving the previous results with an unquantified systematic.

Having demonstrated the efficacy and reliability of two-nucleon calculations with the sLapH method, it is time to move to LQCD calculations with lighter pion masses.  In particular, calculations with $m_\pi\lesssim200$~MeV are desired, such that extrapolations to the physical pion mass are feasible.
The results of this work represent a significant step toward deriving nuclear physics from first principles.

\bigskip
\noindent{DATA AVAILABILITY:}
The two point correlation functions as well as the ground state energies we extracted from them are available at OLCF through Constellation~\cite{nn_c103_data}.  Similarly, the $T_{1g}$ and $A_{1g}$ HAL QCD potential data are provided.  
Both data sets, as well as fit results are also available at NERSC via \url{https://portal.nersc.gov/cfs/m2986/cosmon/nn_c103_2505.05547/}.
The full HAL QCD potential data can be made available to interested parties with $\sim1$TB of space.  The analysis software used to generate the results presented in this work can be found at the git repo: \url{https://github.com/cosmon-collaboration/nn_c103_analysis}.

%-------------------------------------------------------------------------------
%  Acknowledgements
\acknowledgments

We thank Sinya Aoki and Jeremy Green for comments on the manuscript and results.

The results presented here utilized:
the Summit Supercomputer at Oak Ridge Leadership Computing Facility at the Oak Ridge National Laboratory, which is supported by the Office of Science of the U.S. Department of Energy under Contract No. DE-AC05-00OR22725;
the Lassen Supercompuer through the LLNL Multiprogrammatic and Institutional Computing program for Grand Challenge allocations;
the Frontera Supercomputer~\cite{frontera} at the Texas Advanced Computing Center (TACC);
the Perlmutter Supercomputer at the National Energy Research Scientific Computing Center (NERSC), a Department of Energy Office of Science User Facility using NERSC awards NP-ERCAP0021598, NP-ERCAP0017492, and NP-ERCAP0015497.

The \texttt{chroma\_laph} and \texttt{last\_laph} libraries were used to perform the sLapH calculations, making use of \texttt{QDP++}~\cite{qdpxx} and 
CHROMA~\cite{Edwards:2004sx} software 
libraries.
The quark propagator solves were performed with the QUDA library~\cite{Clark:2009wm,Babich:2011np,quda:software} and the final contractions used \texttt{contraction\_optimizer}~\cite{contraction_optimizer}.
The LALIBE library~\cite{lalibe}, branch \texttt{feature/mp\_nn} was used to perform computations of the HAL QCD potential.  The correlation function analysis code makes use of \texttt{lsqfit}~\cite{lsqfit}, \texttt{gvar}~\cite{gvar} and TwoHadronsInBox~\cite{Morningstar:2017spu}.

This work was supported in part by the U.S. National Science Foundation (NSF) under awards 
PHY-1913158 and PHY-2209167 (C.M., S.S.), under award PHY-2209185 (A.S.), the NSF Faculty Early Career Development Program (CAREER) under award PHY-2047185 (A.N.), and by the Graduate Research Fellowship Program under Grant
No. DGE-2040435 (J.M.).
Any opinions, findings, and conclusions or recommendations expressed in this material are those of the author(s) and do not necessarily reflect the views of the NSF.
This work was also supported in part by the U.S. Department of Energy (DOE), Office of Science, Office of Nuclear Physics, under grant contract numbers DE-AC02-05CH11231 (A.WL., J.M.), DE-SC0020250 (A.S.M.), DE-SC0004658 (K.M.), 
the DOE Topical Collaboration “Nuclear Theory for New Physics”, award No. DE-SC0023663 (A.N.,A.S.,A.WL,C.M.,J.M.),
and U.S. DOE, Office of Science, Office of Workforce Development for Teachers and Scientists, Office of Science Graduate Student Research (SCGSR) program. The SCGSR program is administered by the Oak Ridge Institute for Science and Education (ORISE) for the DOE. ORISE is managed by ORAU under contract number DESC0014664 (J.M.). 
This work is supported by Lawrence Livermore National Security, LLC DE-AC52-07NA27344 (P.M.V., A.S.M.), Neutrino Theory Network Program Grant DE-AC02-07CHI11359 (A.S.M.).
This research used resources of the Oak Ridge Leadership Computing Facility at the Oak Ridge National Laboratory, which is supported by the Office of Science of the U.S. DOE under Contract No. DE-AC05-00OR2272 (H.MC.). 
This work was supported in part by the Deutsche Forschungsgemeinschaft (DFG, German Research Foundation) 
through grant 513989149 (A.S.).
This work was supported in part by ERC grant StrangeScatt-101088506 (J.B.).

\appendix
\section{Correlator Analysis Systematics\label{sec:correlator_analysis}}

\begin{figure*}
\begin{tabular}{cc}
\includegraphics[width=0.49\textwidth]{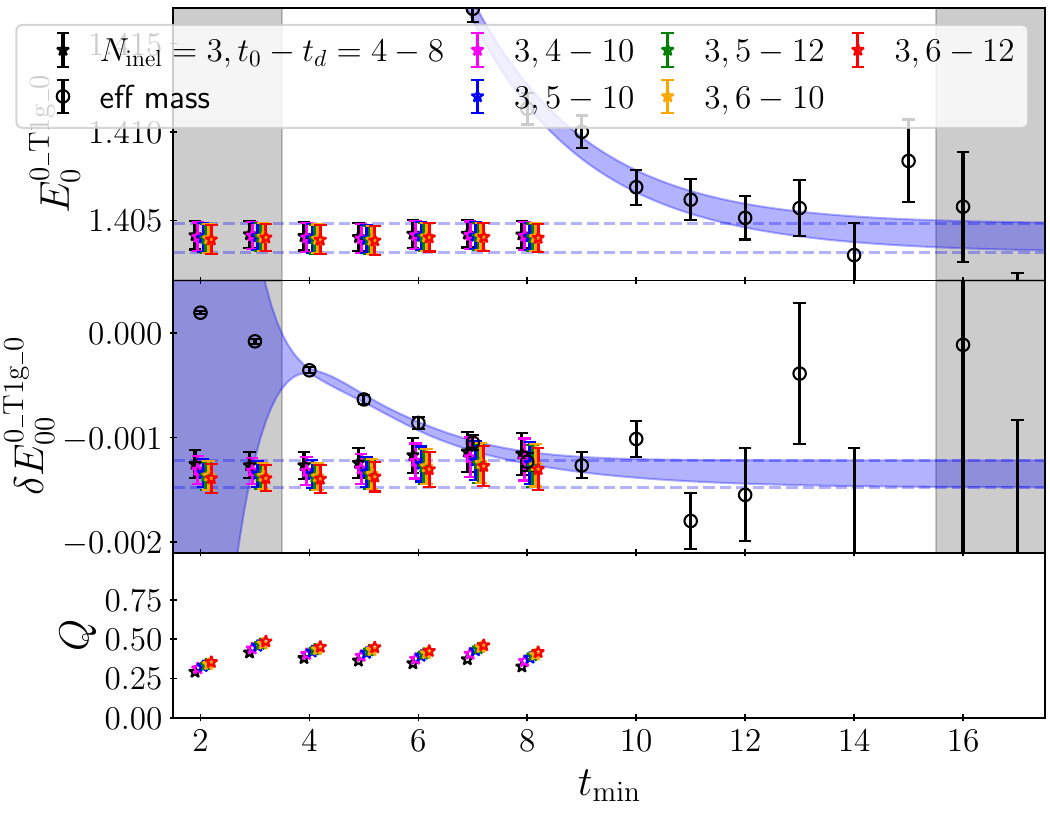}
&
\includegraphics[width=0.49\textwidth]{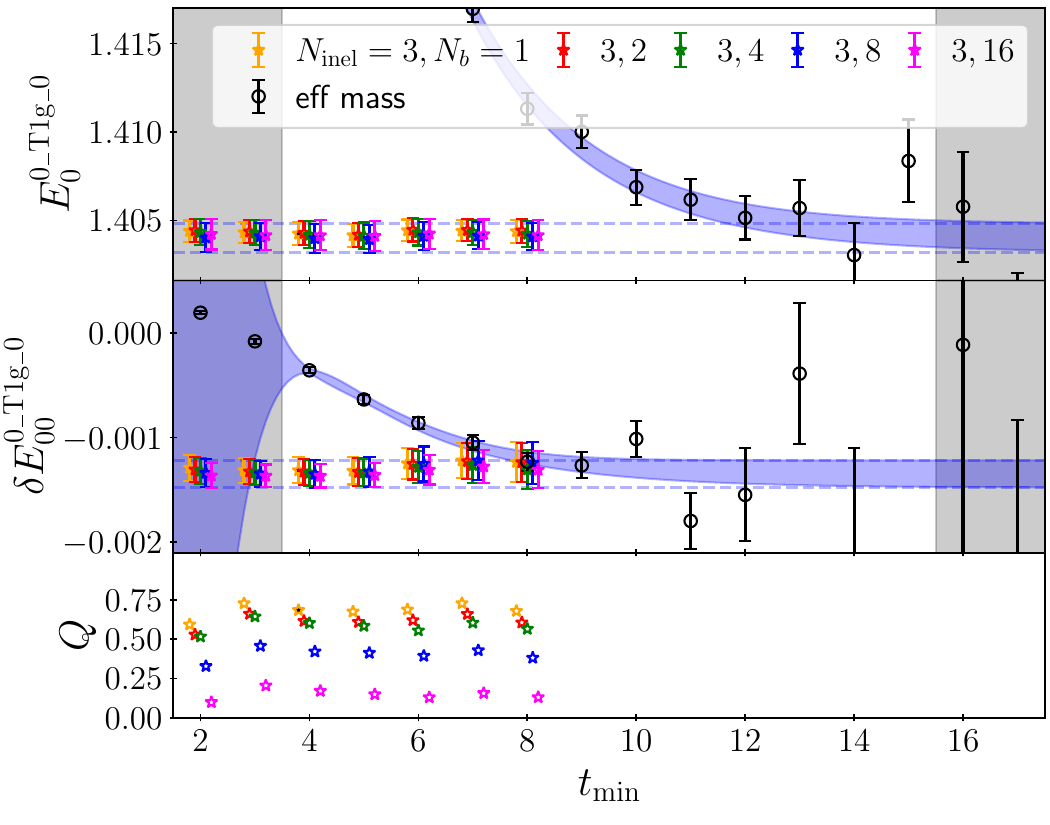}
\\
(GEVP times, $t_0-t_d$)& (blocking size, $N_b$)\\ \\
\includegraphics[width=0.49\textwidth]{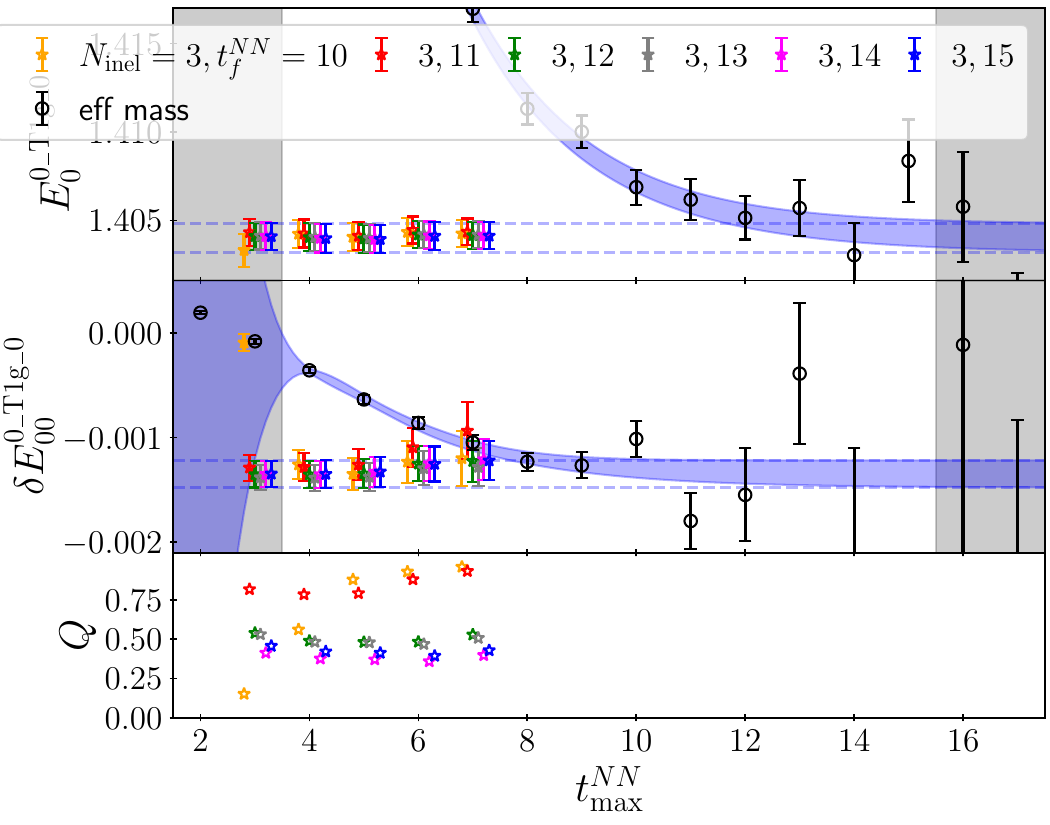}
&
\includegraphics[width=0.49\textwidth]{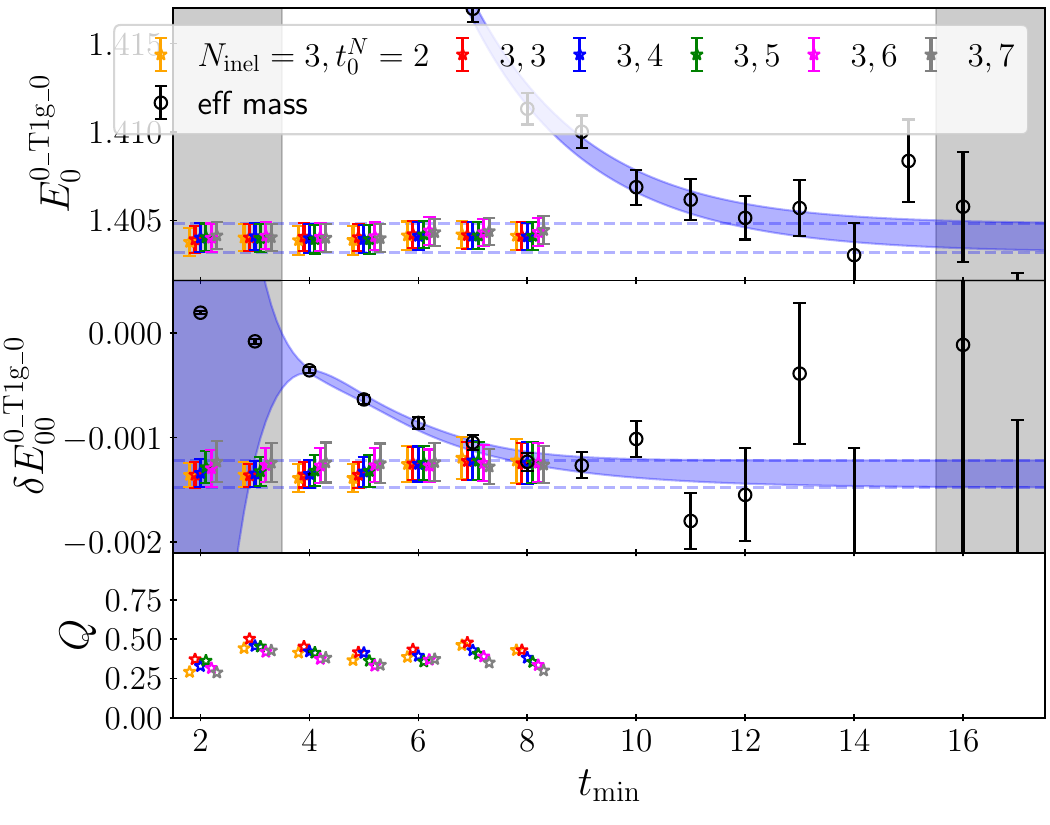}\\
(the final NN time $t_f^{\rm NN}$)& (the minimum N time $t_{\rm min}^{\rm N}$)
\end{tabular}
\caption{\label{fig:gevp_nstate}
Stability of the ground state energy and interaction energy in the $T_{1g}$ irrep versus the choice of GEVP times, the number of states to parameterize the single-nucleon correlators, the final NN time, and the minimum N time.  The posterior of the chosen fit with $t_0=5$, $t_d=10$, and $N_{\rm inel}=3$ is plotted to guide the eye in both cases.
}
\end{figure*}

In \figref{fig:gevp_nstate}, we show the sensitivity of the extracted ground state total NN energy as well as the interaction energy versus the choice of $t_0$ and $t_d$ used to perform the GEVP, the blocking/binning size used ($N_b$), the choice of the final time used for analyzing the NN correlators ($t_f^{\rm NN}$), and the minimum time used for the single nucleon correlator ($t_{\rm min}^N$).  For the blocking, $N_b=1$ indicates no block averaging leaving us with all 1490 configurations while $N_b=8$ leaves us with 187 ``configurations''.

%------------------------------------------------------------------------------
%------------------------------------------------------------------------------

%------------------------------------------------------------------------------
\begin{figure*}
\begin{tabular}{cc}
\includegraphics[width=\columnwidth]{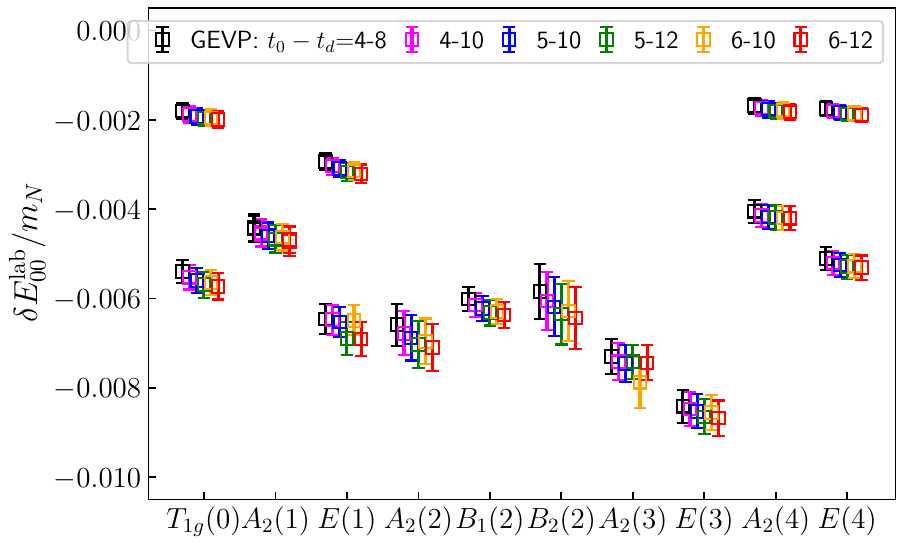}
&
\includegraphics[width=\columnwidth]{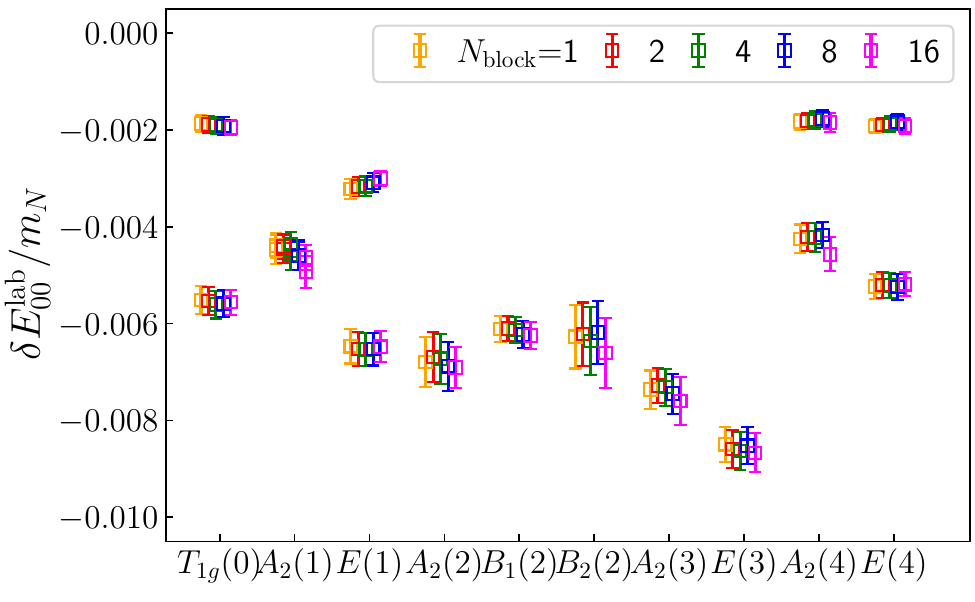}
\\
(GEVP times, $t_0-t_d$)& (blocking size, $N_b$)\\ \\
\includegraphics[width=\columnwidth]{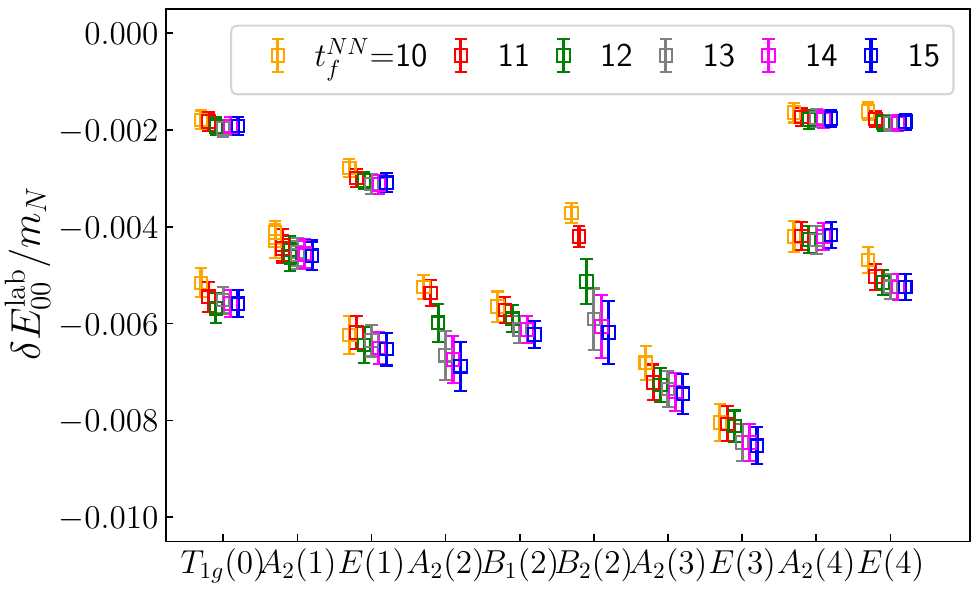}
&
\includegraphics[width=\columnwidth]{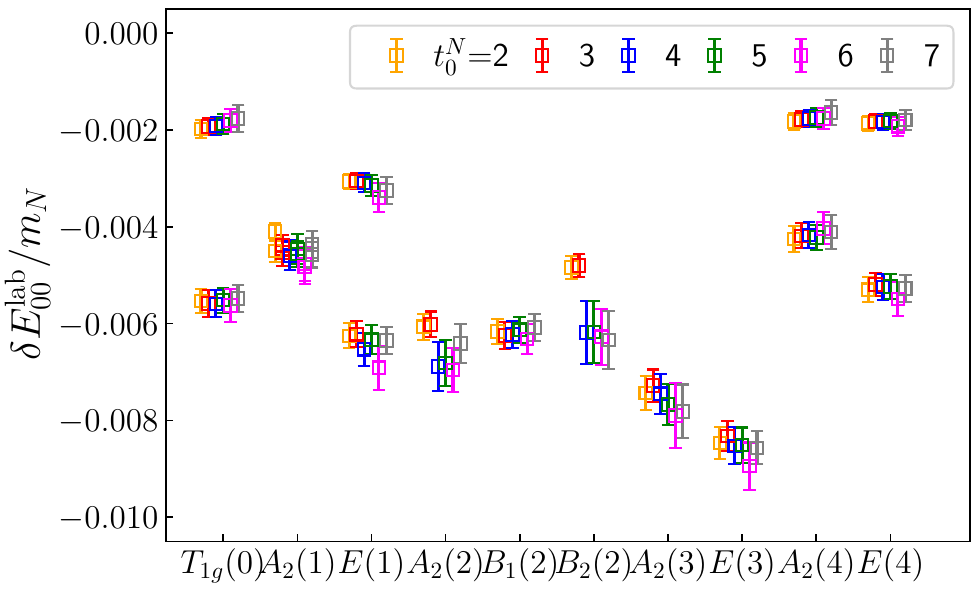}
\\
(the final NN time $t_f^{\rm NN}$)& (the minimum N time $t_{\rm min}^{\rm N}$)\
\end{tabular}
\caption{\label{fig:deuteron_dE_more}
Stability of the ground state interaction energy , $\d E_{00}$ for all 15 irreps/levels used in this work for the deuteron.
Note that the two levels in $A_2(1)$ have overlapping values of $\d E_{00}$.
}
\end{figure*}
%------------------------------------------------------------------------------

%------------------------------------------------------------------------------
\begin{figure*}
\begin{tabular}{cc}
\includegraphics[width=\columnwidth]{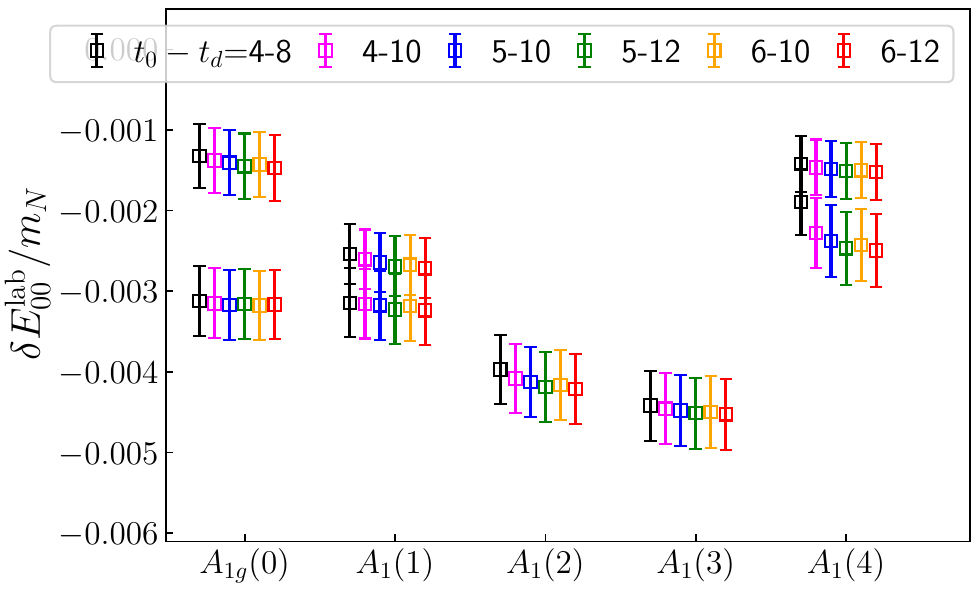}
&
\includegraphics[width=\columnwidth]{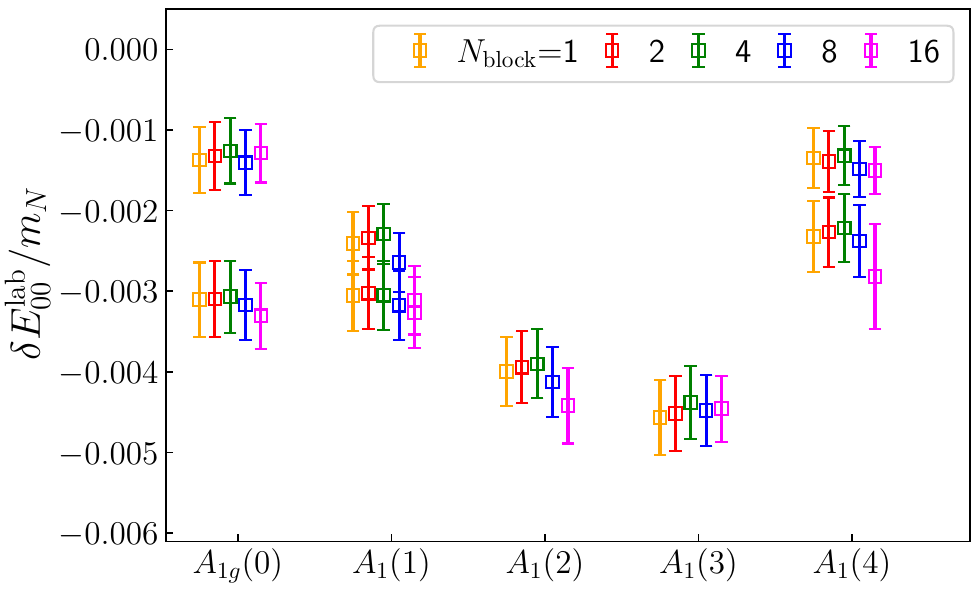}
\\
(GEVP times, $t_0-t_d$)& (blocking size, $N_b$)\\ \\
\includegraphics[width=\columnwidth]{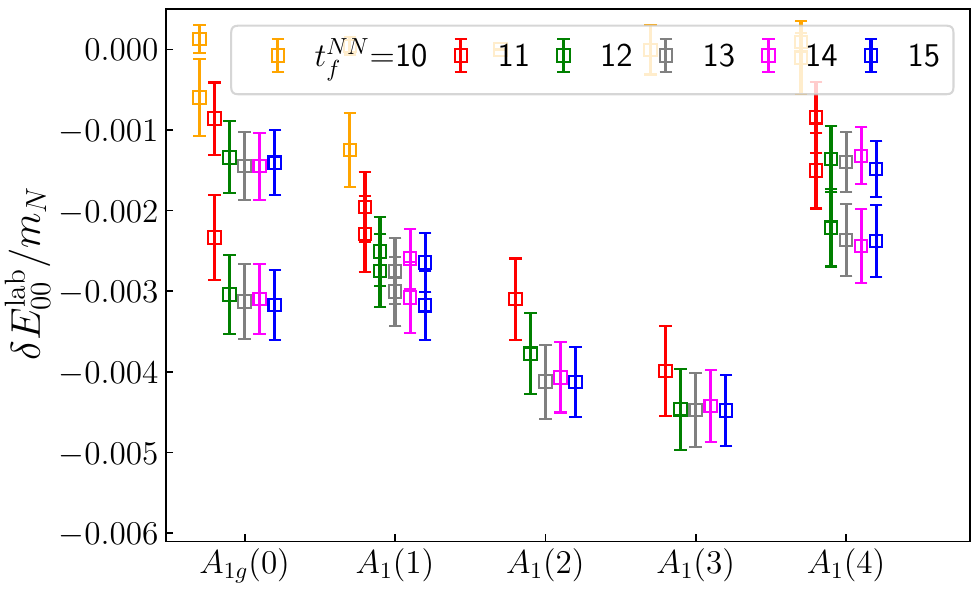}
&
\includegraphics[width=\columnwidth]{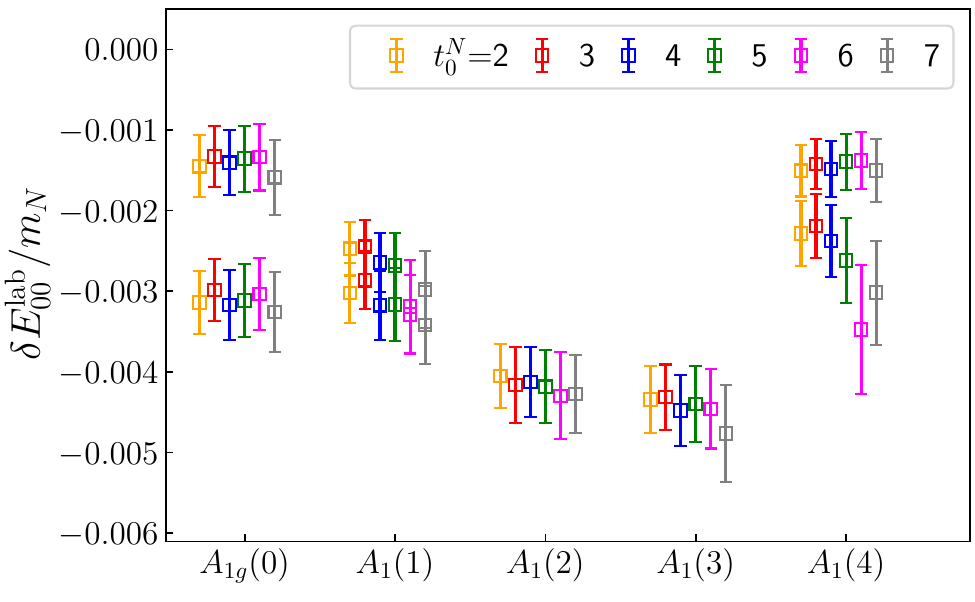}
\\
(the final NN time $t_f^{\rm NN}$)& (the minimum N time $t_{\rm min}^{\rm N}$)\
\end{tabular}
\caption{\label{fig:dineutron_dE_more}
Stability of the ground state interaction energy , $\d E_{00}$ for all 8 irreps/levels used in this work in the di-neutron channel.
}
\end{figure*}
%------------------------------------------------------------------------------

\section{HAL QCD Potential Analysis Details}

\subsection{Inelastic Excited State Contamination\label{sec:HAL_inel_es}}

The single nucleon correlation functions are described by
\begin{equation}
C_{\rm N}(t) = A_0 e^{-M_N t}\left(
    1 + \sum_{n=1}\tilde{\alpha}_n e^{-\D^{\rm N}_n t}    
\right)\, ,
\end{equation}
with $\D^{\rm N}_n=E_n-M_N$ and $\tilde{\alpha}_n$ is the overlap of the interpolating operators on the $n^{th}$ state normalized by $A_0$.
We can similarly parameterize the NN correlation function while separating the sum over states into elastic scattering modes of the two ground state nucleons from the inelastic states arising either from single-nucleon inelastic states, or inelastic states intrinsic to the NN system, such as $\D\D$ scattering,
\begin{equation}
C_{\rm NN}(t,\textbf{r}) = C_{\rm NN}^{el}(t,\textbf{r})
    + C_{\rm NN}^{inel}(t,\textbf{r})\, ,
\end{equation}
where the two contributions are
\begin{align}
C_{\rm NN}^{el}(t,\textbf{r}) &= e^{-2M_N t} 
    \sum_{n=0}
    \psi_{n}(\mathbf{r})Z^E_{n} e^{-\D^{\rm NN,E}_{n}t}\, ,
\nonumber\\
C_{\rm NN}^{inel}(t,\textbf{r}) &= e^{-2M_N t} 
    \sum_{n=1}
    \Psi_{n}(\mathbf{r})Z^I_{n} e^{-\D^{\rm NN,I}_{n}t}\, ,
\end{align}
where the energy gaps are $\D^{\rm NN,E}_{n} = 2\sqrt{M_N^2 + \mathbf{q}_n^2}-2M_N$ and $\D^{\rm NN,I}_{n} = E^{\rm NN,I}_{n} - 2M_N$.
Inserting these expressions into the ratio correlation functions and retaining only the leading inelastic excited state contributions, one has
\begin{align}
R(t,\mathbf{r}) &= \frac{C_{\rm NN}(t,\textbf{r})}{C_{\rm N}(t)^2}
    = R^E(t,\mathbf{r}) + R^I(t,\mathbf{r})
\nonumber\\&\approx
    \left[\sum_{n=0} \psi_{n}(\mathbf{r}) Z^E_{n}e^{-\D^{\rm NN,E}_{n} t} \right]\left[
        1 - 2 \tilde{\alpha}_1 e^{-\D^{\rm N}_1 t}\right]
\nonumber\\&\phantom{\approx}
    + \Psi_1(\mathbf{r}) Z^I_1 e^{-\D^{\rm NN,I}_1 t}\, .
\end{align}
Under this approximation, the elastic ratio function
\begin{equation}\label{eq:R_HAL_elastic}
R^E(t,\mathbf{r}) = \sum_{n=0} \psi_{n}(\mathbf{r}) Z^E_{n}e^{-\D^{\rm NN,E}_{n} t}\, ,
\end{equation}
gives rise to the time-independent central potential $V_C(\mathbf{r})$, while the inelastic terms
\begin{align}\label{eq:R_I}
R^I(t,\mathbf{r}) &= 
    R^I_1(t,\mathbf{r})
    -2\tilde{\alpha}_1 R^I_{\rm N}(t,\mathbf{r})
\nonumber\\
R^I_1(t,\mathbf{r}) &=
\Psi_1(\mathbf{r}) Z^I_1 e^{-\D^{\rm NN,I}_1 t}
\nonumber\\
R^I_{\rm N}(t,\mathbf{r}) & =
    \sum_{n=0} 
    \psi_{n}(\mathbf{r}) Z^E_{n}e^{-(\D^{\rm N}_1+\D^{\rm NN,E}_{n}) t}\, ,
\end{align}
will give rise to time-dependence in the determination of $V_C(t,\mathbf{r})$.
Each state in $R^E(t,\mathbf{r})$ will satisfy the relations
\begin{align}\label{eq:elastic_relations}
\D_n^{\rm NN,E} &= \frac{\mathbf{q}_n^2}{M_N}
    -\frac{(\D_n^{\rm NN,E})^2}{4M_N}\, ,
\nonumber\\
\frac{\mathbf{q}_n^2}{M_N} \psi_n(\mathbf{r}) &=
    \left[H_0 + V_C(\mathbf{r})\right]\psi_n(\mathbf{r})\, .
\end{align}
Under the approximation that the first inelastic state, $\Psi_1(\mathbf{r})$ is an asymptotic scattering state that does not couple to other states, it will satisfy similar relations with a different potential,
\begin{equation}
V_C^I(\mathbf{r}) = V_C(\mathbf{r}) +\d V_C(\mathbf{r})\, .
\end{equation}
The latter terms in \eqnref{eq:R_I} depend upon the elastic NBS wave functions, $\psi(\mathbf{r})$, enabling some simplifications.
Applying the RHS of \eqnref{eq:V_central} to the expression above leads to,
\begin{align}\label{eq:V_t_extrap}
V_C(t,\mathbf{r}) &= \frac{1}{R(t,\mathbf{r})} \bigg[
    V_C(\mathbf{r}) R^E(t,\mathbf{r})
    +V_C^I(\mathbf{r}) R^I_1(t,\mathbf{r})
\nonumber\\&\phantom{=}
    -2\tilde{\alpha}_1 R^I_{\rm N}(t,\mathbf{r})\bigg[
        V_C(\mathbf{r}) 
%\nonumber\\&\phantom{=}\quad
    +\D_1^{\rm N}\left(1 
        +\frac{\D_1^{\rm N}}{4M_N}\right)
    \bigg]
\nonumber\\&\phantom{=}
    -2\tilde{\alpha}_1 \sum_{n=0} \frac{\D_1^{\rm N} \D^{\rm NN}_n}{2M_N}
    \psi_{n}(\mathbf{r}) Z^E_{n}e^{-(\D^{\rm N}_1+\D^{\rm NN,E}_{n}) t}
    \bigg]
\nonumber\\&\simeq V_C(\mathbf{r})
    +\d V_C(\mathbf{r})\frac{R^I_1(t,\mathbf{r})}{R(t,\mathbf{r})}
\nonumber\\&\phantom{=}
    -2\tilde{\alpha}_1\D_1^{\rm N}
        \left(1 
        +\frac{\D_1^{\rm N}}{4M_N}\right)
    \frac{R^I_{\rm N}(t,\mathbf{r})}{R(t,\mathbf{r})}\, .
\end{align}
To get the second equality, we used $\D^{\rm NN}_n\ll \D_1^{\rm N}$ to drop the last term.
As $\D_1^{\rm N}\gg \D_n^{\rm NN,E}$ and $\D_1^{\rm NN,I}\gg \D_n^{\rm NN,E}$, the latter two terms are exponentially damped as
\begin{align}
\frac{R^I_1(t,\mathbf{r})}{R(t,\mathbf{r})}
&\approx \mathrm{O}\left(e^{-(\D_1^{\rm NN,I}-\D_0^{\rm NN,E})t}\right)
\approx e^{-\D_1^{\rm NN,I}t}\, ,
\nonumber\\
\frac{R^I_{\rm N}(t,\mathbf{r})}{R(t,\mathbf{r})}
&\approx \mathrm{O}\left(e^{-(\D_1^{N}-\D_0^{\rm NN,E})t}\right)
\approx e^{-\D_1^{N}t}\, .
\end{align}
This leads to \eqnref{eq:VC_excitedStates}.

\subsection{Dispersion relation modification\label{sec:hal_dispersion}}

Consider the HAL QCD potential equation, \eqnref{eq:time_dependent_HAL}, with only elastic scattering states contributing to $R(t,\mathbf{r})$ and use of the leading approximation to $U(\mathbf{r},\mathbf{r}^\prime)$.  Consider this equation for a single elastic mode, $n$
\begin{align}\label{eq:hal_potential_elastic}
\left[ \frac{\partial_t^2}{4M_N} - \partial_t -H_0 \right] R_n(t,\mathbf{r})
&=
\int d^3 r'\, U(\mathbf{r},\mathbf{r'}) R_n(t,\mathbf{r'})
\nonumber\\
\left[\frac{E_n^2}{4M} +E_n -\frac{\nabla^2}{M}\right]R_n(t,\mathbf{r}) &= V_C(\mathbf{r})R_n(t,\mathbf{r})\, .
\end{align}
Define the energy of this state as
\begin{align}
E_n &= 2\sqrt{q_n^2 +\d q_n^2 + M^2} - 2M\, ,
\nonumber\\
q_n^2 &= n \left(\frac{2\pi}{L}\right)^2\, ,
\end{align}
such that $q_n$ is the quantized momentum from the periodic boundary conditions and $\d q_n^2$ is the distortion induced by the interactions.  Then the kinetic operator $\nabla^2 R_n(t,\mathbf{r}) = -q_n^2 R_n(t,\mathbf{r})$ and one can show that 
\begin{equation}
\frac{\partial_t^2}{4M} R_n(t,\mathbf{r}) = 
    \left[\frac{q_n^2 + \d q_n^2}{M} -E_n\right] R_n(t,\mathbf{r})\, ,
\end{equation}
and thus from \eqnref{eq:hal_potential_elastic}, we have the relation
\begin{equation}\label{eq:hal_pot_dqsq}
\frac{\d_q^2}{M} R_n(t,\mathbf{r}) = V_C(\mathbf{r}) R_n(t,\mathbf{r})\, .
\end{equation}

Now consider the situation where the single nucleons have a modified dispersion relation due to the lattice discretization effects, \eqnref{eq:nucleon_dispersion}.  From the derivation above, we see this would impact the energy, such that 
\begin{equation}
    E_n^\xi = 2\sqrt{\xi q_n^2 +\d q_n^2 + M^2} -2M\, .
\end{equation}
Therefore, in order to maintain the relation between the interacting energy, $\d q_n^2/M$ and the potential, we need to modify the HAL QCD potential equation as
\begin{multline}\label{eq:hal_potential_dispersion}
\left[ \frac{\partial_t^2}{4M_N} - \partial_t -\xi\frac{\nabla^2}{M} \right] R_n(t,\mathbf{r})
\\=
\int d^3 r'\, U(\mathbf{r},\mathbf{r'}) R_n(t,\mathbf{r'})\, 
\end{multline}
such that the $\xi q_n^2$ term cancels as above leaving the relation \eqnref{eq:hal_pot_dqsq}.

\begin{figure*}
\includegraphics[width=\columnwidth]{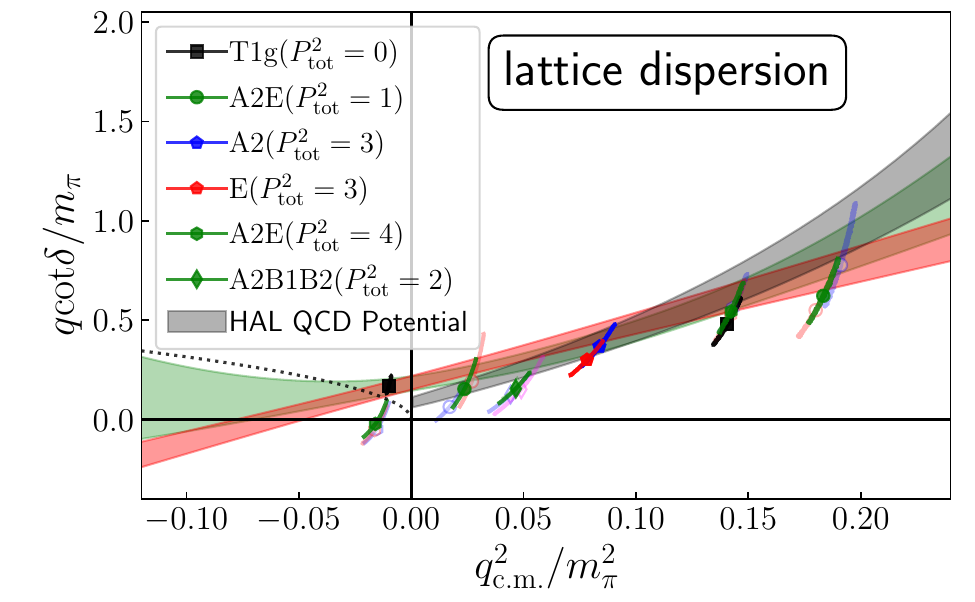}
\includegraphics[width=\columnwidth]{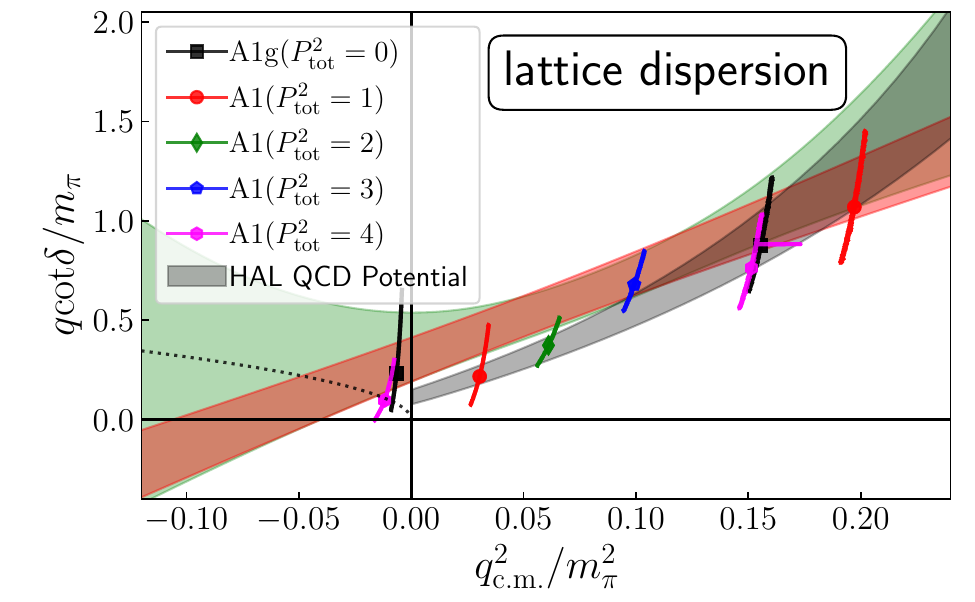}
\caption{Same as \figref{fig:luscher_hal} except using the lattice dispersion relation in the QC2. 
\label{fig:luscher_hal_latticeDisp}}
\end{figure*}

As mentioned in \secref{sec:hal_discretization}, we did not observe statistically significant changes to the medium and long-range part of the potential, or the resulting $q\cot\d$, with this modification, and so it does not seem to be the source of discrepancy shown in \figref{fig:luscher_hal}.  In contrast, if we use the lattice dispersion relation in the QC2 analysis, thereby introducing discretization errors, we observe a shift in the resulting values of $q\cot\d$ that are on the order of the discrepancy between the HAL QCD Potential result and the QC2 results.  This is shown in \figref{fig:luscher_hal_latticeDisp}.

%----------------------------------
% FIGURE: Luscher and HAL
%----------------------------------

\subsection{Di-neutron potential analysis\label{seq:hal_nn}}

In this appendix, we collect the same study of systematics presented in \secref{sec:potential} except for the di-neutron system.

%-----------------------------------------
% HAL QCD di-neutron potential figures
\begin{figure*}
\includegraphics[width=\columnwidth]{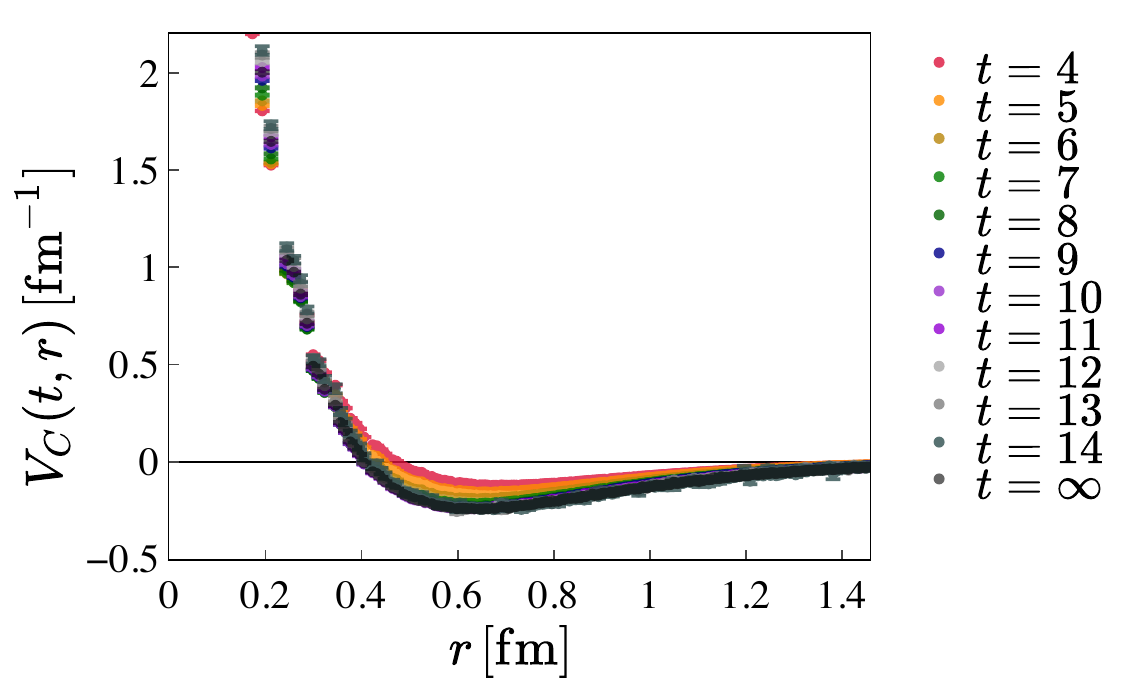}
\includegraphics[width=\columnwidth]{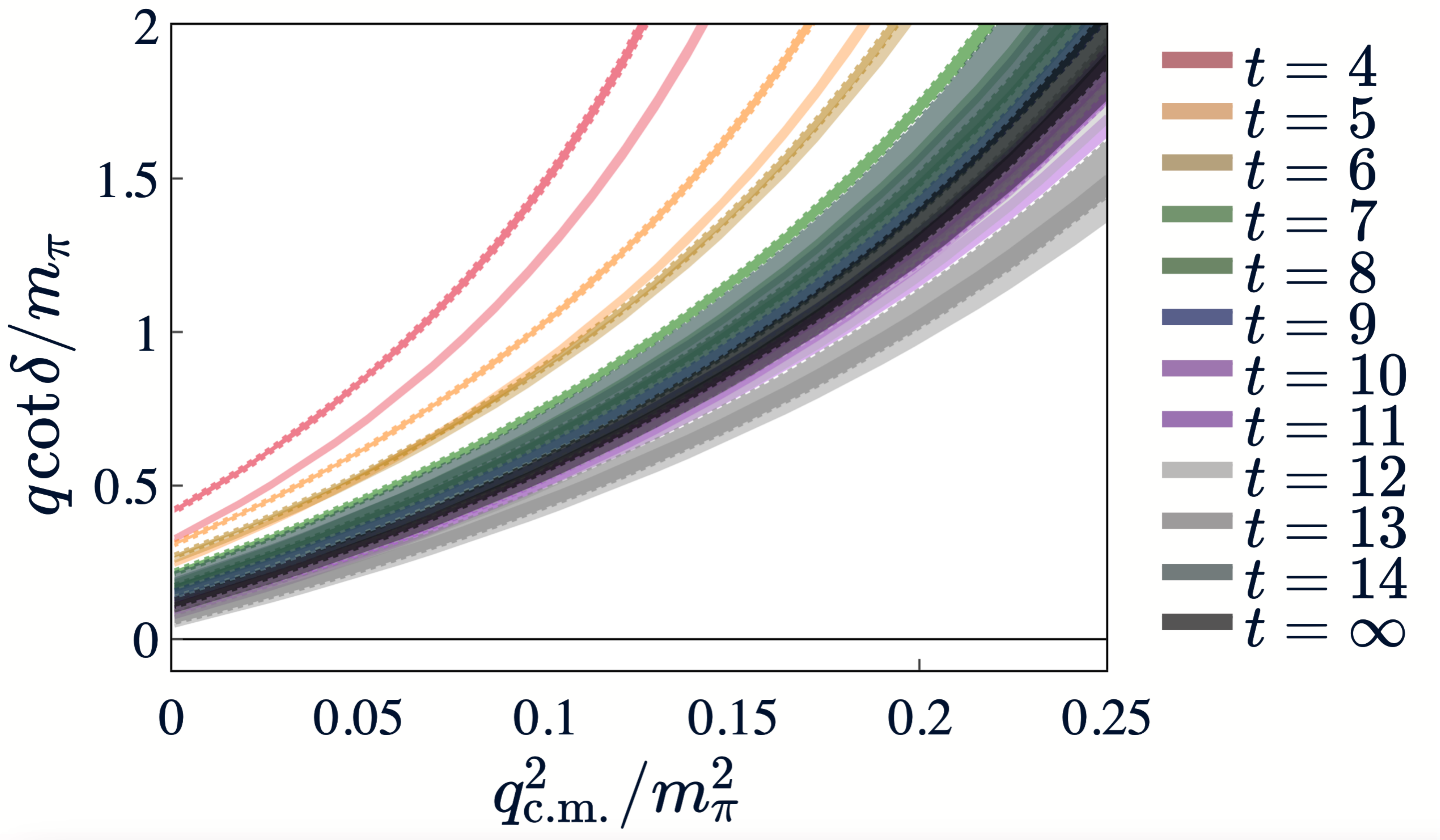}
\includegraphics[width=\columnwidth]{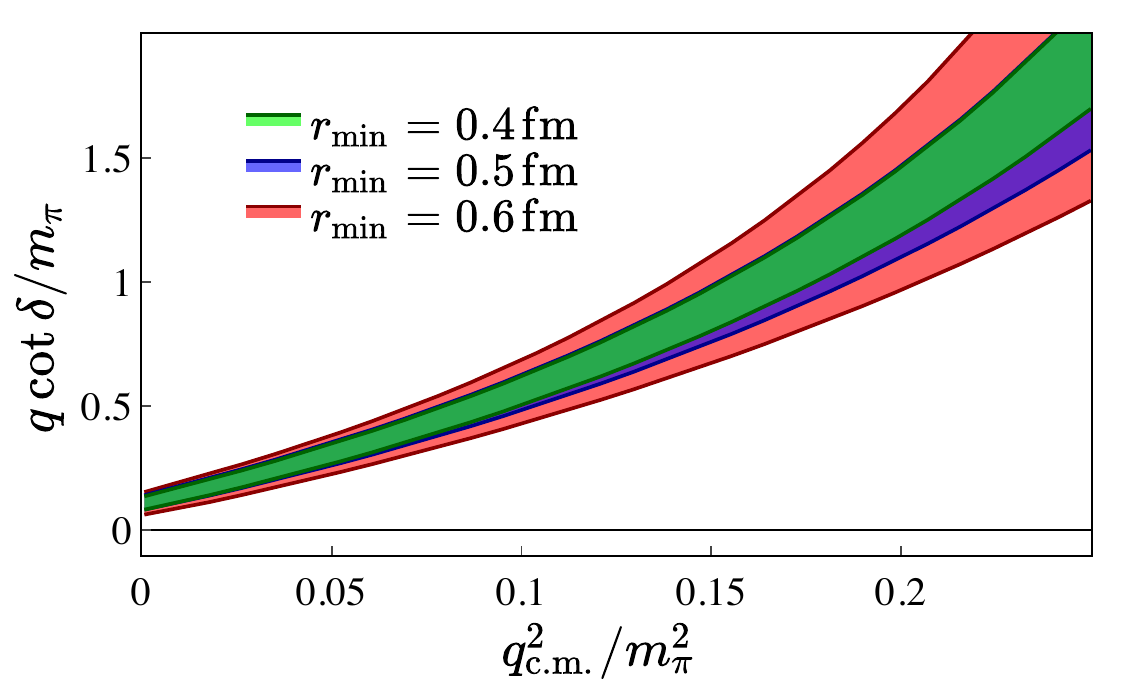}
\includegraphics[width=\columnwidth]{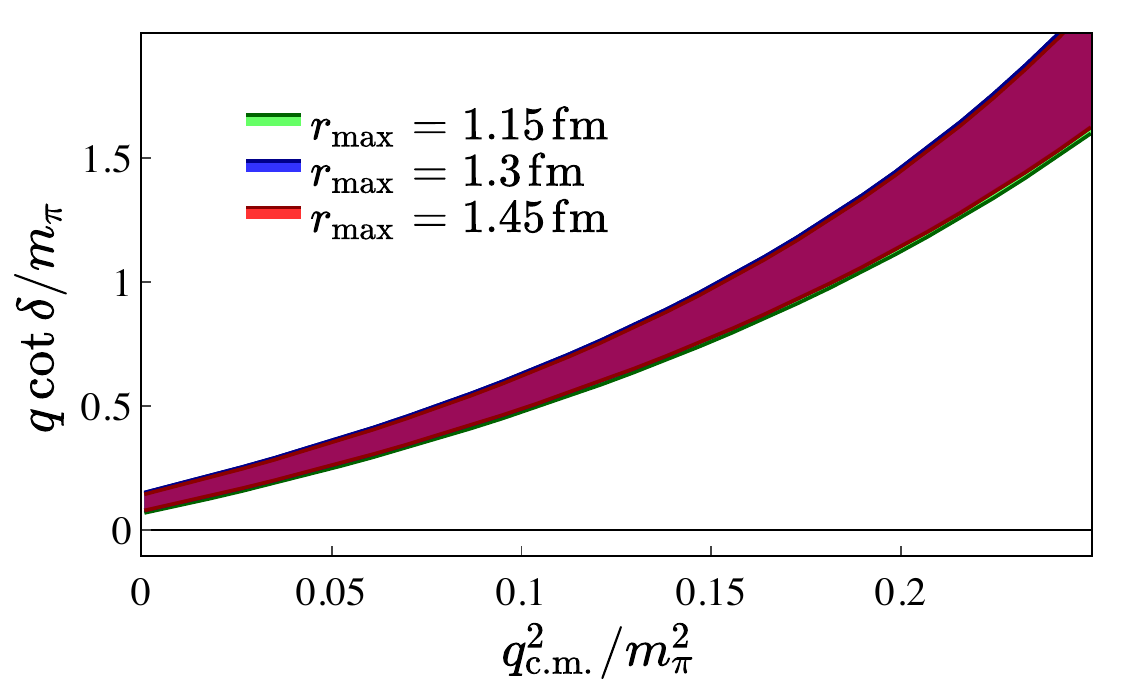}
\includegraphics[width=\columnwidth]{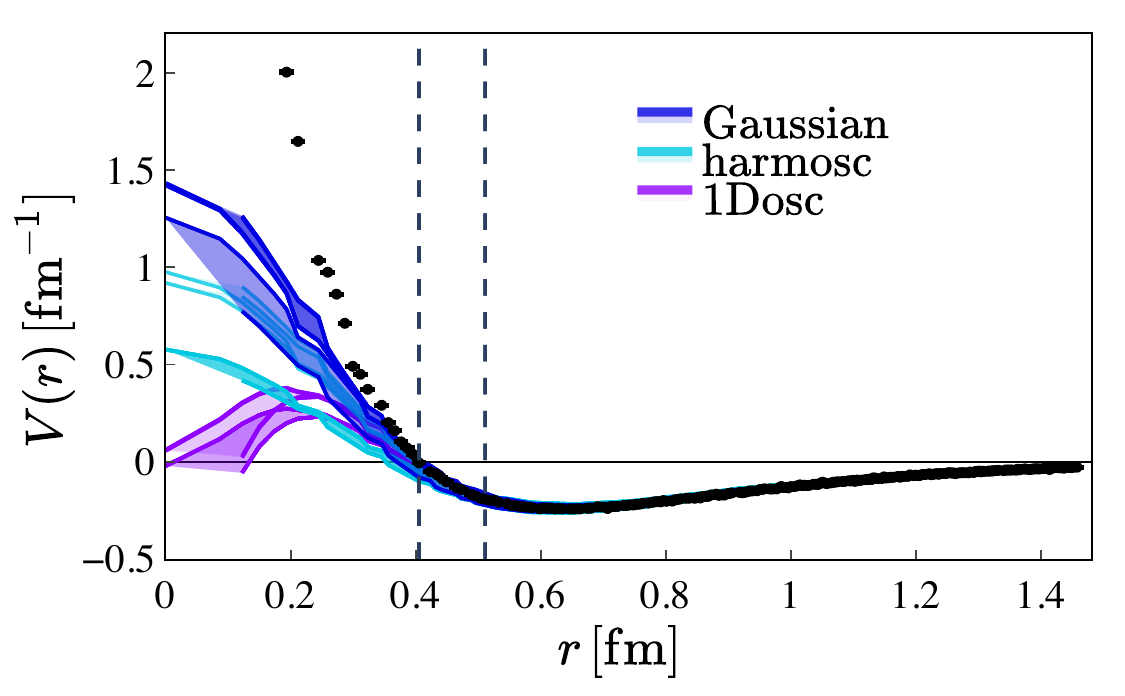}
\includegraphics[width=\columnwidth]{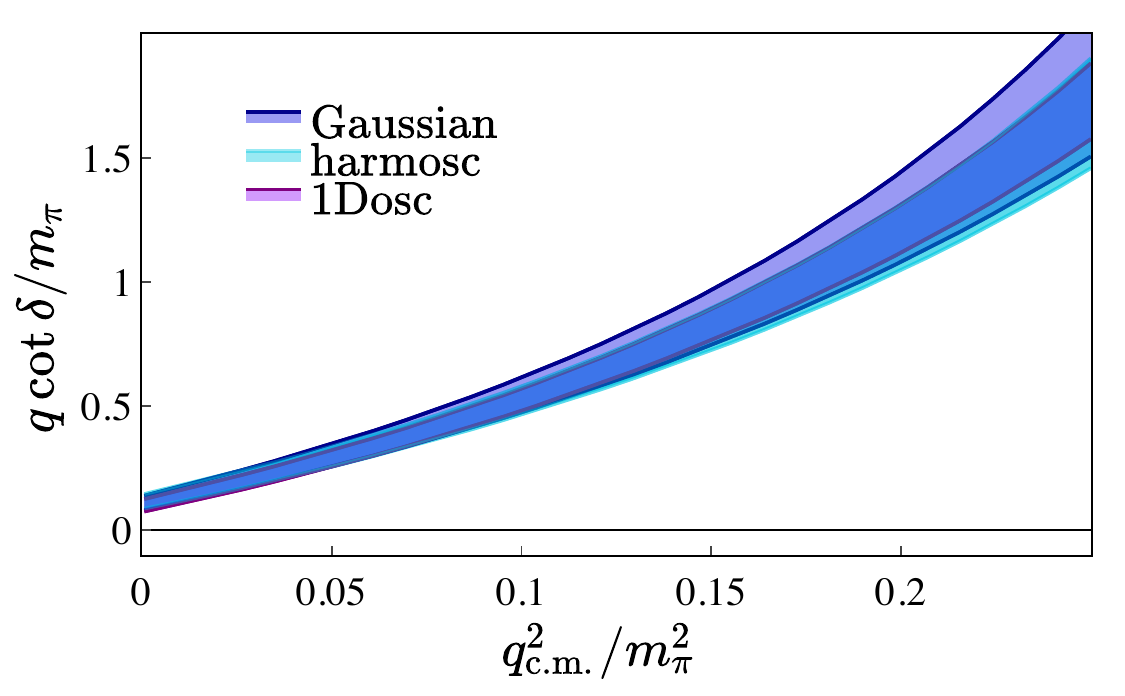}
\includegraphics[width=\columnwidth]{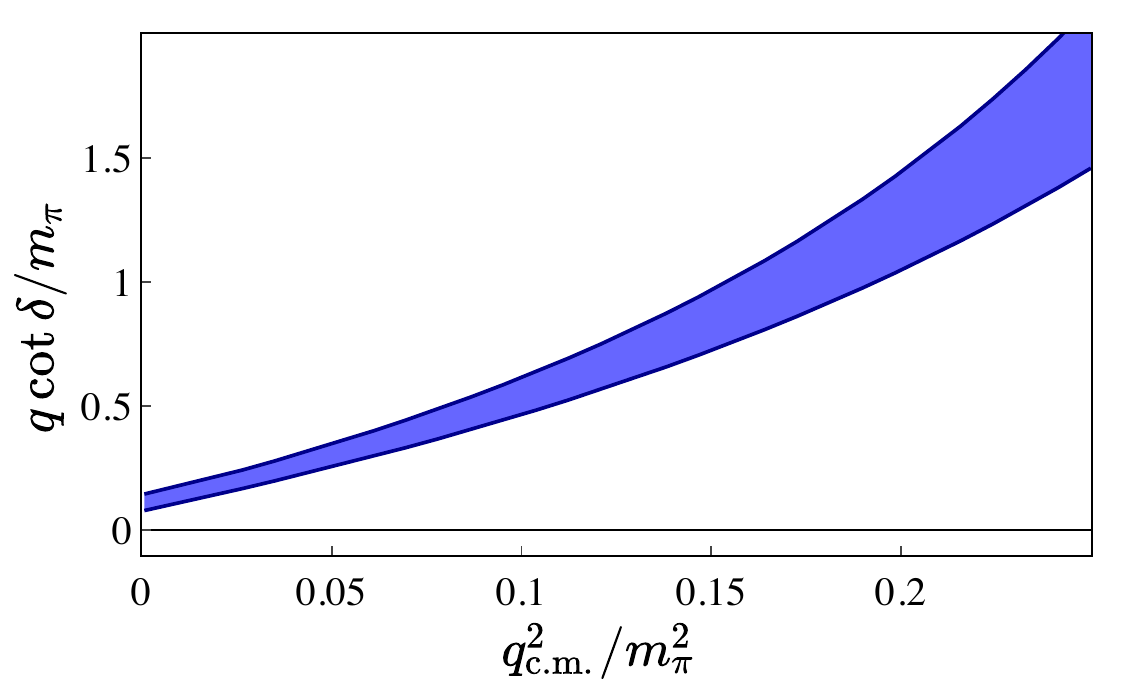}
\caption{\label{fig:hal_dineutron_plots}We show the resulting potential and phase shift analysis as in Figs.~\ref{fig:pot_t}, \ref{fig:phase_t}, \ref{fig:phase_rmin}, \ref{fig:phase_rmax}, \ref{fig:pot_fits}, and \ref{fig:phase_tot}, except for the di-neutron channel.}
\end{figure*}

\clearpage
\subsection{Constructing the potential \label{sec:make_hal_potential}}

The code we use to compute the HAL QCD potential, the LALIBE library~\cite{lalibe}, branch \texttt{feature/mp\_nn}, produces NN correlation functions in relative momentum space with $\mathbf{P}_{\rm tot}=0$.  As discussed in the main text, they were generated with zero-momentum wall sources for each nucleon.

\subsubsection{$A_{1g}$ projection}
We first need to eliminate all partial waves other than $\ell=0$.  Most of them can be quickly removed by projection to $A_{1g}$, leaving only $\ell=0,4,8,10,\ldots$.
We perform the projection on the momentum space correlation function 
 in a two step process by averaging over all the symmetry transformations on the data that leave an element of $A_{1g}$ invariant:
\begin{lstlisting}[language=Python]
def A1g_projection(NN):
    # pseudo code
    # positive parity projection
    NN_g  = NN( p_x, p_y, p_z)
    NN_g += NN(-p_x, p_y, p_z)
    NN_g += NN( p_x,-p_y, p_z)
    NN_g += NN( p_x, p_y,-p_z)
    NN_g += NN(-p_x,-p_y, p_z)
    NN_g += NN(-p_x, p_y,-p_z)
    NN_g += NN( p_x,-p_y,-p_z)
    NN_g += NN(-p_x,-p_y,-p_z)

    NN_g  = NN_g / 8.0

    # equivalent axes swap
    # 0=p_x, 1=p_y, 2=p_z
    NN_A1g  = NN_g
    NN_A1g += NN_g.swap(0,1)
    NN_A1g += NN_g.swap(1,2)
    NN_A1g += NN_g.swap(2,0)
    NN_A1g += (NN_g.swap(0,1)).swap(1,2)
    NN_A1g += (NN_g.swap(0,2)).swap(2,1)

    NN_A1g  = NN_A1g / 6.0

    return NN_A1g
\end{lstlisting}
% Given this $A_{1g}$ projected full momentum space correlation function, the HAL QCD potential in coordinate space is determined by performing a Fourier Transform back to relative coordinate space to produce the numerical data for \eqnref{eq:CNN_r}.
Later steps in the extraction of the HAL QCD potential require a position space representation \eqnref{eq:CNN_r} which is produced by a Fourier Transform back to relative coordinate space.
This correlation function is then divided by the single nucleon one created with the same wall-source and projected to zero momentum at the sink, creating \eqnref{eq:R_t_r}, which is then fed into \eqnref{eq:time_dependent_HAL} to solve for the central potential, \eqnref{eq:V_central}.

\subsubsection{ Projection to $\ell=0$}
We initially expected that $\ell=4,6,8,\ldots$ components of the $A_{1g}$ projected $R(t,r)$ would be small, but decided to perform a projection to $\ell=0$ to be sure. 

Our projection method is to first create an interpolator from the lattice values of $R(t,r)$ using the python function \texttt{scipy.interpolate.RegularGridInterpolator}.   Then we perform integration on spherical shells against $Y^\ell_m(\hat{r})$ to extract the radial function in the partial wave expansion at a set of radii spaced at $a/2$.   The spherical integration is implemented using an order 9 Lebedev quadrature rule provided by \texttt{scipy.integrate.lebedev{\textunderscore}rule}.
Order 9 means that it will exactly integrate up to an $\ell{=}9$  $Y^\ell_m(\hat{r})$, eliminating $\ell=4,6,8$ components exactly.

A 1D interpolator is then constructed from the radial function samples and used to update the $R(t,r)$ values on the lattice sites.  The update happens in a ball that does not reach the faces of the volume.  Later steps will not use data in the corners of the volume.

This projection method is equivalent to the Misner method~\cite{Misner:1999ab} used by the HAL QCD Collaboration~\cite{Miyamoto:2019jjc}.

\begin{figure*}
\includegraphics[width=\columnwidth]{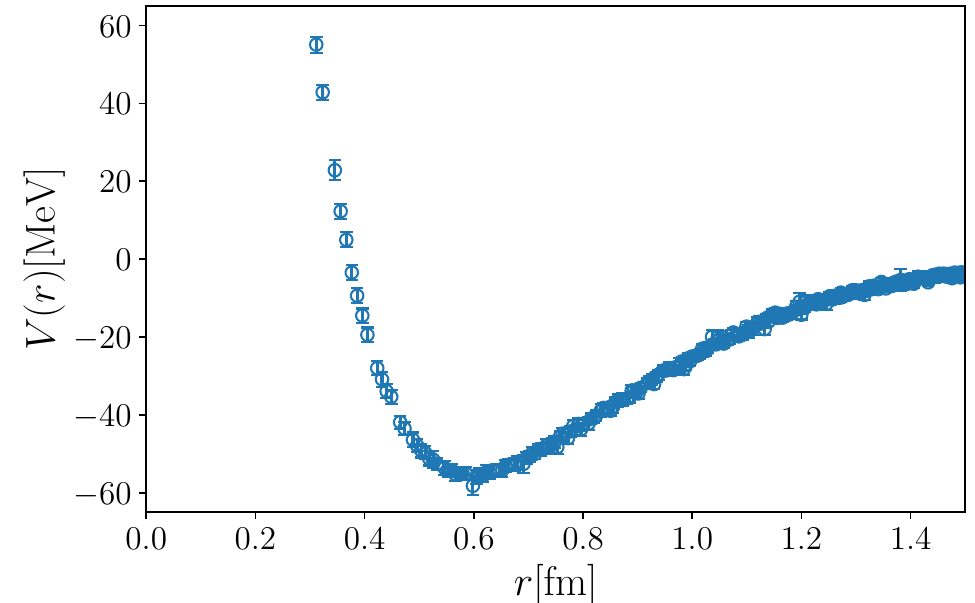}
\includegraphics[width=\columnwidth]{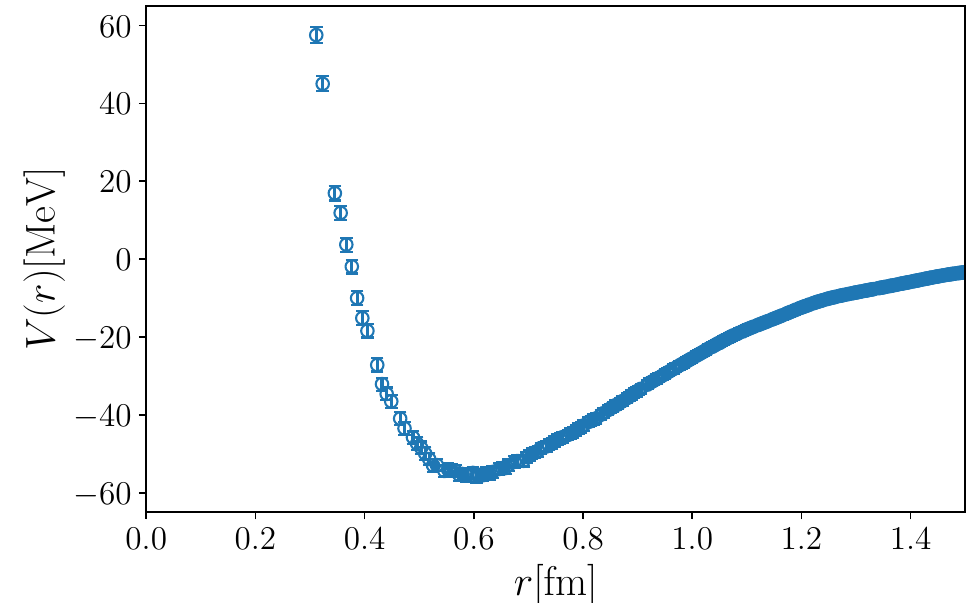}
\caption{\label{fig:hal_swave}
The potential at $t=12$ without (left) and with (right) the $\ell=0$ projection described in the text.
}
\end{figure*}

The observed variation between the $\ell=0$ projected $R(t,r)$ and the $A_{1g}$ projected version it was derived from was less than $1\%$ and we observed no significant changes in the resulting phase shifts derived from the potential.  Using the $\ell=0$ projection  slightly reduces uncertainties from small changes in r that require large changes in direction to remain the lattice.  Essentially the small $\ell=4$ contribution acts as noise.
\figref{fig:hal_swave} shows the potential at $t=12$ with and without the $\ell=0$ projection.

\subsubsection{Improved Time Derivatives}
Last, we provide detail on the extraction of the time derivatives needed for \eqnref{eq:time_dependent_HAL}.  At each time and space point we sampled $R(t+\Delta t, r)$ with $\Delta t$ from $\left\{-2, -1, 0, 1, 2 \right\}$. This 5-element vector is called $\mathbf{v}$.  We then fit a quartic polynomial with 5 coefficients $\mathbf{c}$ to reproduce $\mathbf{v}$.
Let
\begin{equation}
\mathbf{M} = \left( \begin{array}{rrrrr}
1 & {-2} & 4 & -8 & 16 \\
1 & {-1} &  1 & -1 & 1 \\
1 & 0 & 0 & 0 & 0 \\
 1 & \space\space 1 & \space\space 1 & \space\space 1 & \space\space 1 \\
1 & 2 & 4 & 8 & 16
\end{array} \right).
\end{equation}
Each row of $\mathbf{M}$ contains powers of $\Delta t$ begining with $-2$ in the top row, eg. $M_{0i}=(-2)^i$, $M_{1i}=(-1)^i$, etc.  Then
\begin{equation}
    \mathbf{M} \mathbf{c} = \mathbf{v}.
\end{equation}
Each row of $\Delta t$ powers is dotted with the as yet unknown coefficients $\mathbf{c}$ to produce the samples $R(t+\Delta t, r)$ in $\mathbf{v}$.  Multiplication on both sides by $\mathbf{M}^{-1}$ produces the coefficients
\begin{equation}
    \mathbf{c} = \mathbf{M}^{-1} \mathbf{v}.
\end{equation}
The improved derivatives implemented in the code can be determined from derivatives of the polynomial evaluated at $\Delta t=0$.
\begin{equation}
\begin{aligned}
-\partial_t R(t)  &= -c_1 + \mathcal{O}\left(a^4\right) \\
 &= \frac{1}{12}\left[R(t+2) - R(t-2)\right]\\
 &\quad -\frac{2}{3}\left[R(t+1)-R(t-1)\right] +\mathcal{O}(a^4)\\
\partial_t^2 R(t) &=  2 c_2 +\mathcal{O}\left(a^4\right)\\
    &= -\frac{1}{12}\left[R(t+2)+R(t-2)\right]
        - \frac{5 R(t)}{2}\\
    &\quad +\frac{4}{3}\left[R(t+1)+R(t-1)\right]+\mathcal{O}\left(a^4\right)
\end{aligned}
\end{equation}
The spacing dependence of the error for the second derivative benefits from a symmetry induced cancellation, yielding the same order error as the first derivative.

\clearpage
\bibliography{su3}

\end{document}

%% file: preamble.tex
\usepackage{graphicx}  % needed for figures
\usepackage{dcolumn}   % needed for some tables
\usepackage{bm}        % for math
\usepackage{amssymb}   % for math
\usepackage{standalone}
\usepackage{enumitem}
\usepackage[pdftex]{color}
\usepackage[utf8]{inputenc}
\usepackage{xcolor}
\usepackage{slashed}
\usepackage{booktabs}
\usepackage{multirow}
\usepackage{amsmath}
\usepackage{bbm}
\usepackage{stackrel}
\usepackage{rotating}
\usepackage{CJKutf8}
\usepackage{pifont}
\usepackage{mathtools}
\usepackage[caption=false]{subfig}
\usepackage{listings}
\usepackage{hyperref}
\hypersetup{
    colorlinks=true,       % false: boxed links; true: colored links
    linkcolor=blue,          % color of internal links
    citecolor=blue,        % color of links to bibliography
    filecolor=blue,      % color of file links
    urlcolor=blue           % color of external links
}
\usepackage{simplewick}
\usepackage{float} % Forces placement of figures
\usepackage{tikz} % adds /foreach (looping) command
\usepackage{xspace}
\usepackage[normalem]{ulem}

\usepackage{comment}

\allowdisplaybreaks

\AtBeginDocument{%
    \newwrite\bibnotes
    \def\bibnotesext{Notes.bib}
    \immediate\openout\bibnotes=\jobname\bibnotesext
    \immediate\write\bibnotes{@CONTROL{REVTEX41Control}}
    \immediate\write\bibnotes{@CONTROL{%
    apsrev41Control,author="08",editor="1",pages="1",title="0",year="1"}}
     \if@filesw
     \immediate\write\@auxout{\string\citation{apsrev41Control}}%
    \fi
}%

%% file: def.tex
\def\eqref#1{{(\ref{#1})}}
\newcommand{\eqnref}[1]{Eq.~\eqref{#1}}

\newcommand{\figref}[1]{Fig.~\ref{#1}}

\newcommand{\secref}[1]{Sec.~\ref{#1}}
\newcommand{\appref}[1]{App.~\ref{#1}}
\newcommand{\tabref}[1]{Table~\ref{#1}}

\definecolor{kngrey}{HTML}{A6AAA9}
\definecolor{knred}{HTML}{EC5D57}
\definecolor{knorange}{HTML}{F39019}
\definecolor{knyellow}{HTML}{F5D328}
\definecolor{kngreen}{HTML}{70BF41}
\definecolor{knblue}{HTML}{51A7F9}
\definecolor{knpurple}{HTML}{B36AE2}

\def\d{{\delta}}
\def\D{{\Delta}}

\def\l{{\lambda}}

\def\O{{\Omega}}

\def\s{{\sigma}}

\def\luscher{{QC2}\xspace}

\newcommand{\be}{\begin{equation}}
\newcommand{\ee}{\end{equation}}
\newcommand{\bee}{\begin{equation*}}
\newcommand{\eee}{\end{equation*}}
\newcommand{\bea}{\begin{eqnarray}}
\newcommand{\eea}{\end{eqnarray}}
\newcommand{\beaa}{\begin{eqnarray*}}
\newcommand{\eeaa}{\end{eqnarray*}}

%% file: affiliations.tex
\newcommand{\aachen}{
	Institute for Theoretical Particle Physics and Cosmology, 
	TTK, RWTH Aachen University, Germany
}
\newcommand{\bern}{
	Institute for Theoretical Physics, 
	Albert Einstein Center for Fundamental Physics, 
	University of Bern, Switzerland
}
\newcommand{\bochum}{
	Fakult\"at f\"ur Physik und Astronomie, 
	Institut f\"ur Theoretische Physik II, 
	Ruhr-Universit\"at Bochum, 44780 Bochum, Germany
}
\newcommand{\cmu}{
    Department of Physics, Carnegie Mellon University,
    Pittsburgh, Pennsylvania 15213, USA
}
\newcommand{\intelDEUTCH}{
	Intel Deutschland GmbH, 
	Dornacher Str. 1, 85622 Feldkirchen, Germany
}
\newcommand{\kentState}{
	Department of Physics, Kent State University, 
	Kent, OH 44242, USA
}
\newcommand{\lblnsd}{
    Nuclear Science Division,
    Lawrence Berkeley National Laboratory,
	Berkeley, CA 94720, USA
}
\newcommand{\lblnersc}{
    NERSC,
    Lawrence Berkeley National Laboratory,
    Berkeley, CA 94720, USA
}
\newcommand{\llnl}{
	Physics Division,
	Lawrence Livermore National Laboratory,
	Livermore, CA 94550, USA
}
\newcommand{\llnldesign}{
	Design Physics Division,
	Lawrence Livermore National Laboratory,
	Livermore, CA 94550, USA
	}

\newcommand{\nvidia}{
    NVIDIA Corporation,
    2701 San Tomas Expressway, Santa Clara, CA 95050, USA
    }
\newcommand{\ornl}{
	National Center For Computational Sciences
	Oak Ridge National Laboratory, 
	Oak Ridge, Tennessee, USA
}
\newcommand{\ucb}{
	Department of Physics,
	University of California,
	Berkeley, California 94720, USA
	}
\newcommand{\unc}{
	Department of Physics and Astronomy,
	University of North Carolina,
	Chapel Hill, NC 27516-3255, USA
	}